%% file: 0_main.tex
\documentclass[format=acmsmall, review=false, screen=true]{acmart}

\usepackage{booktabs} % For formal tables
\usepackage{multirow}
\usepackage{graphicx}
\usepackage{color}
\usepackage{url}
\usepackage{footnote}
\makesavenoteenv{tabular}
\makesavenoteenv{table}
\usepackage{hyperref}
\usepackage{amsmath}   % From the American Mathematical Society
\usepackage{array}
\usepackage{{inputenc}}
\usepackage{balance}
\usepackage{amsmath}
\usepackage{amsfonts}
\usepackage{subfigure}
\usepackage{enumitem}
\usepackage{ragged2e}
\usepackage{longtable}

\acmJournal{TOIS}
\setcopyright{rightsretained}
\acmDOI{10.475/123_4}

\newcommand{\ie}{\emph{i.e., }}
\newcommand{\eg}{\emph{e.g., }}
\newcommand{\etal}{\emph{et al. }}

\newcommand{\wrt}{\emph{w.r.t. }}

\newcommand{\aka}{\emph{aka. }}

\newcommand{\mb}[1]{\mathbf{#1}}
\newcommand{\mbb}[1]{\mathbb{#1}}
\newcommand{\mc}[1]{\mathcal{#1}}
\newcommand{\tb}[1]{\textbf{#1}}

\begin{document}
\title{Temporal Relational Ranking for Stock Prediction}
\author{Fuli Feng}
\affiliation{%
  \institution{National University of Singapore}
  \streetaddress{13 Computing Drive}
  \postcode{117417}
  \country{Singapore}}
\email{fulifeng93@gmail.com}

\author{Xiangnan He}
\affiliation{%
  \institution{University of Science and Technology of China}
  \streetaddress{443 Huangshan Road, Hefei}
  \postcode{230031}
  \country{China}
}
\email{xiangnanhe@gmail.com}

\author{Xiang Wang}
\affiliation{%
 \institution{National University of Singapore}
 \streetaddress{13 Computing Drive}
 \postcode{117417}
 \country{Singapore}
}
\email{xiangwang@u.nus.edu}

\author{Cheng Luo}
\affiliation{%
  \institution{Tsinghua University}
  \streetaddress{30 Shuangqing Rd}
  \city{Haidian}
  \state{Beijing}
  \country{China}
}
\email{chengluo@tsinghua.edu.cn}

\author{Yiqun Liu}
\affiliation{%
  \institution{Tsinghua University}
  \streetaddress{30 Shuangqing Rd}
  \city{Haidian}
  \state{Beijing}
  \country{China}
}
\email{yiqunliu@tsinghua.edu.cn}

\author{Tat-Seng Chua}
\affiliation{%
  \institution{National University of Singapore}
  \streetaddress{13 Computing Drive}
  \postcode{117417}
  \country{Singapore}
}
\email{dcscts@nus.edu.sg}

\begin{abstract}
Stock prediction aims to predict the future trends of a stock
in order to help investors to make good investment decisions. 
%as the reference of their investments.
Traditional solutions for stock prediction are based on time-series models. 
%With the ability to represent large amount of factors, deep neural networks have recently become promising solutions for stock prediction. 
With the recent success of deep neural networks in modeling sequential data, deep learning has become a promising choice for stock prediction.

However, most existing deep learning solutions are not optimized towards the target of investment, \ie selecting the best stock with highest expected revenue. Specifically, they typically formulate stock prediction as a classification (to predict stock trend) or a regression problem (to predict stock price). More importantly, they largely treat the stocks as independent of each other. The valuable signal in the rich relations between stocks (or companies), such as two stocks are in the same sector and two companies have a supplier-customer relation, is not considered.

%ignore the valuable signal in the stock (or company) relations.
%It is straightforward to apply graph-based learning techniques for encoding such stock relations. However, existing graph-based learning techniques tends to miss the temporal evolution properties of these relations.

In this work, we contribute a new deep learning solution, named \textit{Relational Stock Ranking} (RSR), for stock prediction. Our RSR method advances existing solutions in two major aspects: 1) tailoring the deep learning models for stock ranking, and 2) capturing the stock relations in a time-sensitive manner. The key novelty of our work is the proposal of a new component in neural network modeling, named \textit{Temporal Graph Convolution}, which jointly models the temporal evolution and relation network of stocks. To validate our method, we perform back-testing on the historical data of two stock markets, NYSE and NASDAQ. Extensive experiments demonstrate the superiority of our RSR method. It outperforms state-of-the-art stock prediction solutions achieving an average return ratio of $98\%$ and $71\%$ on NYSE and NASDAQ, respectively. 

%In this work, we address these limitations by formulating stock prediction as a ranking task and proposing an end-to-end solution, named \textit{Relational Stock Ranking}, which encodes the temporal stock relations with a \textit{Temporal Graph Convolutional Operation}. Specifically, we feed the sequential data through an embedding layer and revise the sequential embeddings with temporal stock relations for predicting the ranking score of stocks. 
%We apply our proposed method to predict ranking list of stocks in two markets, New York Stock Exchange (NYSE) and NASDAQ Stock Market (NASDAQ).
%Extensive experiments demonstrate the superiority of our proposed relational stock ranking, which consistently improve over conventional solutions.
\end{abstract}




\maketitle

\input{1_introduction}
\input{3_preliminary}
\input{4_method}
\input{5_data}
\input{6_experiment}
\input{2_related}
\input{7_conclusion}
\input{8_appendix}

\bibliographystyle{ACM-Reference-Format}
\bibliography{sigproc}

\end{document}

%% file: 1_introduction.tex
\section{Introduction}
According to the statistics reported by the World Bank in 2017, the overall capitalization of stock markets worldwide has exceeded 64 trillion U.S. dollars\footnote{\href{https://data.worldbank.org/indicator/CM.MKT.LCAP.CD}{https://data.worldbank.org/indicator/CM.MKT.LCAP.CD/.}}. With the continual increase in stock market capitalization, trading stocks has become an attractive investment instrument for many investors. However, whether an investor could earn or lose money depends heavily on whether he/she can make the right stock selection. Stock prediction, which aims to predict the future trend and price of stocks, is one of the most popular techniques to make profitable stock investment \cite{preethi2012stock}, although there are still debates about whether the stock market is predictable (\aka the Efficient Markets Hypothesis) among financial economists \cite{musgrave1997random, lo2002non}. Some recent evidences indicate the predictability of stock markets, which motivates further exploration of stock prediction techniques \cite{tu2016investment, li2016tensor, zhang2017stock, hu2018listening,schumaker2009textual}.

Traditional solutions for stock prediction are based on time-series analysis models, such as Kalman Filters \cite{xu2015application}, Autoregressive Models and their extensions \cite{adebiyi2014comparison}. Given an indicator of a stock (\eg stock price), this kind of models represents it as a stochastic process and takes the historical data of the indicator to fit the process. We argue that such mainstream solutions for stock prediction have three main drawbacks: 1) The models heavily rely on the selection of indicators, which is usually done manually and is hard to optimize without special knowledge of finance. 2) The hypothesized stochastic processes are not always compatible with the volatile stock in the real world. 3) These models can only consider a few indicators since their inference complexity typically increases exponentially with the number of indicators. As such, they lack the capability to comprehensively describe a stock that could be influenced by a plethora of factors. Towards these drawbacks, advanced techniques like deep neural networks, especially the recurrent neural networks (RNNs), have become a promising solution to substitute the traditional time-series models to predict the future trend or exact price of a stock \cite{bao2017deep, zhao2017constructing, zhang2017learning, zhang2017stock}.
%\begin{table}[t]
%	\centering
%	\caption{An example that two methods with the same stock classification and regression performance lead to different stock selections. Method 1 selects stock A and Method 2 selects stock B.}
%	\vspace{-0.3cm}
%	\label{tab:formulation}
%	\resizebox{0.77\textwidth}{!}{%
%	\begin{tabular}{|c||c|c|c|c||c|c|c|c|}
%		\hline
%		%\multirow{2}{*}{} & \multicolumn{3}{c|}{Predictions of Method 1} & \multirow{2}{*}{Performance} & \multicolumn{3}{c|}{Predictions} & \multirow{2}{*}{Performance} \\ \cline{2-4} \cline{6-8}
%		%& A & B & C &  & A & B & C &  \\ \hline
%		\multirow{3}{*}{} & \multicolumn{4}{c||}{Method 1} & \multicolumn{4}{c|}{Method 2} \\ \cline{2-9}
%		& \multicolumn{3}{c|}{Predictions} & \multirow{2}{*}{Performance} & \multicolumn{3}{c|}{Predictions} & \multirow{2}{*}{Performance} \\ \cline{2-4} \cline{6-8}
%		& A & B & C &  & A & B & C &  \\ \hline \hline
%		Classification & $\mathbf{\uparrow}$ & $\downarrow$ & $\downarrow$ & 0.67 (Acc.) & $\downarrow$ & $\mathbf{\uparrow}$ & $\downarrow$ & 0.67 (Acc.) \\ \hline
%		Regression & \textbf{+5} & -1 & -5 & 4.33 (MSE) & -1 & \textbf{+3} & -5 & 4.33 (MSE) \\ \hline
%		Ground Truth & \textbf{+2} & +1 & -5 & - & \textbf{+2} & +1 & -5 & - \\ \hline
%	\end{tabular}
%	}
%	\vspace{+2pt}
%	\justify
%	\scriptsize{A, B, C denote three stocks; $\uparrow$/$\downarrow$ denotes increase/decrease; numbers (+2) are the true/predicted price change of stocks; values in bold correspond to suggested selections.}
%	%\vspace{-0.1cm}
%\end{table}
\begin{table}[]
	\caption{{\color{blue}An intuitive example that one method predicting the price change of stocks more accurately (\ie smaller MSE) leads to a less profitable stock selection (\ie smaller profit). Method 1 selects stock A (30) while Method 2 selects stock B (10).}}
	\vspace{-0.3cm}
	\label{tab:formulation}
	%\resizebox{0.8\textwidth}{!}{%
		\begin{tabular}{|c|c|c||c|c|c|c|c||c|c|c|c|c|}
			\hline
			\multicolumn{3}{|c|}{\multirow{2}{*}{Ground Truth}} & \multicolumn{5}{c|}{Method 1} & \multicolumn{5}{c|}{Method 2} \\ \cline{4-13} 
			\multicolumn{3}{|c|}{} & \multicolumn{3}{c|}{Prediction} & \multicolumn{2}{c|}{Performance} & \multicolumn{3}{c|}{Prediction} & \multicolumn{2}{c|}{Performance} \\ \hline
			A & B & C & \textbf{A} & B & C & MSE & Profit & A & \textbf{B} & C & MSE & Profit \\ \hline
			+30 & +10 & -50 & \textbf{+50} & -10 & -50 & 266 & 30 & +20 & \textbf{+30} & -40 & 200 & 10 \\ \hline
		\end{tabular}%
	%}
	\vspace{+2pt}
	\justify
	\scriptsize{A, B, C denote three stocks; numbers (+20) are the true/predicted price change of stocks; values in bold correspond to suggested selections.}
\end{table}

% Deep neural networks have achieved great performance on stock prediction, for instance, the State Frequency Memory (SFM) network \cite{zhang2017stock} can accurately predict the daily opening price of fifty U.S. stocks one day ahead with an mean square error (MSE) of less than six dollars. 
A state-of-the-art neural network-based solution is the State Frequency Memory (SFM) network \cite{zhang2017stock}, which models the historical data in a recurrent fashion and captures temporal patterns in different frequencies. This method achieves promising performance of predicting the daily opening price of fifty U.S. stocks one day ahead with a mean square error (MSE) of less than six dollars. 
However, we argue that such prediction methods are suboptimal to guide stock selection, since their optimization target is not at selecting the top stocks with the highest expected revenue. 
%the optimization target of most existing neural network-based solutions is not on selecting the best stocks with the highest expected revenue. 
%Because these solutions formulate 
To be specific, they typically address 
stock prediction as either a classification (on price movement direction) or a regression (on price value) task, 
which would cause a large discrepancy on the investment revenue. {\color{blue}Table \ref{tab:formulation} gives an intuitive example, where a method with better prediction performance (measured by regression MSE) suggests a less profitable stock. This 
implies the possible discrepancy between the actual target of stock selection and the optimized target of regression (classification), such that an optimal method of regression (classification) does not necessarily select the optimal stock to trade.} 
%points to an inherent limitation of formulating the stock prediction task as a classification or regression problem.

Another limitation of existing neural network-based solutions is that they typically treat stocks as independent of each other and ignore the relations between stocks. However, the rich relations between stocks and the corresponding companies may contain valuable clues for stock prediction. For example, stocks under the same sector or industry like GOOGL (Alphabet Inc.) and FB (Facebook Inc.) might have similar long-term trends. Besides, the stock of a supplier company might impact the stock of its consumer companies especially when a scandal of the supplier company is reported, such as the falsification of product quality data. To integrate stock relations into prediction, an intuitive solution is to represent the stock relations as a graph and then regularize the prediction of stocks based on the graph (\ie graph-based learning) \cite{feng2017computational, omari2016novelty, kipf2017semi, jiang2016learning}. However, conventional graph learning techniques cannot capture the temporal evolution property of stock markets (\eg the strength of influence between two given stocks may vary quickly), since the graph is fixed at a particular time.
%As such, it is suboptimal to simply account for stock relations using conventional graph-based learning.

\begin{figure}[]
	\centering
	\includegraphics[width=0.48\textwidth]{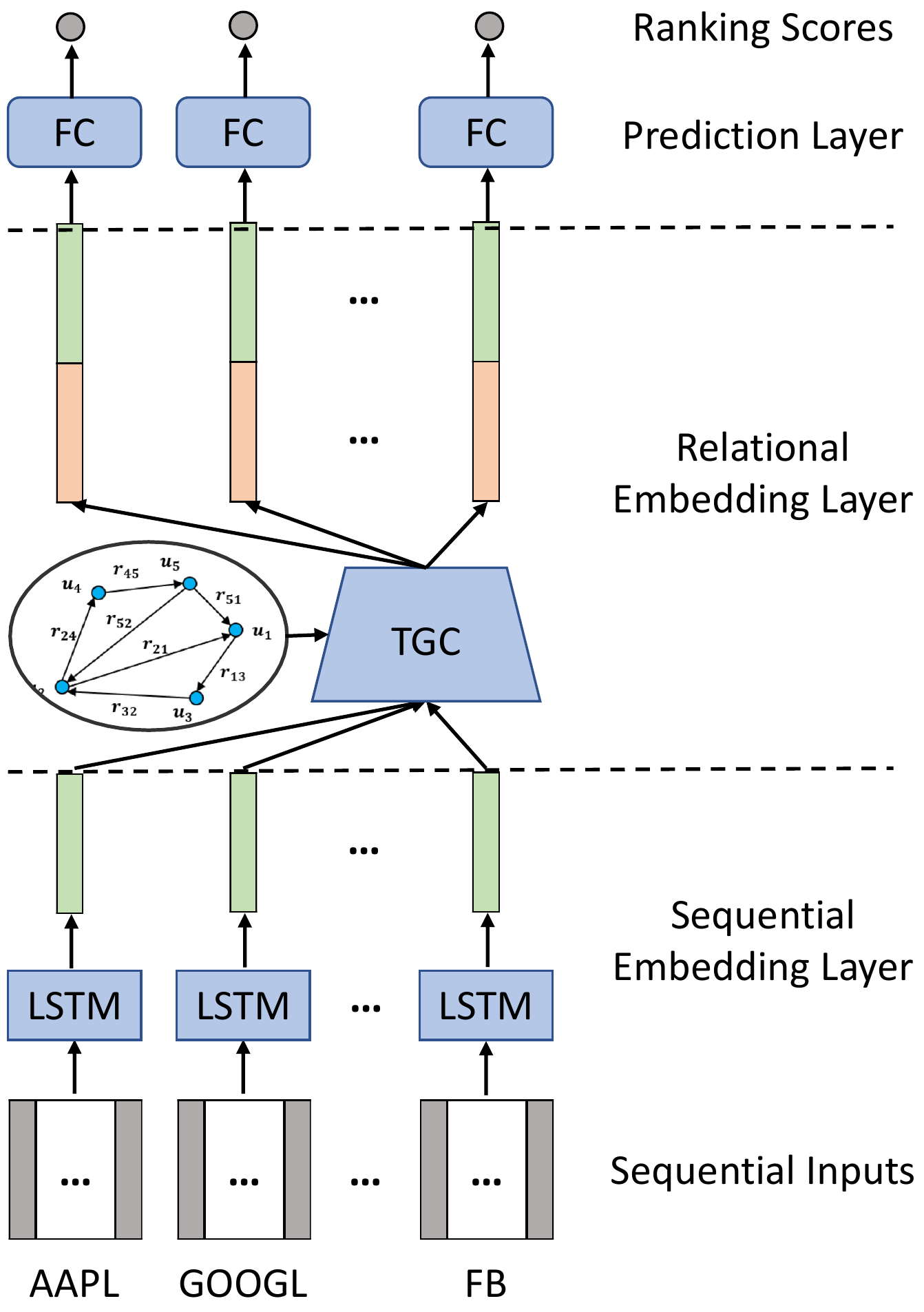} 
	\vspace{-0.3cm}
	\caption{Relational stock ranking framework. It should be noted that the LSTM cells and FC units (Fully Connected layer) depicted in the same layer share the same parameters.
	}%\vspace{-10pt}
	\label{fig:framework}
\end{figure}
To address the aforementioned limitations of existing solutions, we formulate stock prediction as a ranking task, for which the target is to directly predict a stock list ranked by a desired criteria like return ratio. We then propose an end-to-end framework, named \textit{Relational Stock Ranking} (RSR), to solve the stock ranking problem. An illustration of our framework can be found in Figure \ref{fig:framework}. Specifically, we first feed the historical time series data of each stock to a Long Short-Term Memory (LSTM) network to capture the sequential dependencies and learn a stock-wise sequential embedding. By devising a new \textit{Temporal Graph Convolution} (TGC), we next revise the sequential embeddings by accounting for stock relations in a time-sensitive way. Finally, we feed the concatenation of sequential embeddings and relational embeddings to a fully connected layer to obtain the ranking score of stocks. 
% and generate a ranking list according to the ranking scores. 
To justify our proposed method, we employ it on two real-world markets, New York Stock Exchange (NYSE) and NASDAQ Stock Market (NASDAQ). Extensive back-testing results demonstrate that our RSR significantly outperforms SFM \cite{zhang2017stock} with more than 115\% improvements in return ratio.

The key contributions of the paper are summarized as follows. 
\begin{itemize}
	\item We propose a novel neural network-based framework, named \textit{Relational Stock Ranking}, to solve the stock prediction problem in a learning-to-rank fashion.
	\item We devise a new component in neural network modeling, named \textit{Temporal Graph Convolution}, to explicitly capture the domain knowledge of stock relations in a time-sensitive manner.
	% \item To the best of our knowledge, this is the first that formulates stock prediction as a ranking problem and demonstrates the potential of learning-to-rank methods for predicting stocks, which is a well-motivated and highly profitable application.
	% \item We propose a novel \textit{relational stock ranking} framework to rank stocks by considering stock relations with a new \textit{Temporal Graph Convolutional Operation}.
	\item We empirically demonstrate the effectiveness of our proposals on two real-world stock markets, NYSE and NASDAQ.
\end{itemize}

The remainder of this paper is organized as follows. Section~\ref{sec:preliminary} introduces the preliminary knowledge about LSTM and graph-based learning, which forms the building blocks of our method. Section~\ref{sec:method} presents our proposed RSR. Section~\ref{sec:data} and~\ref{sec:experiment} describe the datasets and experiment, respectively. In Section~\ref{sec:related}, we review related work, followed by conclusion in Section~\ref{sec:conclusion}.

%% file: 3_preliminary.tex
\section{Preliminaries}
\label{sec:preliminary}
In this paper, we use bold capital letters (\eg $\mathbf{X}$), bold lowercase letters (\eg $\mathbf{x}$), and capital {\color{blue}script letters} (\eg $\mathcal{X}$) to denote matrices, vectors, and tensors, respectively. 
Scalars and hyperparameters are respectively represented as normal lowercase letters (\eg $x$) and Greek letters (\eg $\lambda$). If not otherwise specified, all vectors are in a column form, and $X_{ij}$ denotes the entry at the $i$-th row and the $j$-th column of $\mathbf{X}$.
The symbols $\sigma$, $tanh$, and $\odot$ stand for the \textit{sigmoid} function, hyperbolic tangent function, and element-wise production operation, respectively.

\subsection{Long Short-Term Memory}
\label{ss:lstm}
LSTM \cite{hochreiter1997long} networks have been widely used to process sequential data, such as natural language \cite{yan2016learning}, voice \cite{graves2013speech}, and video \cite{srivastava2015unsupervised}. LSTM is a special kind of Recurrent Neural Networks (RNNs) \cite{goller1996learning} that evolve hidden states through time to capture the sequential pattern of input data, \eg the dependency between words in a sentence. Compared to the vanilla RNN, which is known to suffer from vanishing gradients while trained with Back-Propagation Through Time (BPTT), LSTM adds cell states to store the long-term memory and capture the long-term dependency in a sequence\footnote{{\color{blue}Detailed illustration of LSTM and its comparison against vanilla RNN are referred to: \url{http://colah.github.io/posts/2015-08-Understanding-LSTMs/.}}}. 

Before providing the specific formulation of LSTM, we first describe the terms associated with LSTM. At each time-step $t$, $\mb{x^t} \in \mbb{R}^D$ denotes an input vector (\eg embedding vector of the $t$-th word in a given sentence), where $D$ is the input dimension.
Vectors $\mb{c^t}$ and $\mb{h^t} \in \mbb{R}^U$ denote the cell (memory) state vector and the hidden state vector, respectively, where $U$ is the number of hidden units.
Vector $\mb{z^t} \in \mbb{R}^{U}$ is an information transformation module. 
Vectors $\mb{i^t}$, $\mb{o^t}$, and $\mb{f^t} \in \mbb{R}^U$ denote the input, output, and forget gate, respectively. Formally, the transformation module, state vectors, and controlling gates are defined via the following equations:
\begin{equation}\label{eq:lstm}
    \begin{aligned}
	& \mb{z^t} = tanh (\mb{W_z x^t} + \mb{Q_z h^{t - 1}} + \mb{b_z}) \\
	& \mb{i^t} = \sigma (\mb{W_i x^t} + \mb{Q_i h^{t - 1}} + \mb{b_i}) \\
	& \mb{f^t} = \sigma (\mb{W_f x^t} + \mb{Q_f h^{t - 1}} + \mb{b_f}) \\
	& \mb{c^t} = \mb{f^t} \odot \mb{c^{t - 1}} + \mb{i^t} \odot \mb{z^t} \\
	& \mb{o^t} = \sigma(\mb{W_o x^t} + \mb{W_h h^{t - 1}} + \mb{b_o} ) \\
	& \mb{h^t} = \mb{o^t} \odot tanh(\mb{c^t}),
    \end{aligned}
\end{equation}
where $\mb{W_z}$, $\mb{W_i}$, $\mb{W_f}$, $\mb{W_o} \in \mbb{R}^{U \times D}$, and $\mb{Q_z}$, $\mb{Q_i}$, $\mb{Q_f} \in \mbb{R}^{U \times U}$ are mapping matrices; $\mb{b_z}$, $\mb{b_i}$, $\mb{b_f}$, and $\mb{b_o} \in \mbb{R}^U$ are bias vectors. The updating formulation can be understood as performing the following procedures: (1) calculate the information to be transformed from the input $\mb{x^t}$ to the memory states $\mb{c^t}$ by updating $\mb{z^t}$; 
(2) update the input gate $\mb{i^t}$ to control the information from $\mb{z^t}$ to $\mb{c^t}$; (3) update the forget gate $\mb{f^t}$ to decide how much information should be kept in the memory state; (4) refresh the memory state $\mb{c^t}$ by fusing the information flows from the input gate and memory gate; (5) update the output gate $\mb{o^t}$ to regulate the amount of information that can be outputted; (6) update the hidden state $\mb{h^t}$. As can be seen, the memory state $\mb{h^t}$ only has linear adding interactions, which allows the information to be unchanged during the {\color{blue}BPTT}. Benefited by the memory state, LSTM is capable of capturing the long-term dependency in the sequential data.

\subsection{Graph-based Learning}
\label{ss:pw_l2r}
Graph-based learning has been applied to various machine learning tasks to utilize entity relations~\cite{mei2014multimedia,yu2016survey,feng2017computational,aggarwal2013data,feng2018learning}. 
The general problem setting is to learn a prediction function $\mathbf{\hat{y}} = f(\mathbf{x})$, which maps an entity from the feature space to the target label space. It is usually achieved by minimizing an objective function abstracted as:
\begin{align}
	%\Gamma = \min \limits_{f} \sum \limits_{i = 1}^{N} loss(f(\mathbf{x_i})) +
	%\Gamma = \mathcal{L} + \lambda \mathcal{G},
	\Gamma = \Omega + \lambda \Phi,
	\label{eqn:hyperobj}
\end{align}
where $\Omega$ is a task-specific loss that measures the error between prediction $\mathbf{\hat{y}}$ and ground-truth $\mathbf{y}$, $\Phi$ is a graph regularization term that smooths the prediction over the graph, and $\lambda$ is a hyperparameter to balance the two terms. 
The regularization term typically implements the \textit{smoothness} assumption that similar vertices tend to have similar predictions. 
A widely used $\Phi$ is defined as:
\begin{align}
	\Phi = \sum_{i=1}^{N}\sum_{j=1}^N 
	\underbrace{g(\mathbf{x_i}, \mathbf{x_j})}_{\text{strength of smoothness}} 
	\underbrace{\left\|\frac{f(\mathbf{x_i})}{\sqrt{D_{ii}}}-\frac{f(\mathbf{x_j})}{\sqrt{D_{jj}}}\right\|^2}_{\text{smoothness}},
\end{align}
where $g(\mathbf{x_i}, \mathbf{x_j})$ is the similarity between the feature vectors of an entity pair (\eg the edge weight between the corresponding vertices); $D_{ii} = \sum_{j=1}^{N} g(\mathbf{x_i}, \mathbf{x_j})$ is the degree of vertex $i$. The regularization term operates smoothness on each pair of entities, enforcing their predictions (after normalized by their degrees) to be close to each other. The strength of smoothness is determined by the similarity over their feature vectors $g(\mathbf{x_i}, \mathbf{x_j})$. It can be equivalently written in a more concise matrix form: 
\begin{align}
	\mathcal{G} = trace(
		\mathbf{\hat{Y}} \mathbf{L} \mathbf{\hat{Y}}^{T}
	),
	\label{eqn:hyperreg}
\end{align}
where $\mathbf{\hat{Y}} = [\mathbf{\hat{y_1}}, \mathbf{\hat{y_2}}, \cdots, \mathbf{\hat{y_N}}]$, $\mathbf{L}$ is defined as $ \mathbf{L} = \mathbf{D}^{-1/2} (
\mathbf{D} - \mb{A}
) \mathbf{D}^{-1/2}$, also known as the \textit{graph Laplacian matrix}, and each element of $\mathbf{A}$ is $A_{ij} = g(\mathbf{x_i}, \mathbf{x_j})$.

\subsubsection{Graph Convolutional Networks}
\label{sss:gcn}
Graph Convolutional Network (GCN) is a special kind of graph-based learning methods, which integrates the core idea of graph-based learning (\ie the \textit{smoothness} assumption over graphs) with advanced convolutional neural networks (CNNs) \cite{kipf2017semi, defferrard2016convolutional, donnat2017spectral, hammond2011wavelets}. The core idea of standard CNNs \cite{krizhevsky2012imagenet} is using convolutions (\eg $3 \times 3$ filter matrices) to capture the local patterns in input data (\eg oblique lines in an image). Following the idea of CNNs, the aim of GCN is to capture the local connection patterns on graphs. However, intuitive solutions like directly applying convolution operations on the adjacency matrix of a graph are not feasible.
%XN: BUG I can not understand this sentence. please rephrase.
Because the filtering output of convolutions might change when we switch two rows of the adjacency matrix, while the switched adjacency matrix still represent the same graph structure. An alternative solution is to use spectral convolutions to capture the local connections in the Fourier domain, such as:
\begin{align}
	f(\mb{F}, \mb{X}) = \mb{UFU}^T \mb{X},
	\label{eqn:graph_conv}
\end{align}
where $f$ denotes the filtering operation of a convolution parameterized by a diagonal matrix $\mb{F}$, and $\mb{U}$ is the eigenvector matrix of the graph Laplacian matrix, \ie $\mb{L} = \mb{U \Lambda U}^T$.

Suffering from the overhead of computing the eigendecomposition of $\mb{L}$, it is suggested to treat $\mb{F}$ as a function of $\mb{\Lambda}$. Then it can be approximated by the Chebyshev polynomials $T_k(x)$ of up to the $K$-th order,
\begin{align}
	\mb{F} \approx \sum_{k = 0}^{K} \theta_k T_k(\hat{\Lambda}),
\end{align}
where $\hat{\Lambda} = \frac{2}{\lambda_{max}} \mb{\Lambda} - \mb{I}$ with $\lambda_{max}$ denotes the largest eigenvalue of $\mb{L}$; $\theta_k$ represents the Chebyshev coefficient; $T_k(x) = 2xT_{k - 1}(x) - T_{k - 2}(x)$ with $T_1{x} = x$ and $T_0{x} = 0$. 
In \cite{kipf2017semi}, the authors proved that the GCN performed well enough while setting $K$ to 1. As such, they reduced Equation (\ref{eqn:graph_conv}) to $f(\mb{F}, \mb{X}) = \mb{AX}$ and injected the convolution into a fully connected layer as $\mb{A}(\mb{XW} + \mb{b})$, which is the state-of-the-art formulation of GCN\footnote{Note that in the reduced form of GCN, the input diagonal matrix $\textbf{F}$ is omitted due to the Chebyshev approximation.}.

%% file: 4_method.tex
\section{Relational Stock Ranking}
\label{sec:method}
The typical problem setting of stock prediction (\ie price movement classification and price regression) is to learn a prediction function $\hat{y}^{t+1} = f(\mb{X^t})$ which maps a stock from the feature space to the target label space at time-step $t$. 
Matrix $\mb{X^t} = [\mb{x^{t - S + 1}}, \cdots, \mb{x^{t}}]^T \in \mbb{R}^{S \times D}$ represents the sequential input features, where $D$ is the dimension of features at each time-step and $S$ is the length of the sequence. Distinct from the typical problem setting of stock prediction, which treats different stocks as independent sequences, our target is to learn a ranking function $\mb{\hat{r}^{t+1}} = f(\mc{X}^t)$, which simultaneously maps a bunch of stocks to a ranking list. In the learned ranking list, stocks with higher ranking scores are expected to achieve higher investment revenue at time-step $t + 1$. Assuming we have $N$ stocks, then $\mc{X}^t \in \mbb{R}^{N \times S \times D} = [\mb{X_1^t}, \cdots, \mb{X_N^t}]^T$ is the collected features. In addition, we further associate the problem with a set of explicit stock relations (\eg supplier-consumer relations), which reflect the potential influence between different stocks. Given $K$ {\color{blue}types of relations}, we encode the pairwise relation between two stocks as a multi-hot binary vector $\mb{a_{ij}} \in \mbb{R}^K$ {\color{blue}and represent the relation of all stocks as a tensor $\mc{A} \in \mbb{R}^{N \times N \times K}$, of which the entry at the $i$-th row and $j$-th column is $\mb{a_{ij}}$.}
%(\ie $\mb{a_{ij}}$) denotes the relation between stock $i$ and $j$.}

In what follows, we first present the overall solution. We then elaborate our proposed \textit{Temporal Graph Convolution} for handling stock relations, followed by discussing its connections to existing graph-based learning methods. {\color{blue}In Table~\ref{tab:notations}, we summarize some of the terms and notations.}

\begin{table}[]
	\centering
	\caption{{\color{blue}Terms and notations.}}
	\vspace{-0.4cm}
	\label{tab:notations}
	\resizebox{\textwidth}{!}{%
		\begin{tabular}{|c|l|}
			\hline
			Symbol & Definition \\ \hline \hline
			$\mc{X}^t \in \mbb{R}^{N \times S \times D} = [\mb{X_1^t}, \cdots, \mb{X_N^t}]^T$ & historical prices of $N$ stocks on trading day $t$. \\ %\hline
			$\mc{A} \in \mbb{R}^{N \times N \times K}$ & binary encoding of stock relations. \\ %\hline
			$\mb{E^t} = [\mb{e_1^t}, \cdots, \mb{e_N^t}]^T \in \mbb{R}^{N \times U}$ & sequential embedding of $N$ stocks learned from historical prices. \\ %\hline
			$\overline{\mb{E^t}} = [\mb{\overline{e_1^t}}, \cdots, \mb{\overline{e_N^t}}]^T \in \mbb{R}^{N \times U}$ & relational embedding of all stocks learned from $\mb{E^t}$ and $\mc{A}$. \\ %\hline
			$\mb{r^{t + 1}}$, $\mb{\hat{r}^{t + 1}} \in \mbb{R}^{N}$ & ground-truth and predicted ranking scores of $N$ stocks. \\ %\hline
			$\mb{w}$, $b$ & weights and bias to be learned. \\ \hline
		\end{tabular}%
	}
\vspace{-0.3cm}
\end{table}
\subsection{Framework}
As illustrated in Figure \ref{fig:framework}, RSR contains three layers, named a sequential embedding layer, a relational embedding layer, and a prediction layer, which are elaborated as follows. \vspace{+5pt}

\noindent \textbf{Sequential Embedding Layer}. Considering the strong temporal dynamics of stock markets, it is intuitive to regard the historical status of a stock as the most influential factor to predict its future trend. As such, we first apply a sequential embedding layer to capture the sequential dependencies in the historical data. 
Since RNN has achieved significant performance to process sequential data \cite{yan2016learning, srivastava2015unsupervised, graves2013speech} and demonstrated to be effective in recent stock prediction research \cite{bao2017deep, zhang2017stock}, 
we opt for RNN to learn the sequential embeddings. 
Among various RNN models, such as vanilla RNN, LSTM, and Gated Recurrent Unit (GRU)~\cite{cho2014learning}, we choose LSTM owing to its ability to capture long-term dependency, which is of great importance to stock prediction. 
This is because that many factors have long-term effects on a stock, such as the rise of interest rates, the release of annual reports, a rapid drop in its price, among others. 
For example, if a stock has experienced a very rapid drop in its price, after that, the stock's price tends to exhibit an upward trend in the following days or weeks (\aka the mean reversion phenomenon). 
%it usually exhibits an upward trend towards the price before the jump in the following weeks. 
As such, we feed the historical time series data of stock $i$ at time-step $t$ ($\mb{X_i^t}$) to a LSTM network and take the last hidden state ($\mb{h_i^t}$) as the sequential embedding ($\mb{e_i^t}$) of a stock (note that $\mb{e_i^t} = \mb{h_i^t}$), \ie we have,
\begin{align}
	\mb{E^t} = LSTM(\mc{X}^t),
\end{align}
where $\mb{E^t} = [\mb{e_1^t}, \cdots, \mb{e_N^t}]^T \in \mbb{R}^{N \times U}$ denotes the sequential embeddings of all stocks, and $U$ denotes the embedding size (\ie $U$ is the number of hidden units in LSTM). \vspace{+5pt}

\begin{figure}[tbp!]
	\centering
	\mbox{
		%\hspace{-0.1in}
		\subfigure[Sector-industry relation]{
			\label{fig:google_micro}
			\includegraphics[width=0.42\textwidth]{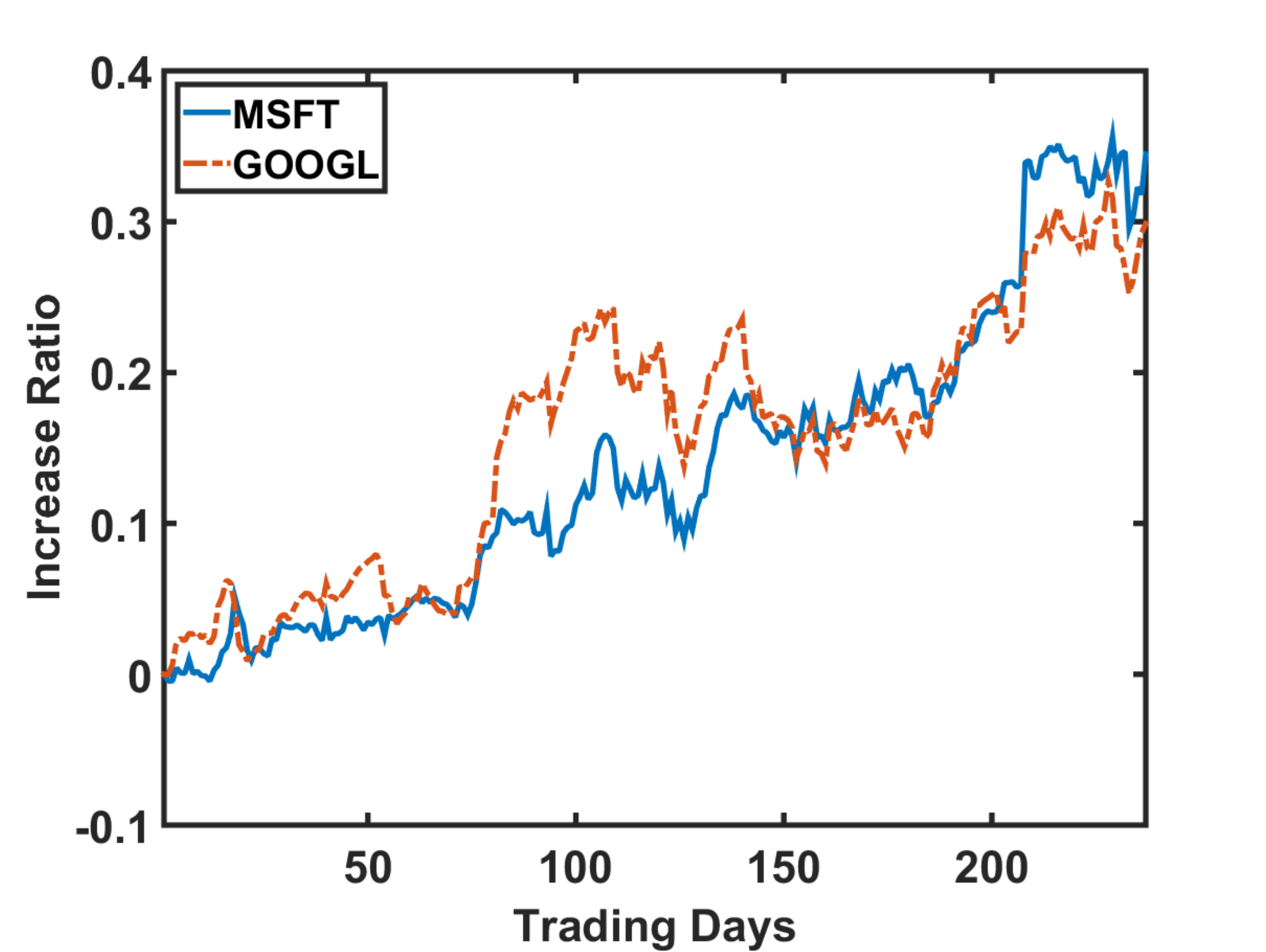}
		}
		%\hspace{-0.25in}
		\subfigure[Supplier-consumer relation]{
			\label{fig:lens}
			\includegraphics[width=0.42\textwidth]{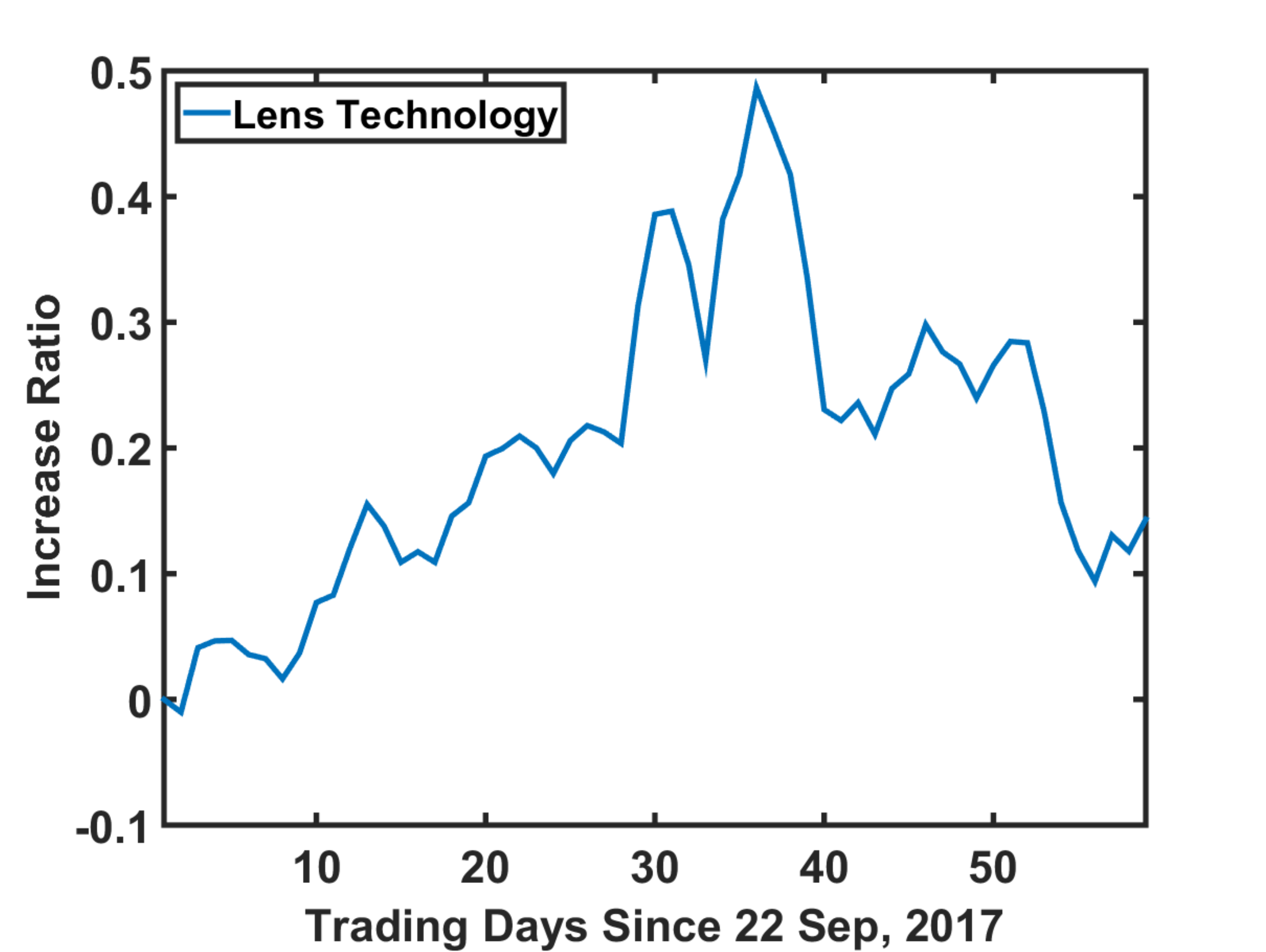}
		}
	}
	\vspace{-0.5cm}
	\caption{{\color{blue}Two examples of stock price history (normalized as increase ratio as compared to the first depicted trading day) to illustrate the impact of company relations on the stock price.
	}}
	\label{fig:curparatune}
	\vspace{-0.3cm}
\end{figure}

\noindent\textbf{Relational Embedding Layer}. 
We now consider how to model the influence between different stocks, especially the ones with explicit relations. Note that it can be seen as an {\color{blue}injection of explicit domain knowledge} (\ie stock relations) into the data-driven approach for sequential embedding learning. Here we give two cases for illustration:
\begin{itemize}[leftmargin=*]
    \item If two companies are in the same sector or industry, they may exhibit similar trends in their stock prices, since they tend to be influenced by similar external events. Figure \ref{fig:google_micro} shows two example stocks, MSFT (Microsoft Inc.) and GOOGL (Alphabet Inc.), both of which are in the same sector (Technology) and industry (Computer Software)\footnote{\href{http://www.nasdaq.com/screening/companies-by-industry.aspx}{http://www.nasdaq.com/screening/companies-by-industry.aspx}}. As can be seen in Figure \ref{fig:google_micro}, the two stocks exhibit quite similar trends in terms of the change on price in {\color{blue}2017}. Note that we normalize the prices of each stock separately by calculating the increase ratio at each trading day according to the price on the first day to reflect the price changes.
    \item If two companies are partners in a supply chain, then the events of the upstream company may affect the stock price of the downstream company. 
    Figure \ref{fig:lens} shows an example to demonstrate the impact of such supplier-consumer relation, which shows the stock price change of Lens Technology Co Ltd {\color{blue}after the release of iPhone 8 (09/22/2017)\footnote{\href{https://www.techradar.com/reviews/iphone-8-review}{https://www.techradar.com/reviews/iphone-8-review}}.} 
    %since 22/09/2017 (the day that iPhone 8 was released)\footnote{\href{https://www.techradar.com/reviews/iphone-8-review}{https://www.techradar.com/reviews/iphone-8-review}}. 
    Since the Lens Technology Co Ltd is the supplier of the screen of iPhone, which was expected to be selling well, its stock price kept increasing in the following several weeks of 09/22/2017.
\end{itemize}
%and industry tends to have similar trends since they are equally influenced by the general development status and usually traded simultaneously. 

\noindent To capture such patterns in stock historical data, we devise a new component of neural network modeling, named \textit{Temporal Graph Convolution} to revise the sequential embeddings according to stock relations. It generates the relational embeddings $\overline{\mb{E^t}} \in \mbb{R}^{N \times U}$ in a time-sensitive (dynamic) way, which is a key technical contribution of this work and will be elaborated later in Section \ref{ss:tgco}. \vspace{+5pt}

\noindent \textbf{Prediction Layer}. Lastly, we feed the sequential embeddings and revised relational embeddings to a fully connected layer to predict the ranking score of each stock; the ranked list of stocks recommended to buy is then generated based on the prediction scores. 

%Specifically, after concatenating the sequential embeddings and the relational embeddings, we apply a fully connected layer to predict the ranking score of stocks.

%During the training phase, the predictions and the corresponding ground truth are fed into the following loss function,
%XN: BUG 1. Add more discussion on the pairwise loss. Why design the loss like this? Any connection/backup with/from existing work? 2. What's the meaning of the groundtruth r? 3. Fix the hat and bold issues in the equation. 
To optimize the model, we propose an objective function that combines both pointwise regression loss and pairwise ranking-aware loss: 
%\begin{align}
%	l(\mb{\hat{r}^{t+1}}, \mb{r^{t+1}}) = &
%	\left\|\mb{\hat{r}^{t+1}} - \mb{r^{t+1}} \right\|^2 \\
%	+& \alpha \sum_{i=0}^{N} \sum_{j=0}^{N} \notag
%	max(0, -(
%		\hat{r_i}^{t + 1} - 
%		\hat{r_j}^{t + 1}
%		)(
%		r_i^{t + 1} - 
%		r_j^{t + 1}
%		)
%	),
%	\label{eq:loss_function}
%\end{align}
\begin{align}
l(\mb{\hat{r}^{t+1}}, \mb{r^{t+1}}) = 
\left\|\mb{\hat{r}^{t+1}} - \mb{r^{t+1}} \right\|^2
+ \alpha \sum_{i=0}^{N} \sum_{j=0}^{N}
max(0, -(
\hat{r_i}^{t + 1} - 
\hat{r_j}^{t + 1}
)(
r_i^{t + 1} - 
r_j^{t + 1}
)
),
\label{eq:loss_function}
\end{align}
where $\mb{r^{t + 1}} = [r_1^{t + 1}, \cdots r_N^{t + 1}]$ and $\mb{\hat{r}^{t + 1}} = [\hat{r}_1^{t + 1}, \cdots, \hat{r}_N^{t + 1}] \in \mbb{R}^{N}$ are ground-truth and predicted ranking scores, respectively, and $\alpha$ is a hyperparameter to balance the two loss terms. 
Since we focus on identifying the most profitable stock to trade, {\color{blue}we use the 1-day return ratio of a stock as the ground-truth rather than the normalized price used in previous work~\cite{zhang2017stock}}. We will provide more details on computing the ground-truth in Section \ref{ss:sequential} in our data collection. 

The first regression term
%is a typical point-wise ranking loss \cite{cossock2006subset}, which can ~\cite{Weimer:2007} ~\cite{NIPS2013_5028}
punishes the difference between the scores of ground-truth and prediction. The second term is pair-wise max-margin loss \cite{zheng2007regression}, which encourages the predicted ranking scores of a stock pair to have the same relative order as the ground-truth. The similar max-margin loss has been used in several applications such as recommendation \cite{Weimer:2007} and knowledge based completion \cite{socher2013reasoning} and demonstrated good performance in ranking tasks.
%(\ie $\hat{r_i}^{t + 1} - \hat{r_j}^{t + 1}$ and $r_j^{t + 1} - r_i^{t + 1}$ have the same sign).
Minimizing our proposed combined loss will force the prediction ranking scores to be close to both 1) the return ratios of stocks in terms of absolute values, and 2) the relative orders of return ratios among stocks, {\color{blue}so as to facilitate investors making better investment decisions. On one hand, correct relative order of stocks could help to select the investment targets (\ie the top ranked stocks). On the other hand, the accurate prediction of return ratio would facilitate deciding the timing of investment since the top ranked stocks are valuable targets only when the return ratios would largely increase.}

\subsection{Temporal Graph Convolution}
\label{ss:tgco}
Given $N$ stocks with their sequential embeddings $\mb{E^t} \in \mbb{R}^{N \times U}$ (\ie the output of sequential embedding layer) and their multi-hot binary relation encodings $\mc{A} \in \mbb{R}^{N \times N \times K}$, the aim of \textit{Temporal Graph Convolution} is to learn revised embeddings $\overline{\mb{E^t}} \in \mbb{R}^{N \times U}$ that encode the relation information. Instead of directly presenting the formulation of TGC, we detail how we design the component to shed some lights on its rationale. Lastly, we discuss its connection with existing graph-based learning methods.  \vspace{+5pt}

\noindent \textbf{a) Uniform Embedding Propagation}. 
Our first inspiration comes from the link analysis research, where in a graph, the impact of a vertex on another one can be captured by propagating information on the graph. A well-known example is the PageRank~\cite{page1999pagerank} method that propagates the importance score of a vertex to its connected vertices. 
%impacts between entities are captured by propagating information through connection structures. 
%Taking the famous PageRank \cite{page1999pagerank} method as an example, a Webpage propagates its importance score to the pages it jumps to. 
Since a stock relation encodes certain similarity information between two connected stocks, we consider relating their embeddings through the similar {\color{blue}propagation process} as in link analysis:
\begin{align}
	\mb{\overline{e_i^t}} = \sum_{\{j | sum(\mb{a_{ji}}) > 0\}} \frac{1}{d_j} \mb{e_j^t},
\end{align}
where $sum(\mb{a_{ji}})$ is the sum of all elements in the relation vector $\mb{a_{ji}}$ (recall that $\mb{a_{ji}}$ is a multi-hot binary vector where each element denotes whether the corresponding {\color{blue} type of} relation exists between $j$ and $i$). The condition $sum(\mb{a_{ji}}) > 0$ ensures that only stocks have at least one relation will be considered. $d_j$ is the number of stocks satisfying the condition $sum(\mb{a_{ji}}) > 0$. After such a propagation in the embedding space, the relational embedding $\mb{\overline{e_i^t}}$ encodes the impacts coming from other stocks that have relations with stock $i$ at time $t$. \vspace{+5pt}

\noindent \textbf{b) Weighted Embedding Propagation}. Consider that different relations between two stocks may have varying impacts on their prices, 
we apply a non-uniform coefficient when propagating the embeddings:
\begin{align}
	\mb{\overline{e_i^t}} = \sum_{\{j | sum(\mb{a_{ji}}) > 0\}} \frac{g(\mb{a_{ji}})}{d_j} \mb{e_j^t},
	\label{eq:tgco}
\end{align}
where $g(\mb{a_{ji}})$ is a mapping function that aims to learn the impact strength of the relations in $\mb{a_{ji}}$, and we term it as the \textit{relation-strength function}. 
As an example, suppose we have two relations named $supplier\_customer$ and $same\_industry$, and three stocks $j$, $i$, and $k$. 
Given that stock $j$ is a supplier of stock $i$ while stock $k$ is in the same industry as stock $i$, we can encode their relations as two different vectors: $\mb{a_{ji}} = [1, 0]$ and $\mb{a_{ki}} = [0, 1]$. 
%(\ie $\mb{a_{ji}} = [1, 0]$) and stock $k$ is in the same industry as stock $i$ (\ie $\mb{a_{ji}} = [0, 1]$). 
%stock $j$ is just a supplier of $i$ (\ie $\mb{a_{ji}} = [1, 0]$), while stock $k$ is both a supplier and under the same industry as $i$ (\ie $\mb{a_{ki}} = [1, 1]$), 
We can see that by feeding different relation vectors into a learnable relation-strength function for different stock pairs, we allow the embedding propagation process to account for both the topology of relation graph and the semantics of relations. \vspace{+5pt}

%the mapping function $g$ tends to encode the difference between impact from $j$ and $k$. As such, the embedding propagation operation jointly encodes the connection topology and difference of relation types. 
\noindent \textbf{c) Time-aware Embedding Propagation}. A limitation of the above weighted propagation process is that the \textit{relation-strength function} returns a fixed weight for a given relation vector $\mb{a_{ji}}$ regardless the evolution across different time-steps. As stock market is highly dynamic such that the status of a stock and the strength of a relation are continuously evolving, assuming a relation vector to have a static weight limits the modeling fidelity. For instance, in the previous example of Figure~\ref{fig:lens}, the $supplier\_customer$ relation between Apple Inc. and Lens Technology Co Ltd has a larger impact on the Lens's stock price in the period of releasing new version of iPhone than usual. 
To address this limitation, we propose to encode the temporal information into the relation-strength function and define the \textit{Time-aware Embedding Propagation} process as follows:
%However, the propagation misses the temporal evolution property of stock markets, \ie the impact strength from stock $j$ to $i$ tends to vary through different timestamps. For example, the impact from AAPL (Apple Inc.) to 300433 (Lens Technology) during the phase of releasing iPhone is stronger than other ordinary times. To address such limitation, we encode the temporal information into the embedding propagation operation,
\begin{align}
	\mb{\overline{e_i^t}} = \sum_{\{j | sum(\mb{a_{ji}}) > 0\}} \frac{g(\mb{a_{ji}}, \mb{e_i^t}, \mb{e_j^t})}{d_j} \mb{e_j^t},
\end{align}
which takes the sequential embeddings (note that they are time-sensitive) into account to estimate the strength of a relation. Besides encoding the temporal information, another benefit of such a design is that sequential embedding also encodes the stock information. This allows the relation-strength function to estimate the impact of a relation vector based on the stocks of concern, which is very desirable. 

%As the sequential embeddings $\mb{e_i^t}$ and $\mb{e_j^t}$ encodes different information at different timestamps, the mapping function $g$ has the ability to assign different strength for the impact from stock $j$ to $i$ at different time.

%We then devise two formulations for the impact strength mapping function regarding whether it considers the interaction between stocks explicitly:
Next we describe two designs of the time-sensitive relation-strength function, which differ in whether to model the interaction between two stocks in an explicit or implicit manner.

\begin{itemize}[leftmargin=*]
    \item \textbf{Explicit Modeling}. For the explicit way, we define the relation strength function as:
\begin{align}
	g(\mb{a_{ji}}, \mb{e_i^t}, \mb{e_j^t}) = \underbrace{\mb{e_i^t}^T\mb{e_j^t}}_{\text{similarity}} \times \underbrace{\phi(\mb{w}^T \mb{a_{ji}} + b)}_{\text{relation importance}},
	\label{eq:explicit}
\end{align}
where $\mb{w}\in \mbb{R}^K$ and $b$ are model parameters to be learned; $\phi$ is an activation function\footnote{Note that we employ the \textit{leaky rectifier} \cite{maas2013rectifier} with a slope of 0.2 as the activation function in our implementation.}. The relation strength of $\mb{a_{ji}}$ is determined by two terms -- \textit{similarity} and \textit{relation importance}. Specifically, the first term measures the similarity between the two stocks at the current time-step. The intuition is that the more similar the two stocks are at the current time, it is more likely that their relations will impact their prices in the near future. We use inner product to estimate the similarity, inspired by its effectiveness in modeling the similarity (interaction) between two entities (embeddings) in Collaborative Filtering~\cite{he2017neural}. The second term is a nonlinear regression model on the relations, where each element in $\textbf{w}$ denotes the weight of a relation in general and $b$ is a bias term. Since both terms of this function are directly interpretable, we call it as \textit{Explicit Modeling}. 
%The strength of impact is determined by \textit{interaction strength} and \textit{relation strength}. The interaction strength between sequential embeddings encodes whether the corresponding stocks are strongly connected in the past few timestamps. We formulate the interaction strength as the inner-production of sequential embeddings, inspired by encoding the interaction between use and item embeddings with inner-product in Collaborate Filtering \cite{he2017neural}. The relation strength is a bias for different relation patterns.
    \item \textbf{Implicit Modeling}. In this design, we feed the sequential embeddings and the relation vector into a fully connected layer to estimate the relation strength:
\begin{align}
	g(\mb{a_{ji}}, \mb{e_i^t}, \mb{e_j^t}) = \phi (\mb{w}^T [\mb{e_i^t}^T, \mb{e_j^t}^T, \mb{a_{ji}}^T]^T + b),
\end{align}
where $\mb{w}\in \mbb{R}^{2U + K}$ and $b$ are model parameters to be learned; $\phi$ is an activation function same as the one in Equation \ref{eq:explicit}. Then we normalize the outputs using a softmax function, which also endows it with more non-linearities. Since this way of interaction is implicitly captured by the parameters, we call it as \textit{Implicit Modeling}. 
\end{itemize}

%Finally, we normalized the calculated impact strength with a softmax function,
%\begin{align}
%	w_{ji} = \frac{exp(g(\mb{a_{ji}}, \mb{e_i^t}, \mb{e_j^t}))}{\sum_{i | max(\mb{a_{ji}}) = 1} exp(g(\mb{a_{ji}}, \mb{e_i^t}, \mb{e_j^t}))},
%\end{align}
%and obtain the final version of temporal embedding propagation operation,
%\begin{align}
%	\mb{\overline{e_i^t}} = \sum_{\{j | max(\mb{a_{ji}}) = 1\}} w_{ji} \mb{e_j^t},
%\end{align}

\subsubsection{Connection with Graph-based Learning}
The embedding propagation is equivalent to the graph convolutional network (GCN). 
To show the relation, let us first construct a graph based on the stock relation encodings $\mc{A}^t$, where 
%With the stock relation encodings $\mc{A}^t$, we can construct a simple graph where 
vertices represent stocks and edges connect vertices with at least one relation, \ie we connect vertex $i$ and $j$ if they satisfy the condition $sum(\mb{a_{ij}}) > 0$. 
If we represent the graph with an adjacency matrix $\mb{A}$ and normalize it by column, 
the Uniform Embedding Propagation (\ie Equation \ref{eq:tgco}) has exactly the same effect as the state-of-the-art graph convolutional operation (\ie $f(\mb{F}, \mb{X}) = \mb{AX}$; details see Section~\ref{sss:gcn}). 
%XN: cannot understand
%If the edges in the graph are weighted, an embedding propagation operation with same effect as graph convolution can be easily obtained by assigning the same weights for corresponding propagation. 
However, GCN cannot capture the temporal evolution properties as designed in our TGC, since the adjacency matrix $\mb{A}$ has to be fixed in GCN. As such, our proposed operation can be seen as 
generalizing the GCN by specifically modeling the temporal patterns, thus we term it as the \textit{Temporal Graph Convolution}.

%XN: cannot understand
%The conventional graph-based learning, which regulates the learning target with a graph Laplacian encoding the entity (stock) relations, is roughly a special case of TGCO by injecting TGCO into the prediction layer. However, the conventional graph-based learning also suffers from the limitation of missing temporal evolution properties, which is of importance under applications like stock prediction.

%% file: 5_data.tex
\section{Data Collection}
\label{sec:data}
Most existing works evaluate stock prediction on dozens of stocks, and there lacks a large stock dataset for an extensive evaluation. {\color{blue}As such, we consider constructing data by ourselves, which is accessible through: \href{https://github.com/hennande/Temporal_Relational_Stock_Ranking}{https://github.com/hennande/Temporal\_Relational\_Stock\_Ranking}.} Specifically, we collect the stocks from the NASDAQ and NYSE markets that have transaction records between 01/02/2013 and 12/08/2017, obtaining $3,274$ and $3,163$ stocks respectively. 
%Typically, existing research evaluates stock prediction with dozens of selected stocks, which leads to the lack of public stock prediction benchmark for large scale evaluation. 
%As such, we collect data of 3,274 and 3,163 stocks between Jan-02-2013 to Dec-08-2017 from NASDAQ and NYSE markets, respectively. 
Note that we select these two markets for their representative properties that NASDAQ is more volatile whereas NYSE is more stable \cite{schwert2002stock}.  %Among the stocks from NASDAQ and NYSE, we selected 1,026 and 1,737 stocks satisfying two conditions: 
Furthermore, we perform a filtering on the stocks by retaining the stocks satisfying the two conditions: 
1) have been traded on more than 98\% of trading days since 01/02/2013; 2) have never been traded at less than five dollars per share during the collection period. It should be noted that the first condition is based on concerns that intermittent sequences may bring abnormal patterns; the second condition ensures that the selected stocks are not penny stocks\footnote{\href{https://www.sec.gov/fast-answers/answerspennyhtm.html}{https://www.sec.gov/fast-answers/answerspennyhtm.html}}, which are too risky for general investors as suggested by the U.S. Securities and Exchange Commission. 
This results in $1,026$ NASDAQ and $1,737$  NYSE stocks for our experiments. 
For these stocks, we collect three kinds of data: 1) historical price data, 2) sector-industry relations, and 3) Wiki relations between their companies such as supplier-consumer relation and ownership relation. Next, we present the details of these data.
%We then successively detail the collection procedure of these three kinds of data.
% using the Industrial Classification Benchmark\footnote{\href{http://www.ftserussell.com/financial-data/industry-classification-benchmark-icb}{http://www.ftserussell.com/financial-data/industry-classification-benchmark-icb}}

\subsection{Sequential Data}\label{ss:sequential}
Following \cite{zhang2017stock}, we set the prediction frequency as daily-level. Under our problem formulation, we aim to predict a ranking list of stocks for the following trading day, based on the daily historical data in the last $S$ trading days. As the return ratio of a stock indicates the expected revenue of the stock, we set the ground-truth ranking score of stock $i$ as its {\color{blue}$1$-day} return ratio $r_i^{t + 1} = (p_i^{t + 1} - p_i^{t}) / p_i^{t}$ where $p_i^{t}$ is the closing price at day $t$. 
% To calculate the ground-truth, we first collect the daily closing price of stocks in NASDAQ and NYSE, launched before Jan-02-2013. 
To calculate the ground-truth, we first collect the daily closing price of each stock ranging from 01/02/2013 and 12/08/2017.
After the collection, we normalize the price of each stock via dividing it {\color{blue}by its maximum value throughout the entire 2013-2017 dataset.} In addition to the normalized closing price, we calculate four more sequential features: 5, 10, 20, and 30 days moving averages which represent the {\color{blue}weekly and monthly trends}. {\color{blue}Following the existing work of stock prediction~\cite{zhang2017stock}, we chronologically separate the sequential data into three time periods for training (2013-2015), validation (2016), and evaluation (2017), respectively, and summarize the basic statistics in Table \ref{tab:seq_data}.} As can be seen, there are 756, 252, and 237 trading days in training, validation, and evaluation, respectively.
\begin{table}[t]
	\centering
	\caption{Statistics of the sequential data.}
	\label{tab:seq_data}
	\vspace{-0.4cm}
	\resizebox{0.7\textwidth}{!}{%
		\begin{tabular}{|c||c|c|c|c|}
			\hline
			Market & Stocks\# & \begin{tabular}[c]{@{}c@{}}Training Days\#\\ 01/02/2013\\ 12/31/2015\end{tabular} & \begin{tabular}[c]{@{}c@{}}Validation Days\#\\ 01/04/2016\\ 12/30/2016\end{tabular} & \begin{tabular}[c]{@{}c@{}}Testing Days\#\\ 01/03/2017\\ 12/08/2017\end{tabular} \\ \hline \hline
			NASDAQ & 1,026 & 756 & 252 & 237 \\ \hline
			NYSE & 1,737 & 756 & 252 & 237 \\ \hline
		\end{tabular}%
	}
	\vspace{-0.3cm}
\end{table}

\subsection{Stock Relation Data}

\subsubsection{Sector-Industry relations}
\begin{figure}[]
	\centering
	\includegraphics[width=0.75\textwidth]{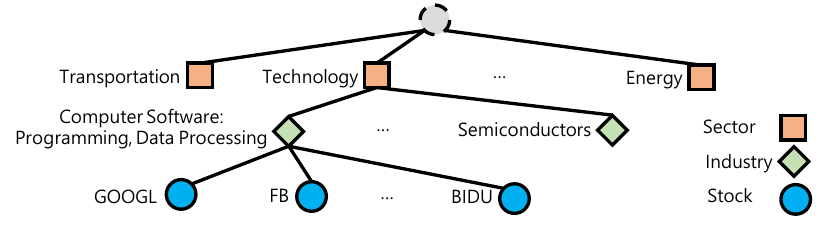}
	\vspace{-0.3cm}
	\caption{Illustration of sector-industry hierarchy of companies in NASDAQ and NYSE.}
	\label{fig:industry}
	\vspace{-0.3cm}
\end{figure}

Observing the trends that stocks under the same industry are similarly influenced by the prospect of the industry, we collect the sector-industry relation between stocks. {\color{blue}In NASDAQ and NYSE, each stock is classified into a sector and an industry as illustrated in Figure \ref{fig:industry} in which stocks in the same industry are organized under the corresponding industry node. We collect the hierarchy structure of NASDAQ and NYSE stocks from the official company list maintained by NASDAQ Inc.\footnote{\url{https://www.nasdaq.com/screening/industries.aspx}}} and extract industry relations for each stock pair under the same industry node, such as (\textit{GOOGL; Computer Software: Programming, Data Processing; FB}). The specific relations extracted are detailed in the Appendix \ref{ss:sec_ind_rel} at the end of this paper. After the extraction, we count the number of industry (\ie relation) types occurred in each market, and the ratio of stock pairs having industry relations and summarize them in Table \ref{tab:industry_relation}. As can be seen, there are 112 and 130 types of industry relations between stock pairs in NASDAQ and NYSE, respectively. Moreover, the industry relation data is sparse since less than 10\% of stock pairs have at least one type of industry relation in both stock markets.
\begin{table}[]
	\centering
	\caption{{\color{blue}Statistics of sector-industry relation and Wiki relation data in the NASDAQ and NYSE datasets.}}
	\vspace{-0.4cm}
	\label{tab:industry_relation}
	\resizebox{\textwidth}{!}{%
		\begin{tabular}{|c||c|c||c|c|}
			\hline
			\multirow{2}{*}{} & \multicolumn{2}{c||}{Sector-Industry Relation} & \multicolumn{2}{c|}{Wiki Relation} \\ \cline{2-5} 
			& Relation Types\# & Relation Ratio (Pairwise) & Relation Types\# & Relation Ratio (Pairwise) \\ \hline \hline
			\textbf{NASDAQ} & 112            & 5.00\% & 42            & 0.21\%                   \\ \hline
			\textbf{NYSE}   & 130            & 9.37\% & 32            & 0.30\%                   \\ \hline
		\end{tabular}
	}
\vspace{-0.1cm}
\end{table}
%\begin{table}[]
%	\centering
%	\caption{Statistics of industry relation data in our NASDAQ and NYSE datasets.}
%	\vspace{-0.3cm}
%	\label{tab:industry_relation}
%	%\resizebox{0.38\textwidth}{!}{%
%		\begin{tabular}{|c||c|c|}
%			\hline
%			& Relation Types\# & Relation Ratio (Pairwise) \\ \hline \hline
%			\textbf{NASDAQ} & 112            & 5.00\%                    \\ \hline
%			\textbf{NYSE}   & 130            & 9.37\%                    \\ \hline
%		\end{tabular}%
%	%}
%\end{table}

\subsubsection{Wiki Company-based Relations}
\begin{figure}[t]
	\centering
	\includegraphics[width=0.75\textwidth]{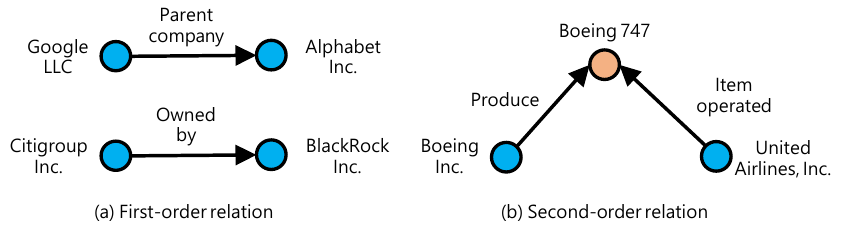}
	\vspace{-0.4cm}
	\caption{Examples of the first-order and second-order company relations extracted from Wikidata.}
	\label{fig:wiki}
	\vspace{-0.2cm}
\end{figure}

As rich sources of entity relations, knowledge bases contain company entities and company relations, which might reflect the impact across stocks. As such, we extract the first-order and second-order company relations from Wikidata \cite{vrandevcic2014wikidata}, one of the biggest and most active open domain knowledge bases with more than 42 million items (\eg \textit{Alphabet Inc.}) and 367 million statements (\eg {\color{blue}\textit{Alphabet Inc.; founded by; Larry Page}}) in the format of (\textit{subject; predicate; object})\footnote{\href{https://www.mediawiki.org/wiki/Wikibase/DataModel/JSON}{https://www.mediawiki.org/wiki/Wikibase/DataModel/JSON}}. As shown in Figure \ref{fig:wiki}, company $i$ has a first-order relation with $j$ if there is a statement that has $i$ and $j$ as the subject and object, respectively. Companies $i$ and $j$ have a second-order relation if they have statements sharing the same object, such as \textit{Boeing Inc.} and \textit{United Airlines, Inc.} have different statements towards \textit{Boeing 747}. After an exhausted exploration of a recent dump of Wikidata (01/05/2018), we obtain 5 and 53 types of first-order and second-order relations, respectively\footnote{We manually filter out less informative relations such as \textit{located at the same timezone}.}. The detailed description of these relations {\color{blue}is} elaborated in the Appendix \ref{ss:wiki_com_rel} at the end of this paper. We then summarize the count of relation types and the ratio of stock pairs with at least one Wiki company-based relation in Table \ref{tab:industry_relation}. As can be seen, there are 42 and 32 types of company relations occurring between stock pairs in NASDAQ and NYSE, respectively.
%\begin{table}[]
%	\centering
%	\caption{Statistics of Wiki company-based relation data in our NASDAQ and NYSE datasets.}
%	\vspace{-0.3cm}
%	\label{tab:wiki_relation}
%	%\resizebox{0.38\textwidth}{!}{%
%		\begin{tabular}{|c||c|c|}
%			\hline
%			& Relation Types\# & Relation Ratio (Pairwise) \\ \hline \hline
%			\textbf{NASDAQ} & 42            & 0.21\%                    \\ \hline
%			\textbf{NYSE}   & 32            & 0.30\%                    \\ \hline
%		\end{tabular}%
%	%}
%\end{table}

%% file: 6_experiment.tex
\section{Experiment}
\label{sec:experiment}
{\color{blue}To the best of our knowledge, our work is the first one to incorporate stock relations into the models for stock prediction, especially neural network-based ones.} As such, in this section, we conduct experiments with the aim of answering the following research questions:
\begin{enumerate}
	\item \textbf{RQ1}: 
	How is the utility of formulating the stock prediction as a ranking task? Can our proposed RSR solution outperform state-of-the-art stock prediction solutions?
	%Does formulating the stock prediction as a ranking task and solving it with method based on learning to rank outperform the state-of-the-art stock prediction solutions?
	\item \textbf{RQ2}: {\color{blue}Do stock relations enhance the neural network-based solution for stock prediction?} How is the effectiveness of our proposed TGC component compared to conventional graph-based learning? 
	%Does the proposed RSR equipped with the devised TGCO layer outperform conventional graph-based learning methods?
	\item \textbf{RQ3}: How does our proposed RSR solution perform under different back-testing strategies?
\end{enumerate}

In what follows, we first present the experimental settings, followed by answering the above three research questions.

\subsection{Experimental Setting}
\subsubsection{Evaluation Protocols} 
Following \cite{dixon2016classification}, we adopt a daily buy-hold-sell trading strategy to evaluate the performance of stock prediction methods regarding the revenue. On each trading day $t + 1$ during the testing period (from 01/03/2017 to 12/08/2017),  we simulate a trader using a stock prediction method to trade in the following way:
%\begin{enumerate}
%	\item \textbf{Before the market opens}: 
%	The trader uses the method to get the prediction score for each stock.  
%	%The method makes prediction for the stocks, which would be the prediction of closing price of stocks for regression methods and the ranking list of stocks regarding the return ratio based on the closing price on day $t$.
%	\item \textbf{When the market opens}: The trader buys the stock with the highest expected revenue.
%	\item \textbf{When the market closes}: The trader sells the purchased stock.
%\end{enumerate}
{\color{blue}\begin{enumerate}
	\item \textbf{When the market closes at trading day $t$}: The trader uses the method to get the prediction, a ranking list with predicted return ratio of each stock. The trader buys the stock with the highest expected revenue (\ie ranked at the top).
	\item \textbf{When the market closes at trading day $t + 1$}: The trader sells the stock purchased at day $t$.
\end{enumerate}}
In calculating the cumulative investment return ratio, we follow several simple assumptions: (1) The {\color{blue}trader} spends the same amount of {\color{blue}money (\eg 50 thousand dollars)} on every trading day. We make this assumption to eliminate the temporal dependency of the testing procedure for a fair comparison. %Otherwise, if we cumulate the earned and lose money during testing, a method that loses money at the beginning cannot earn the same amount of money in the following days, while it makes the same selections as other methods with more capitals. 
(2) The market is always sufficiently liquid such that the buying order gets filled at the closing price of day $t$ and the selling price is the closing price of day $t + 1$. (3) The transaction costs are ignored since the costs for trading US stocks through brokers are quite cheap no matter charging by trades or shares. For instance, Fidelity Investments and Interactive Brokers charge only 4.95 dollars per trade and 0.005 dollar per share, respectively\footnote{\href{https://www.stockbrokers.com/guides/commissions-fees}{https://www.stockbrokers.com/guides/commissions-fees}}.
%The last two assumptions are made for simple calculation.

{\color{blue}Since the target is to accurately predict the return ratio of stocks and appropriately rank the relative order of stocks, we employ three metrics, Mean Square Error (MSE), Mean Reciprocal Rank (MRR), and the cumulative investment return ratio (IRR), to report model performance. MSE has been widely used for evaluating regression tasks such as stock price prediction \cite{kumar2016survey, nassirtoussi2014text, zhang2017stock}. We thus calculate the MSE over all stocks on every trading day within the testing period. MRR~\cite{song2017neurostylist} is a widely used metric for ranking performance evaluation. Here, we calculate the average reciprocal rank of the selected stock over the testing days. Since directly reflecting the effect of stock investment, IRR is our main metric, which is calculated by summing over the return ratios of the selected stock on each testing day. Smaller value of MSE ($\geq 0$) and larger value of MRR ($[0, 1]$) and IRR indicate better performance. For each method, we repeat the testing procedure five times and report the average performance to eliminate the fluctuations caused by different initializations.}

\subsubsection{Methods}
We compare with the following stock price prediction baselines with \textit{regression} formulation:
\begin{itemize}[leftmargin=*]
	\item \textbf{SFM} \cite{zhang2017stock}: This method is the state-of-the-art stock price prediction method. It takes the historical closing prices as input and decomposes the prices into {\color{blue}signals of different frequencies} with a Discrete Fourier Transform (DFT). It then feeds the DFT coefficients into an extended LSTM with {\color{blue}separate memory states for different frequencies to learn the frequency-aware sequential embeddings, which are fed into a FC layer to make the prediction.}
	\item \textbf{LSTM} \cite{bao2017deep}: This method is the vanilla LSTM, {\color{blue}which operates on the sequential data including closing prices and moving averages of 5, 10, 20, and 30 days}, to obtain a sequential embedding; and then a FC layer is used to make prediction of the return ratio.
\end{itemize}
It should be noted that we ignore the potential baselines based on time-series models and shallow machine learning models, since they have been reported to be less effective than \textbf{SFM} and \textbf{LSTM} in several previous works \cite{zhang2017stock, bao2017deep, hu2018listening}. Moreover, we also compare with several methods with \textit{ranking} formulation:
\begin{itemize}[leftmargin=*]
	\item \textbf{Rank\_LSTM}: We remove the relational embedding layer of the proposed RSR to obtain this method, \ie this method ignores stock relations. %which solves the stock ranking task instead of the stock price prediction one.
	\item Graph-based ranking (\textbf{GBR}): According to Equation \ref{eqn:hyperobj}, we add the graph regularization term to the loss function of \textbf{Rank\_LSTM}, which smooths predicted return ratios over the graph of \textit{stock relations}. In the graph, we connect a pair of vertices (\ie stocks) having at least one type of relations.
	\item \textbf{GCN} \cite{kipf2017semi}: GCN is the state-of-the-art graph-based learning method. We obtain this method by replacing the TGC layer of our proposed RSR with a GCN layer. The graph of \textit{stock relations} in \textbf{GBR} is fed into the \textbf{GCN} layer.
	\item \textbf{RSR\_E}: Our proposed RSR with explicit modeling in the TGC.
	\item \textbf{RSR\_I}: Our proposed RSR with implicit modeling in the TGC.
\end{itemize}
% Note that \textbf{GBR} and \textbf{GCN} also encode \textit{stock relations}. {\color{blue}It is worth mentioning that the implementation and parameter setting of each compared method can be accessed through: \href{https://github.com/hennande/Temporal_Relational_Stock_Ranking}{https://github.com/hennande/Temporal\_Relational\_Stock\_Ranking}.}

\subsubsection{Parameter Settings}
{\color{blue}We implement the models with TensorFlow\footnote{\url{https://www.tensorflow.org/}} except \textbf{SFM} of which we use the original implementation\footnote{\url{https://github.com/z331565360/State-Frequency-Memory-stock-prediction}}.} {\color{blue}It is worth mentioning that the implementations can be accessed through: \href{https://github.com/hennande/Temporal_Relational_Stock_Ranking}{https://github.com/hennande/Temporal\_Relational\_Stock\_Ranking}.} We employ grid search to select the optimal hyperparameters regarding IRR for all methods. For \textbf{SFM}, we follow the original setting in \cite{zhang2017stock}, optimizing it by RMSProp with a learning rate of 0.5, and tuning the number of frequencies and hidden units within $\{5, 10, 15\}$ and $\{10, 20, 30\}$, respectively. 
For all other methods,
we apply the Adam \cite{kingma2014adam} optimizer with a learning rate of 0.001. 
We tune two hyperparameters for \textbf{LSTM}, the length of sequential input $S$ and the number of hidden units $U$, within $\{2, 4, 8, 16\}$ and $\{16, 32, 64, 128\}$, respectively. 
Besides $S$ and $U$, we further tune $\alpha$ in Equation \ref{eq:loss_function}, which balances the point-wise and pair-wise terms; specifically, we tune $\alpha$ within
$\{0.1, 1, 10\}$ for \textbf{Rank\_LSTM}, %and use the optimal value for \textbf{GBR},
\textbf{GCN}, \textbf{RSR\_E}, and \textbf{RSR\_I}. 
We further tune the $\lambda$ of the regularization term in \textbf{GBR} within $\{0.1, 1, 10\}$.

% Whether ranking helps
\subsection{\mbox{Study of Stock Ranking Formulation (RQ1)}}
\label{ss:rank_impact}
\begin{table}[tb]
	\centering
	\caption{{\color{blue}Performance comparison between the solutions with regression formulation (\tb{SFM} and \tb{LSTM}) and ranking formulation (\tb{Rank\_LSTM}).}}
	\vspace{-0.4cm}
	\label{tab:rank_per}
	\resizebox{\textwidth}{!}{%
		\begin{tabular}{|c||c|c|c||c|c|c|}
			\hline
			\multirow{2}{*}{} & \multicolumn{3}{c||}{NASDAQ} & \multicolumn{3}{c|}{NYSE} \\ \cline{2-7} 
			& MSE & MRR & IRR & MSE & MRR & IRR \\ \hline \hline
			\textbf{SFM} & 5.20e-4$\pm$5.77e-5 & 2.33e-2$\pm$1.07e-2 & -0.25$\pm$0.52 & 3.81e-4$\pm$9.30e-5 & \textbf{4.82e-2$\pm$4.95e-3} & 0.49$\pm$0.47 \\ \hline
			\textbf{LSTM} & 3.81e-4$\pm$2.20e-6 & 3.64e-2$\pm$1.04e-2 & 0.13$\pm$0.62 & 2.31e-4$\pm$1.43e-6 & 2.75e-2$\pm$1.09e-2 & -0.90$\pm$0.73 \\ \hline \hline
			\textbf{Rank\_LSTM} & \textbf{3.79e-4$\pm$1.11e-6} & \textbf{4.17e-2$\pm$7.50e-3} & \textbf{0.68$\pm$0.60} & \textbf{2.28e-4$\pm$1.16e-6} & 3.79e-2$\pm$8.82e-3 & \textbf{0.56$\pm$0.68} \\ \hline
		\end{tabular}%
	}
	\vspace{-0.7cm}
\end{table}
\begin{figure}[]
	\centering
	\mbox{
		\hspace{-0.1in}
		\subfigure[NASDAQ]{
			\label{fig:rank_nasdaq}
			\includegraphics[width=0.42\textwidth]{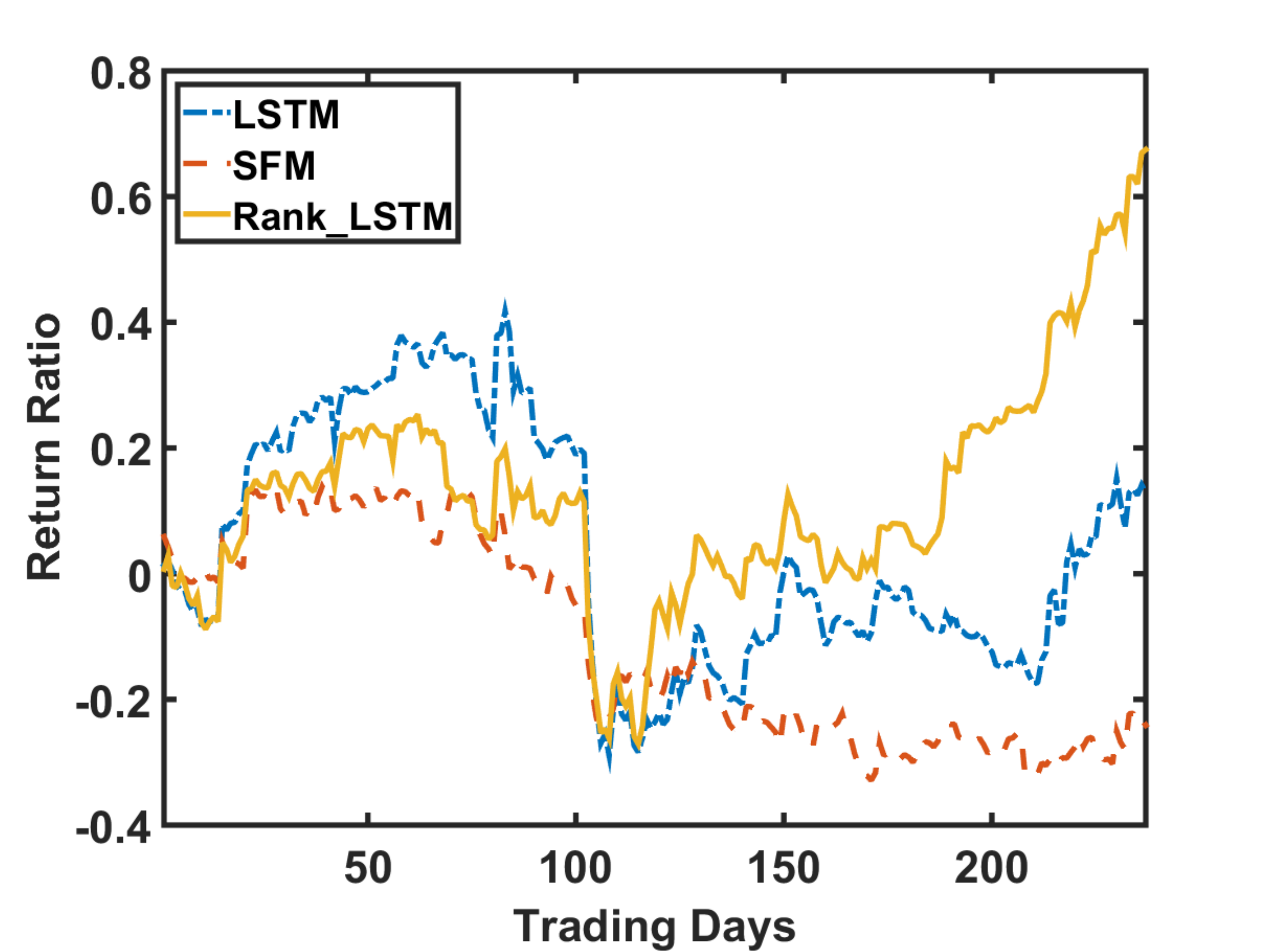}
		}
		\hspace{-0.25in}
		\subfigure[NYSE]{
			\label{fig:rank_nyse}
			\includegraphics[width=0.42\textwidth]{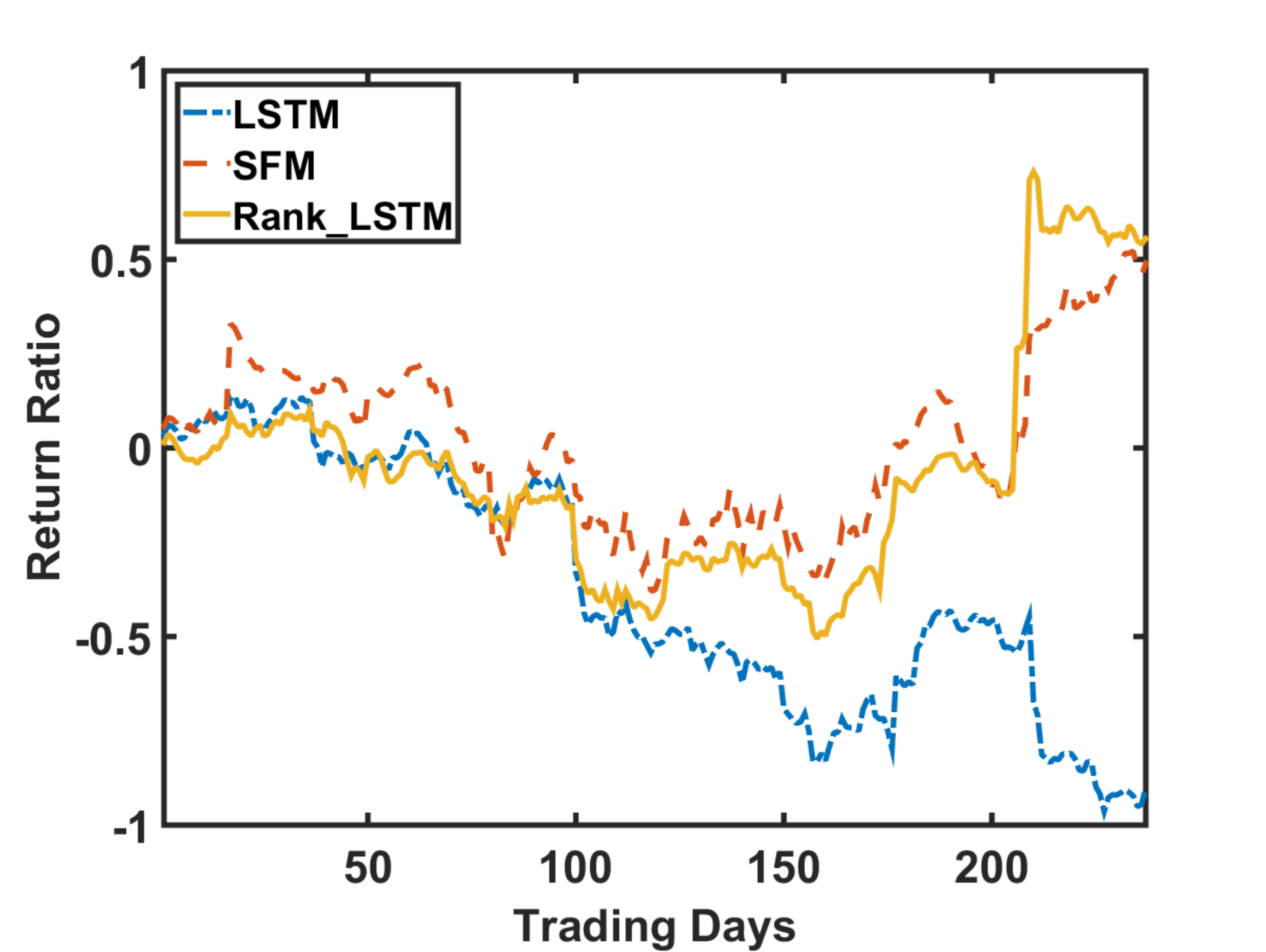}
		}
	}
	\vspace{-0.5cm}
	\caption{Performance comparison of \textbf{Rank\_LSTM}, \textbf{SFM}, and \textbf{LSTM} regarding IRR.
	}
	\label{fig:rank_bt}
	\vspace{-0.5cm}
\end{figure}
{\color{blue}Table \ref{tab:rank_per} summarizes the performance of baselines in \textit{regression} fashion and \tb{Rank\_LSTM} our basic solution of stock ranking \wrt MSE, MRR, and IRR}, from which we have the following observations:
\begin{itemize}[leftmargin=*]
\item \tb{Rank\_LSTM} outperforms both \tb{SFM} and \tb{LSTM} on the two markets with great improvement \wrt IRR ($>$14\%). It verifies the advantage of the stock ranking solutions and answers \tb{RQ1} that stock ranking is a promising formulation of stock prediction. {\color{blue}Moreover, it indicates the potential of advanced learning-to-rank techniques in solving the stock prediction task.}
\item {\color{blue}However, \tb{Rank\_LSTM} fails to consistently beat \tb{SFM} and \tb{LSTM} regarding all evaluation measures, its performance on NYSE \wrt MRR is worse than \tb{SFM}. The reason could be attributed to minimizing the combination of point-wise and pair-wise losses, which would lead to a tradeoff between accurately predicting absolute value of return ratios and their relative order.}
\item The performance \wrt IRR varies a lot under different runs of a method. It is reasonable since the absolute value of daily return ratio varies from 0 to 0.98 in our dataset, which means that a tiny switch of the top 2 ranked stocks may lead to a huge change of the IRR. Such results also indicate that learning to rank techniques emphasizing the top-ranked stocks is worthwhile to be explored in the future.
\item The performance of \tb{LSTM} on the NYSE market \wrt IRR is unexpectedly bad. We repeat the parameter tuning and testing procedure several times and find that \tb{LSTM} could achieve better performance (with IRR value between 0.1 and 0.2) with other settings of hyperparameters. However, the selected setting always beats the others on the validation. {\color{blue}This result indicates the potential difference between the validation and testing.}
\end{itemize}

Figure \ref{fig:rank_bt} illustrates the procedure of back-testing regarding the cumulative return ratios. As can be seen, {\color{blue}in all cases, the curves are volatile, which indicates that selecting only one stock from more than 1,000} is a highly risk operation. Consequently, it also suggests the worth of introducing risk-oriented criteria into stock ranking tasks in the future.

\subsection{Impact of Stock Relations (RQ2)}
\label{ss:rel_impact}
\begin{table}[tb]
	\centering
	\caption{{\color{blue}Performance comparison among relational ranking methods with industry relations.}}
	\vspace{-0.4cm}
	\label{tab:rel_ind}
	\resizebox{\textwidth}{!}{%
		\begin{tabular}{|c||c|c|c||c|c|c|}
			\hline
			\multirow{2}{*}{} & \multicolumn{3}{c||}{NASDAQ} & \multicolumn{3}{c|}{NYSE} \\ \cline{2-7} 
			& MSE & MRR & IRR & MSE & MRR & IRR \\ \hline \hline
			\textbf{Rank\_LSTM} & \textbf{3.79e-4$\pm$1.11e-6} & 4.17e-2$\pm$7.50e-3 & \textbf{0.68$\pm$0.60} & 2.28e-4$\pm$1.16e-6 & 3.79e-2$\pm$8.82e-3 & 0.56$\pm$0.68 \\ \hline \hline
			\textbf{GBR} & 5.80e-3$\pm$1.20e-3 & \textbf{4.46e-2$\pm$5.20e-3} & 0.57$\pm$0.29 & 2.29e-4$\pm$2.02e-6 & 3.43e-2$\pm$6.26e-3 & 0.68$\pm$0.31 \\ \hline
			\textbf{GCN} & 3.80e-4$\pm$2.24e-6 & 3.45e-2$\pm$8.36e-3 & 0.24$\pm$0.32 & 2.27e-4$\pm$1.30e-7 & \textbf{5.01e-2$\pm$5.56e-3} & 0.97$\pm$0.56 \\ \hline \hline
			\textbf{RSR\_E} & 3.82e-4$\pm$2.69e-6 & 3.16e-2$\pm$3.45e-3 & 0.20$\pm$0.22 & 2.29e-4$\pm$2.77e-6 & 4.28e-2$\pm$6.18e-3 & 1.00$\pm$0.58 \\ \hline
			\textbf{RSR\_I} & 3.80e-4$\pm$7.90e-7 & 3.17e-2$\pm$5.09e-3 & 0.23$\pm$0.27 & \textbf{2.26e-4$\pm$5.30e-7} & 4.51e-2$\pm$2.41e-3 & \textbf{1.06$\pm$0.27} \\ \hline
		\end{tabular}%
	}
\vspace{-0.3cm}
\end{table}
\begin{figure}[tb]
	\centering
	\mbox{
		%\hspace{-0.1in}
		\subfigure[NASDAQ]{
			\label{fig:rel_nasdaq_ind}
			\includegraphics[width=0.42\textwidth]{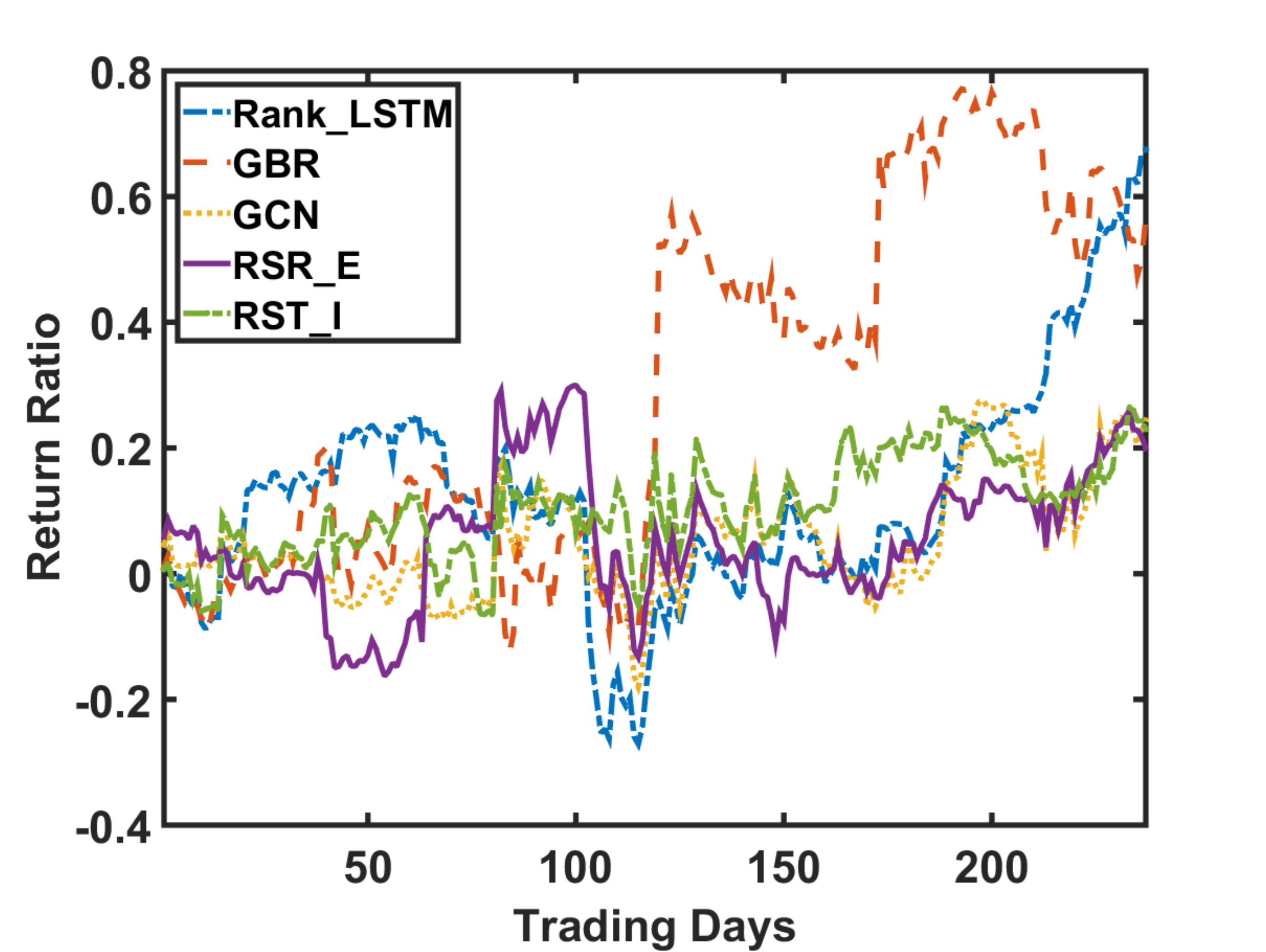}
		}
		\hspace{-0.25in}
		\subfigure[NYSE]{
			\label{fig:rel_nyse_ind}
			\includegraphics[width=0.42\textwidth]{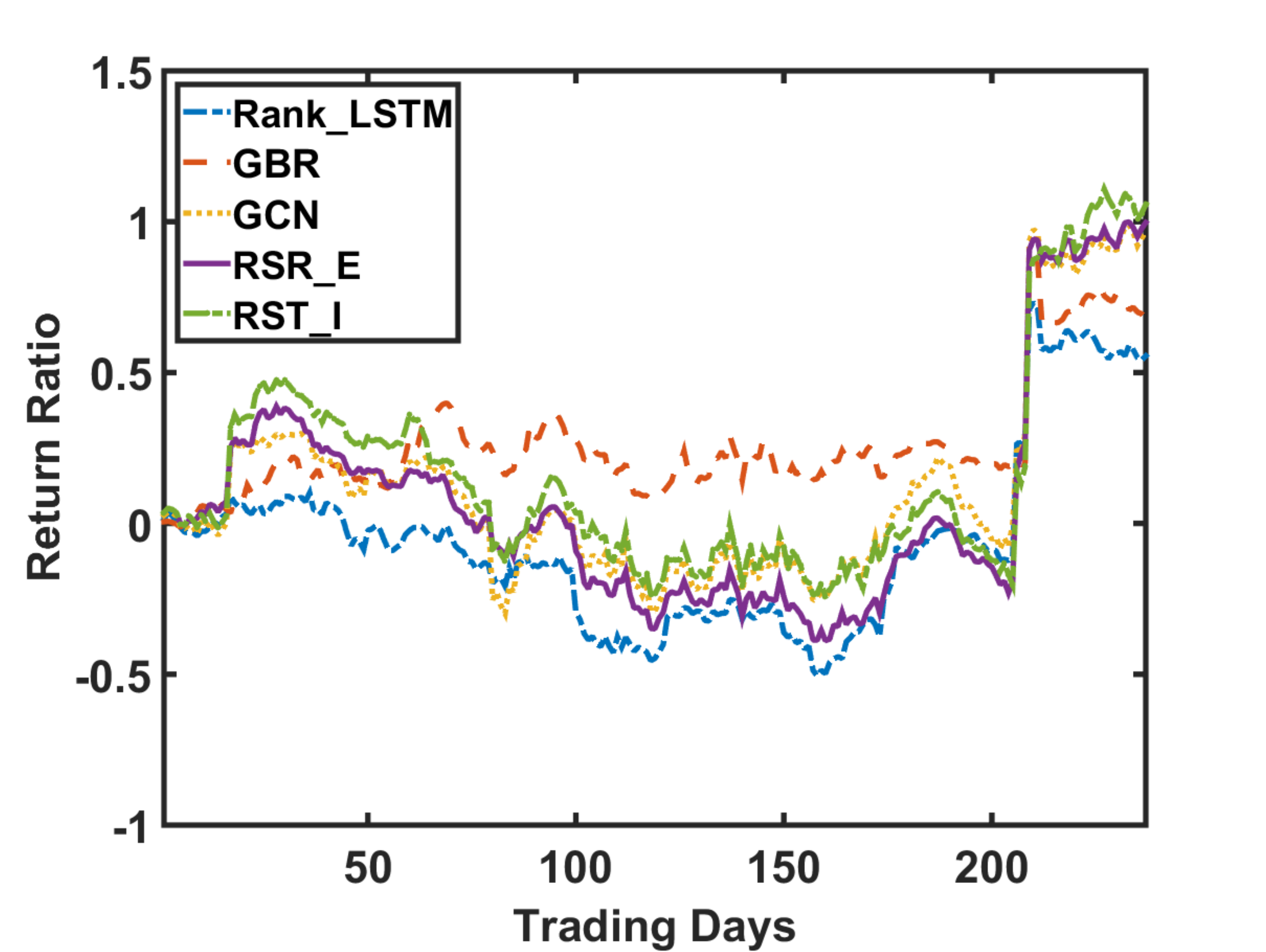}
		}
	}
	\vspace{-0.5cm}
	\caption{Back-testing procedure of relational ranking methods with industry relations regarding IRR.
	}
	\label{fig:rel_ind}
	\vspace{-0.5cm}
\end{figure}
\tb{Effect of Industry Relations}: Table \ref{tab:rel_ind} shows the performance of methods considering the industry relation of stocks. We can see that: 
\begin{itemize}[leftmargin=*]
	\item {\color{blue}Considering industry relations is more beneficial to stock ranking on NYSE as compared to NASDAQ. It could be attributed to that the industry relations reflect more of long-term correlations between stocks}, since NASDAQ is considered as a much more volatile market as compared to NYSE and dominated by short-term factors \cite{schwert2002stock}. 
	\item {\color{blue}On NYSE, all methods considering stock relations, \ie \tb{GBR}, \tb{GCN}, \tb{RSR\_E}, and \tb{RSR\_I}, outperform \tb{Rank\_LSTM} \wrt IRR. Considering that all these methods take \tb{Rank\_LSTM} as the building block, this result verifies the effectiveness of encoding stock relations in stock prediction.} 
	\item {\color{blue}Moreover, \tb{RSR\_E} and \tb{RSR\_I} achieve improvement over \tb{GCN} and \tb{GBR}. This result verifies the effectiveness of the proposed \textit{Temporal Graph Convolution} as compared to the traditional modeling of relational data. Considering that \tb{GCN} and \tb{GBR} utilize a static graph to represent stock relations, the result also indicates the rationale of considering temporal property in stock relation modeling.}
	\item {\color{blue}Again, the performance regarding different evaluation measures is inconsistent. We speculate the reason is that we tune the hyperparameters regarding IRR, which focuses more on correct ranking on testing days with high return ratios. For instance, correct prediction on a trading day with ground truth return ratio of 0.5 would lead to higher IRR than correct predictions in ten trading days with return ratio of 0.01. As such, a model achieves better IRR could achieve suboptimal MSE and MRR.}
\end{itemize}

Figure \ref{fig:rel_ind} illustrates the IRR curve of the compared methods in the back-testing. Again, the curves are volatile of which the reason has been discussed in Section~\ref{ss:rank_impact}. {\color{blue}On NYSE, the IRR of the methods presents huge increases on the 206-th and 209-th trading days when the best-performed stock exhibits return ratios larger than 0.6. This result further highlights the importance of accurately predicting both the return ratio of single stock and the relative order of stocks. Note that capturing rare opportunities by precisely ranking the stocks on trading days with huge change of return ratios would lead to satisfied IRR.}

\begin{figure}[tb]
	\centering
	\mbox{
		%\hspace{-0.1in}
		\subfigure[NASDAQ]{
			\label{fig:rel_nasdaq_wiki}
			\includegraphics[width=0.42\textwidth]{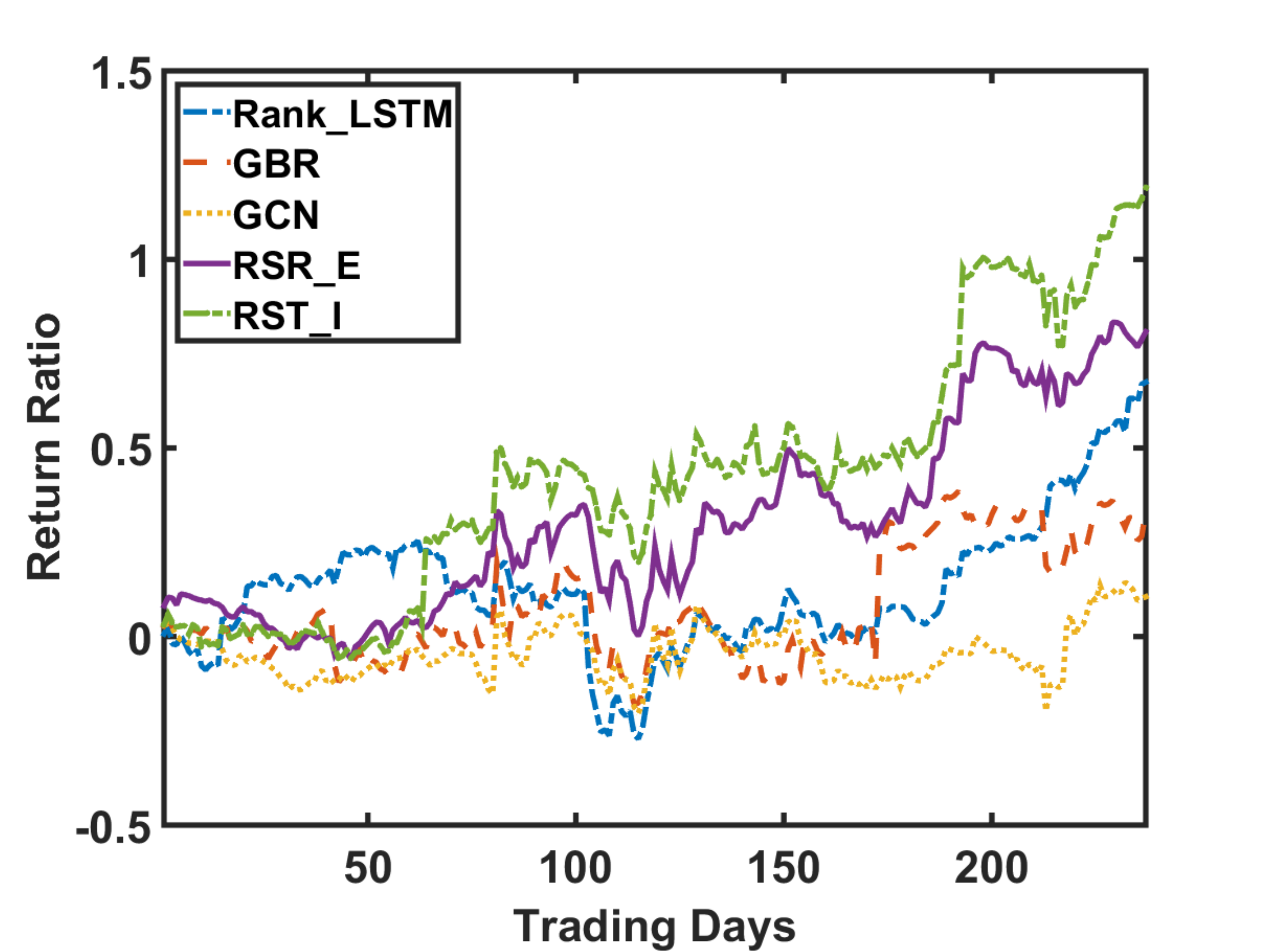}
		}
		\hspace{-0.25in}
		\subfigure[NYSE]{
			\label{fig:rel_nyse_wiki}
			\includegraphics[width=0.42\textwidth]{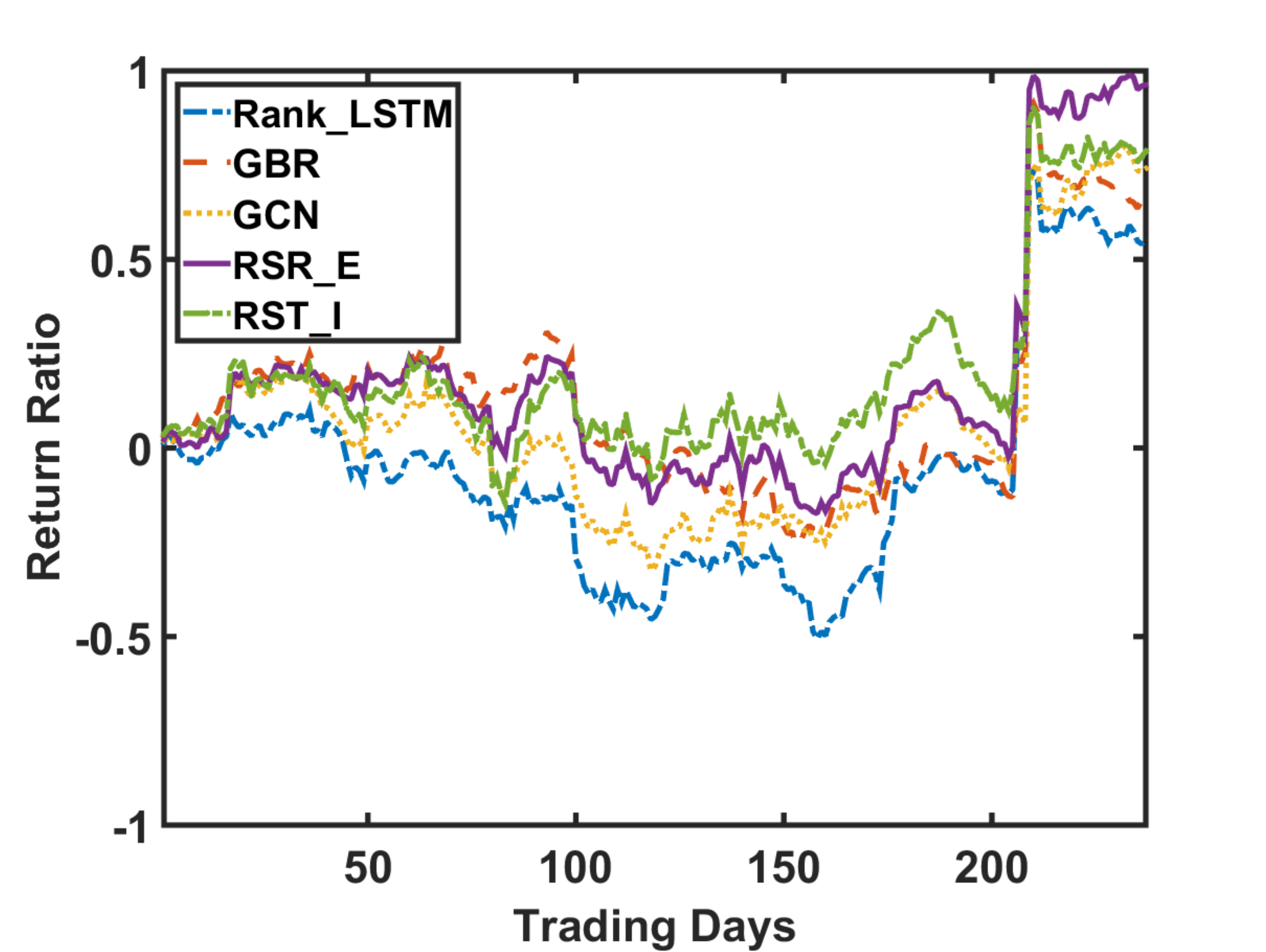}
		}
	}
	\vspace{-0.5cm}
	\caption{Performance comparison of relational ranking methods with Wiki relations regarding IRR.
	}
	\label{fig:rel_wiki}
	\vspace{-0.5cm}
\end{figure}

\tb{Effect of Wiki Relations}: Similarly, Table \ref{tab:rel_wiki} shows the results of considering the Wiki relation of stocks. We observe that: 
\begin{itemize}[leftmargin=*]
	\item In all cases, our proposed \tb{RSR\_E} and \tb{RSR\_I} achieve the best performance \wrt IRR. It further demonstrates the effectiveness of the approach we model stock relations, that is, the \textit{TGC} component. 
	\item All methods considering Wiki relations outperform \tb{Rank\_LSTM} with a significant improvement ($>$0.09) \wrt IRR on NYSE. It again verifies the merit of encoding stock relations in stock prediction and the effectiveness of the RSR framework. 
	% \item \tb{GBR} and \tb{GCN} perform worse than \tb{Rank\_LSTM} on NASDAQ. Recall that we ignore the type of relations while constructing the simple graph for \tb{GBR} and \tb{GCN}, \tb{GBR} and \tb{GCN} hence will equally propagate impacts to a stock $i$ from stocks having relation with $i$. Such operation will leads to noisy propagation when the market is volatile. This result indicates that stock relations should be carefully encoded on volatile stock markets like NASDAQ. 
\end{itemize}
Figure \ref{fig:rel_wiki} shows the associated back-testing procedure, {\color{blue}which presents similar trends as the results of considering industry relations (Figure~\ref{fig:rel_ind}).}

\begin{table}[tb]
	\centering
	%\vspace{-0.2cm}
	\caption{{\color{blue}Performance comparison among relational ranking methods with Wiki relations.}}
	\vspace{-0.4cm}
	\label{tab:rel_wiki}
	\resizebox{\textwidth}{!}{%
		\begin{tabular}{|c||c|c|c||c|c|c|}
			\hline
			\multirow{2}{*}{} & \multicolumn{3}{c||}{NASDAQ} & \multicolumn{3}{c|}{NYSE} \\ \cline{2-7} 
			& MSE & MRR & IRR & MSE & MRR & IRR \\ \hline \hline
			\textbf{Rank\_LSTM} & \textbf{3.79e-4$\pm$1.11e-6} & 4.17e-2$\pm$7.50e-3 & \textbf{0.68$\pm$0.60} & 2.28e-4$\pm$1.16e-6 & 3.79e-2$\pm$8.82e-3 & \textbf{0.56$\pm$0.68} \\ \hline \hline
			\textbf{GBR} & 3.80e-4$\pm$2.40e-7 & 3.32e-2$\pm$4.50e-3 & 0.33$\pm$0.34 & \textbf{2.26e-4$\pm$4.20e-7} & 3.64e-2$\pm$5.35e-3 & 0.65$\pm$0.27 \\ \hline
			\textbf{GCN} & \textbf{3.79e-4$\pm$9.70e-7} & 3.24e-2$\pm$3.21e-3 & 0.11$\pm$0.06 & \textbf{2.26e-4$\pm$6.60e-7} & 3.99e-2$\pm$1.03e-2 & 0.74$\pm$0.30 \\ \hline
			\textbf{RSR\_E} & 3.80e-4$\pm$7.20e-7 & 3.94e-2$\pm$8.15e-3 & 0.81$\pm$0.85 & 2.29e-4$\pm$2.77e-6 & 4.28e-2$\pm$6.18e-3 & \textbf{0.96$\pm$0.47} \\ \hline
			\textbf{RSR\_I} & \textbf{3.79e-4$\pm$6.60e-7} & \textbf{4.09e-2$\pm$5.18e-3} & \textbf{1.19$\pm$0.55} & \textbf{2.26e-4$\pm$1.37e-6} & 4.58e-2$\pm$5.55e-3 & 0.79$\pm$0.34 \\ \hline
		\end{tabular}%
	}
	\vspace{-0.4cm}
\end{table}

{\color{blue}\tb{Sector-wise Performance}: Taking \tb{RSR\_I} as an example, we then investigate whether the performance is sensitive to sectors via evaluating its performance over the stocks in each sector, \ie separately conducting back-testing for each sector. Recall that, on NASDAQ, the performance of \tb{RSR\_I} considering industry relations is not promising (with an IRR of 0.23). We take this case to investigate the sector-wise performance, which is presented in Table~\ref{tab:impact_ind}. Note that we only show the performance on sectors with the top-5 most stocks. We can see that, the method only achieves acceptable performance with an IRR of 1.12 on the \textit{Technology} sector. This result further indicates the less effectiveness of considering industry relation on the NASDAQ market, which is coherent with the results in Table~\ref{tab:rel_ind}. In addition, it also suggests the separate consideration of stocks in each single sector.}

{\color{blue}\tb{Importance of Each Type of Wiki relation}: By comparing the performance of \tb{RSR\_I} when a type of relation is removed, we investigate the importance of different types of Wiki relations. Table~\ref{tab:impact_wiki} shows the relative performance decrease \wrt IRR as compared to \tb{RSR\_I} with all Wiki relations as input (NASDAQ). Note that we only present the relations with the top-5 largest performance decreases. We can see that the most importance relation is the \textit{P1056\_P1056} of which \textit{P1056} denotes the predicate of \textit{product or material produced}. Mainly, two stocks have the relation of \textit{P1056\_P1056} means the associated companies collaborate on producing the same product. The high impact is reasonable, considering that collaborated companies have closely connected revenues and would be affected by similar factors.}
\begin{table}[]
	\centering
	\caption{{\color{blue}Performance of \textbf{RSR\_I} on ranking stocks in different sectors of NASDAQ \wrt IRR.}}
	\vspace{-0.4cm}
	\label{tab:impact_ind}
	\resizebox{0.7\textwidth}{!}{%
		\begin{tabular}{|c||c|c|c|c|c|}
			\hline
			Sector & Finance & Technology & N/A & Consumer Services & Health Care \\ \hline
			\#Stocks & 222 & 182 & 156 & 117 & 91 \\ \hline
			IRR & 0.33 & 1.12 & -0.70 & 0.57 & -0.85 \\ \hline
		\end{tabular}%
	}
	\vspace{-0.2cm}
\end{table}
\begin{table}[]
	\caption{{\color{blue}Impacts of different types of Wiki relation regarding the Relative Performance Decrease (\textbf{RPD}) of \textbf{RSR\_I} on NASDAQ as removing the selected relation.}}
	\vspace{-0.4cm}
	\label{tab:impact_wiki}
	\resizebox{0.8\textwidth}{!}{%
		\begin{tabular}{|c||c|c|c|c|c|}
			\hline
			Relation & P1056\_P1056 & P463\_P463 & P452\_P452 & P361\_P361 & P1056\_P452 \\ \hline
			ID in Table~\ref{tab:second_order_wiki} & 46 & R38 & 35 & 31 & 45 \\ \hline
			\#Occurrences & 130 & 58 & 506 & 1,194 & 10 \\ \hline \hline
			\textbf{RPD} & -144.00\% & -70.13\% & -21.66\% & -17.52\% & -15.71\% \\ \hline
		\end{tabular}%
	}
\vspace{-0.4cm}
\end{table}
\begin{figure*}[tb]
	\centering
	\mbox{
		\hspace{-0.2in}
		\subfigure[NASDAQ-Industry]{
			\label{fig:bt_nasdaq_indi}
			\includegraphics[width=0.27\textwidth]{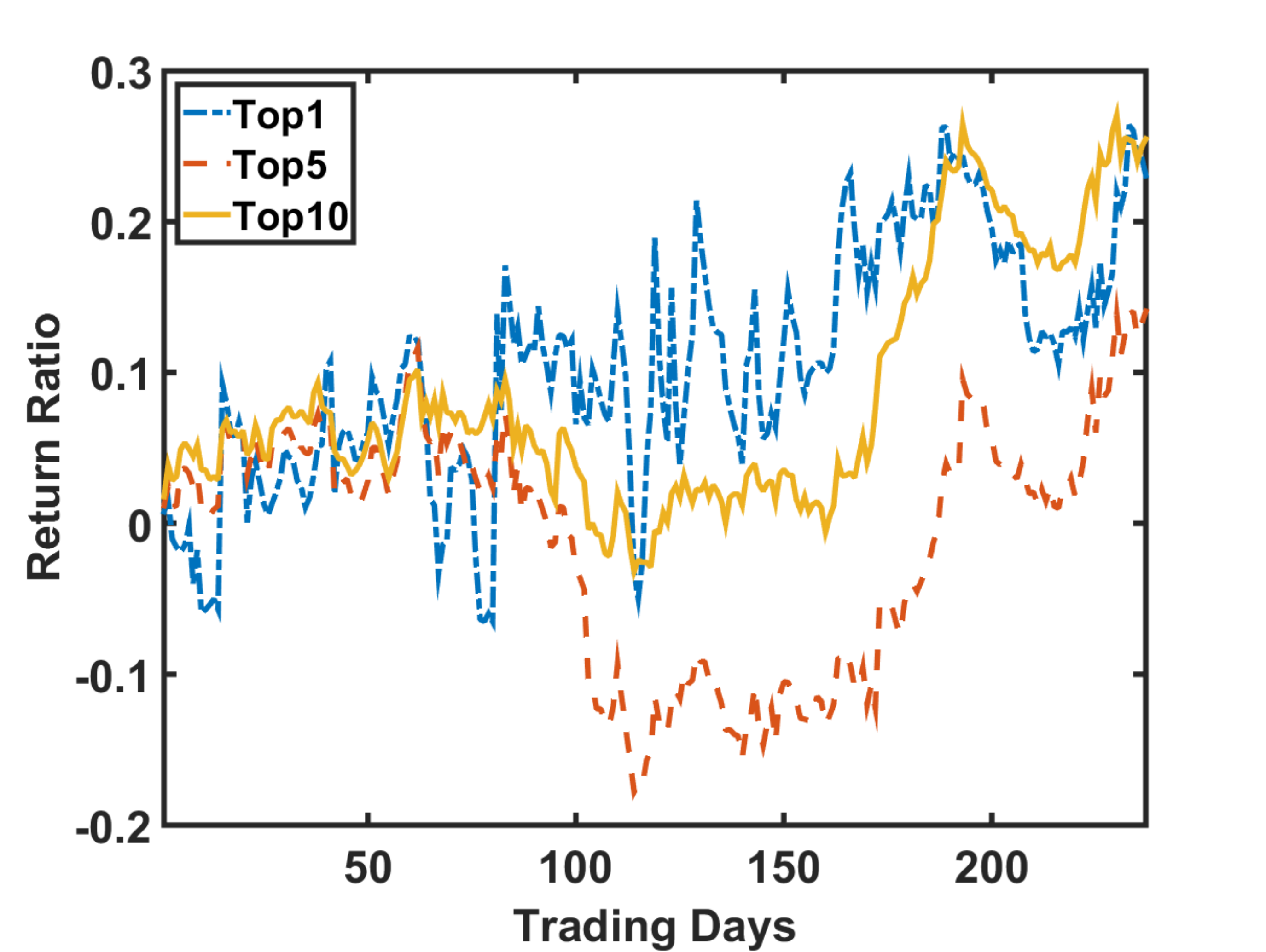}
		}
		\hspace{-0.3in}
		\subfigure[NASDAQ-Wiki]{
			\label{fig:bt_nasdaq_wikii}
			\includegraphics[width=0.27\textwidth]{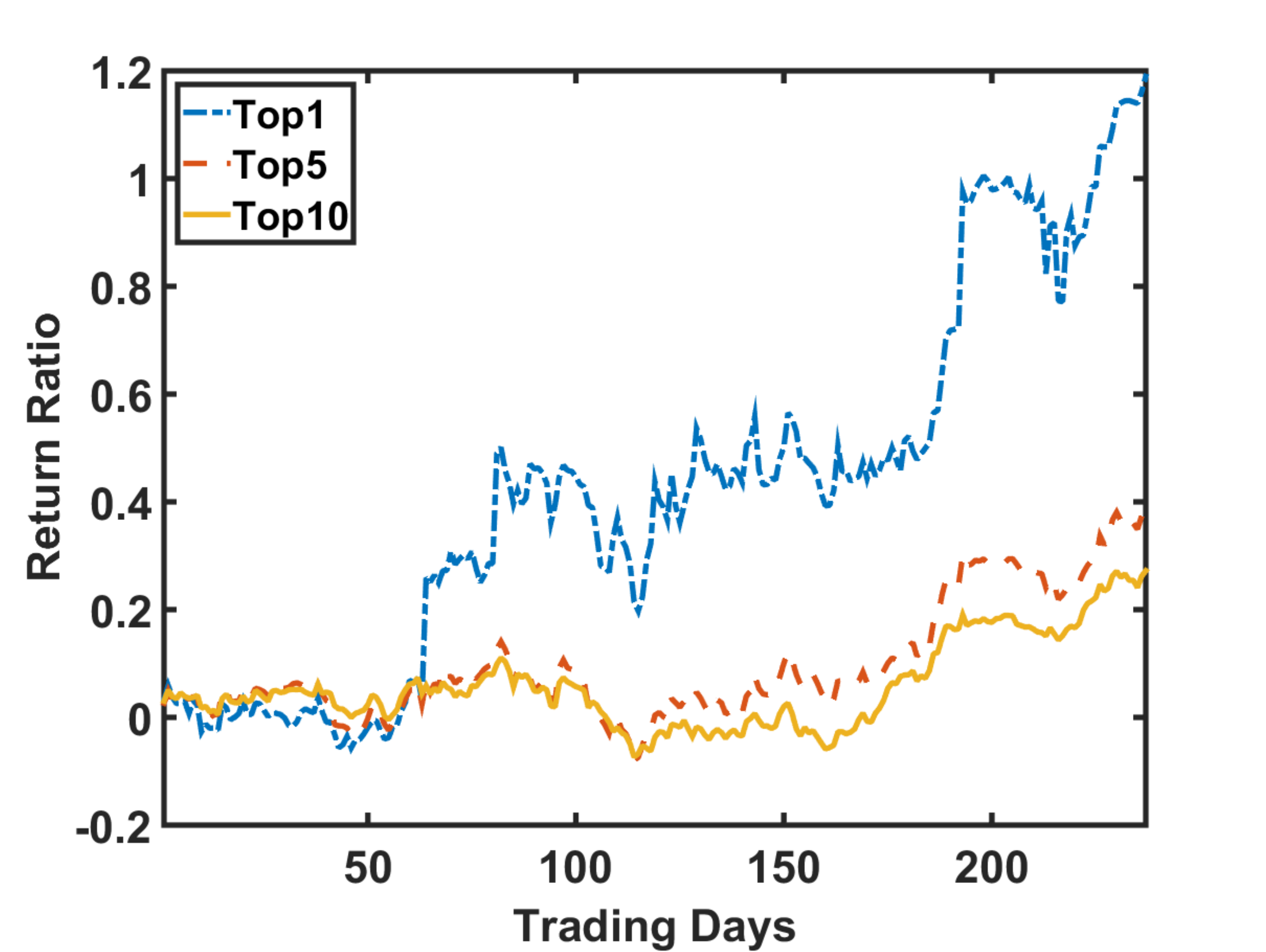}
		}
		\hspace{-0.3in}
		\subfigure[NYSE-Industry]{
			\label{fig:bt_nyse_indi}
			\includegraphics[width=0.27\textwidth]{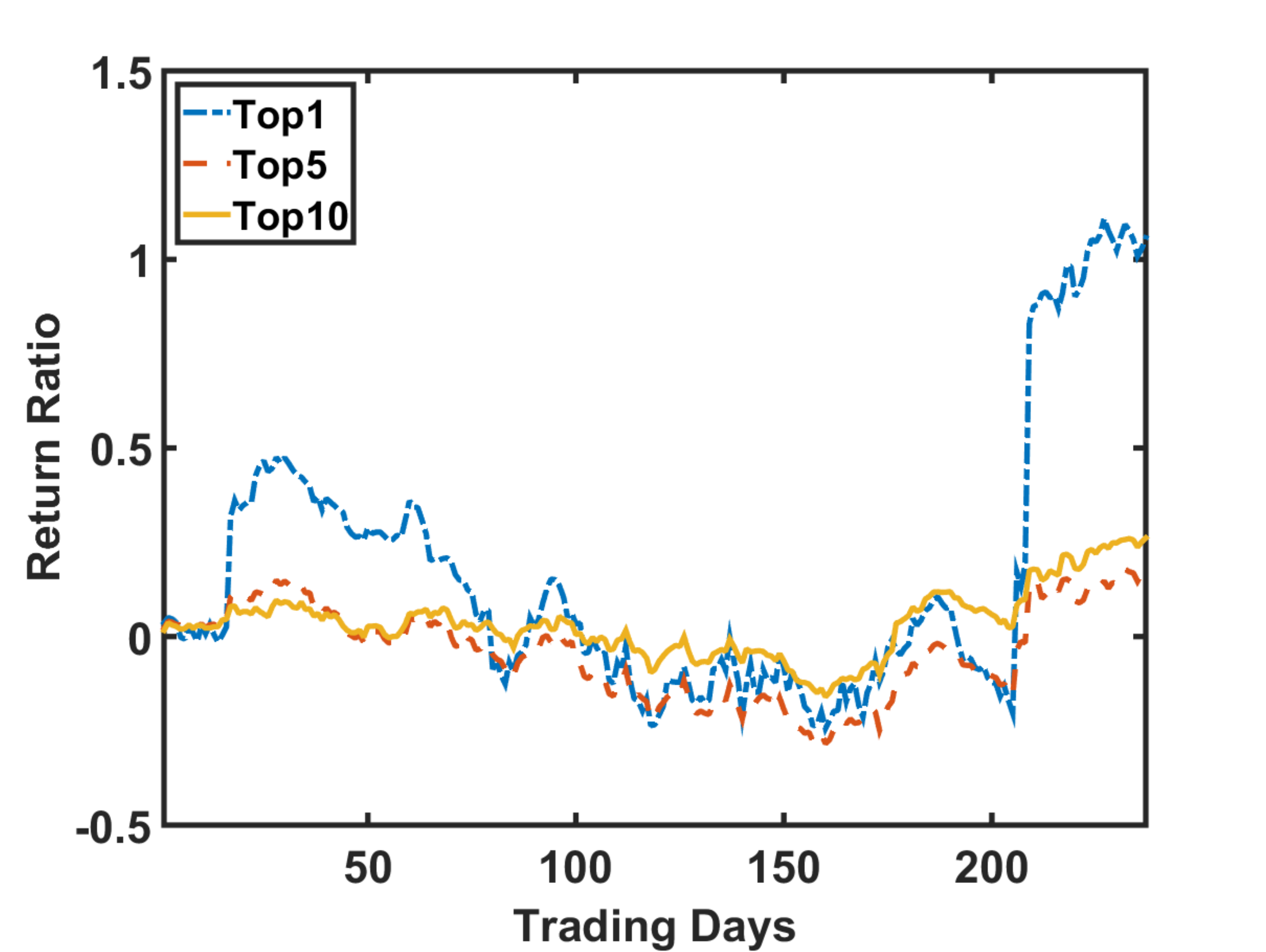}
		}
		\hspace{-0.3in}
		\subfigure[NYSE-Wiki]{
			\label{fig:bt_nyse_wikii}
			\includegraphics[width=0.27\textwidth]{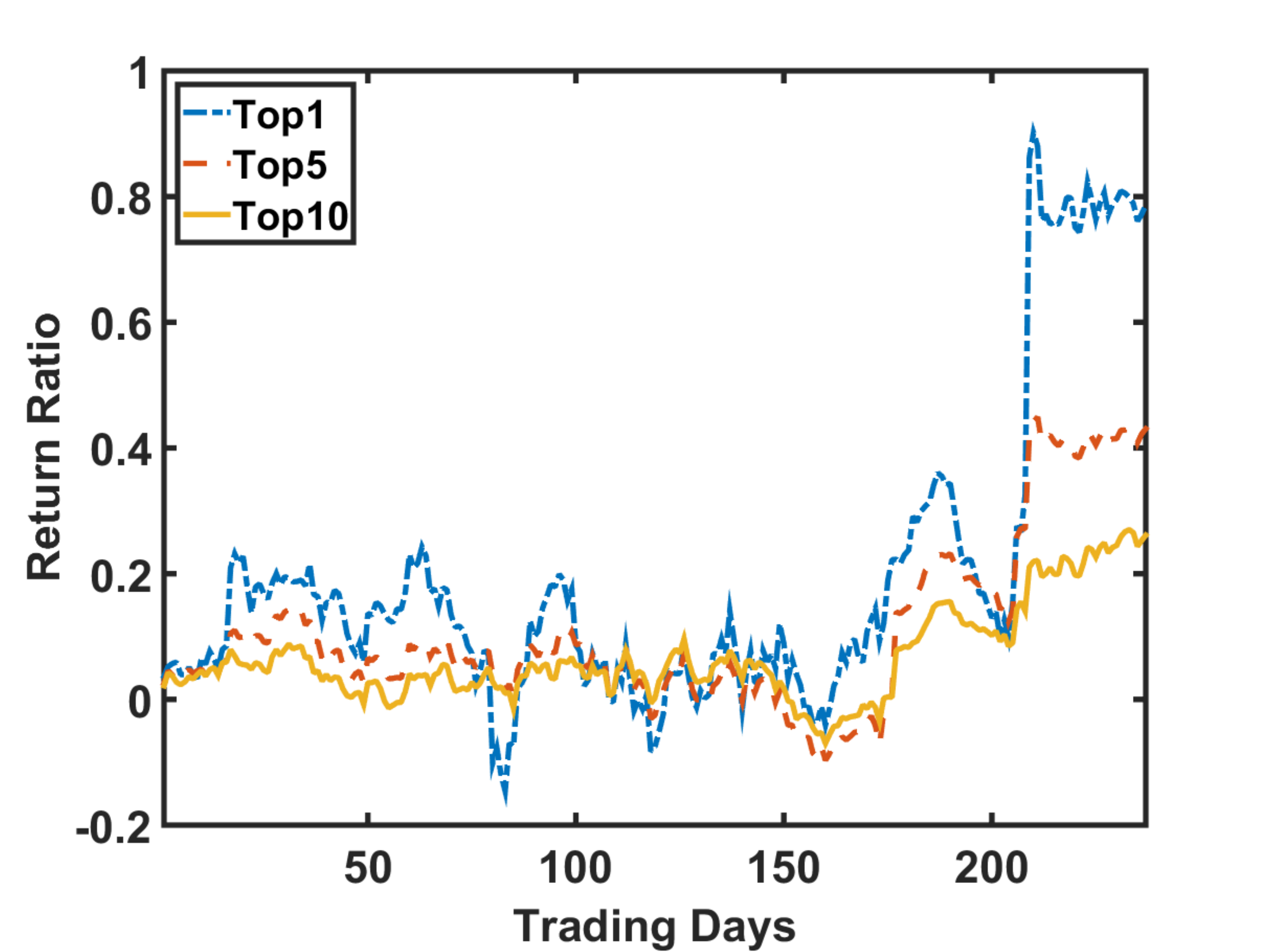}
		}
	}
	\vspace{-0.4cm}
	\caption{Comparison on back-testing strategies (\tb{Top1}, \tb{Top5}, and \tb{Top10}) \wrt IRR based on prediction of \tb{RSR\_I}.
	}
	\label{fig:backtest_i}
	\vspace{-0.2cm}
\end{figure*}
%\begin{figure*}[tb]
%	\centering
%	\mbox{
%		%\hspace{-0.1in}
%		\subfigure[NASDAQ-Industry]{
%			\label{fig:bt_nasdaq_inde}
%			\includegraphics[width=0.42\textwidth]{figure/backtest_NASDAQ_ind_inner}
%		}
%		\hspace{-0.25in}
%		\subfigure[NASDAQ-Wiki]{
%			\label{fig:bt_nasdaq_wikie}
%			\includegraphics[width=0.42\textwidth]{figure/backtest_NASDAQ_wiki_inner}
%		}
%		%\hspace{-0.25in}
%	}
%%\vspace{-0.5cm}
%	\mbox{
%		\subfigure[NYSE-Industry]{
%			\label{fig:bt_nyse_inde}
%			\includegraphics[width=0.42\textwidth]{figure/backtest_NYSE_ind_inner}
%		}
%		\hspace{-0.25in}
%		\subfigure[NYSE-Wiki]{
%			\label{fig:bt_nyse_wikie}
%			\includegraphics[width=0.42\textwidth]{figure/backtest_NYSE_wiki_inner}
%		}
%	}
%	\vspace{-0.4cm}
%	\caption{Comparison on back-testing strategies (\tb{Top1}, \tb{Top5}, and \tb{Top10}) \wrt IRR based on prediction of \tb{RSR\_E}.
%	}
%	\label{fig:backtest_e}
%	%\vspace{-0.2cm}
%\end{figure*}

\tb{Brief Conclusion}: a) Considering stock relations is helpful for stock ranking, especially on the stable markets (\eg NYSE). 2) The proposed \textit{TGC} is a promising solution for encoding stock relations. 3) It is important to consider appropriate relations suitable for the target market, for example, encoding industry relations on NASDAQ is a suboptimal choice.

\subsection{\mbox{Study on Back-testing Strategies (RQ3)}}
%\begin{figure*}[tb]
%	\centering
%	\mbox{
%		%\hspace{-0.1in}
%		\subfigure[NASDAQ-Industry]{
%			\label{fig:bt_nasdaq_indi}
%			\includegraphics[width=0.4\textwidth]{figure/backtest_NASDAQ_ind_add}
%		}
%		\hspace{-0.25in}
%		\subfigure[NASDAQ-Wiki]{
%			\label{fig:bt_nasdaq_wikii}
%			\includegraphics[width=0.4\textwidth]{figure/backtest_NASDAQ_wiki_add}
%		}
%		\vspace{-0.5cm}
%	}
%	%\hspace{-0.25in}
%	\mbox{
%		\subfigure[NYSE-Industry]{
%			\label{fig:bt_nyse_indi}
%			\includegraphics[width=0.4\textwidth]{figure/backtest_NYSE_ind_add}
%		}
%		\hspace{-0.25in}
%		\subfigure[NYSE-Wiki]{
%			\label{fig:bt_nyse_wikii}
%			\includegraphics[width=0.4\textwidth]{figure/backtest_NYSE_wiki_add}
%		}
%	}
%	\vspace{-0.4cm}
%	\caption{Comparison on back-testing strategies (\tb{Top1}, \tb{Top5}, and \tb{Top10}) \wrt IRR based on prediction of \tb{RSR\_E}.
%	}
%	\label{fig:backtest_i}
%	\vspace{-0.2cm}
%\end{figure*}
We then investigate the performance of our proposed methods under three different back-testing strategies, named \tb{Top1}, \tb{Top5}, and \tb{Top10}, buying stocks with top-1, 5, 10 highest expected revenue, respectively. {\color{blue}For instance, with the back-testing strategy of \tb{Top10}, we equally split our budget to trade the top-10 ranked stocks on each testing day. Note that we accordingly calculate the IRR by summing the mean return ratio of the 10 selected stocks on each testing day.} Figure \ref{fig:backtest_i} illustrates the performance comparison of these strategies with the predictions of \tb{RSR\_I}. {\color{blue}Similar trends are observed on the predictions of \tb{RSR\_E}, which are omitted for the consideration of saving space. From the figure, we have the following observations:} 
\begin{itemize}[leftmargin=*]
	\item \tb{RSR\_I} (Figure \ref{fig:bt_nasdaq_indi}) fails to achieve expected performance with different back-testing strategies under the NASDAQ-Industry setting (\ie ranking stocks in NASDAQ and modeling their industry relations). It further indicates the less effectiveness of industry relations on NASDAQ. 
	\item In the other cases, the performance of \tb{Top1}, \tb{Top5}, and \tb{Top10} on most testing days follows the order of \tb{Top1} $>$ \tb{Top5} $>$ \tb{Top10}, \ie the \tb{Top1} and \tb{Top10} achieve the highest and lowest IRR, respectively. {\color{blue}The reason could be that the ranking algorithm could accurately rank the relative order of stocks regarding future return ratios. Once the order is accurate, buying and selling the stock with higher expected profit (\eg, the top-1 ranked one) would achieve higher cumulative return ratio.}
	%It further demonstrates that our proposed methods consistently rank more profitable stocks at the top. 
	% \item \tb{RSR\_E} and \tb{RSR\_I} outperform the S\&P 500 Index and Dow Jones Industrial Average Index under different back-testing strategies (\tb{Top1}, \tb{Top5}, and \tb{Top10}) on the two markets, NASDAQ and NYSE. During the testing period, S\&P 500 Index and Dow Jones Industrial Average Index increase by a ratio of 0.17 (from 2,257.83 to 2,636.98)\footnote{\href{https://finance.yahoo.com/quote/\%5EGSPC/history?period1=1483200000\&period2=1512662400\&interval=1d\&filter=history\&frequency=1d}{https://tinyurl.com/yabfrxx4}} and 0.22 (from 19,881.76 to 24,294.12)\footnote{\href{https://finance.yahoo.com/quote/\%5EDJI/history?period1=1483200000\&period2=1512662400\&interval=1d\&filter=history\&frequency=1d}{https://tinyurl.com/y7389vds}}, respectively. It indicates that our prediction can help stock investors make stock selections.
\end{itemize}

{\color{blue}Considering that it would easier to achieve better performance regarding IRR in a bullish market, we further compare the performance of our method with two market indices, S\&P 500 Index and Dow Jones Industrial Average Index (DJI). Moreover, in order to better judge the achieved performance, we compare two more ideal investment strategies: a) selecting the stocks with highest return ratio (\eg Top10) in the testing period from the whole market; and b) among the stocks traded by the proposed method, selecting the stocks with highest return ratio in the testing period. Table~\ref{tab:bt_all} shows the performance of the compared investment strategies \wrt IRR. From which, we have the following observations:
\begin{itemize}[leftmargin=*]
	\item In the testing period, the stock market is bullish, which suggests future exploration of the proposed method in bearish market. In addition, noting that market indices are competitive portfolios (investment strategies)\footnote{From 2008 to 2017, the S\&P 500 achieved a return ratio of 125.8\%, beating most of the portfolios of funds of hedge funds.}, achieving IRR higher than market indices justifies the effectiveness of the proposed method.
	\item When trading the same number of stocks (\eg \tb{Top5}), the performance of the proposed method presents a significant gap towards the ideal investment strategies. This result is acceptable since accurately selecting the stock performing best in the range of almost one year is non-trivial, but reflects the huge improvement space for stock prediction methods.
	\item The \tb{Top1} version of our method, \ie trading the top-1 ranked stock on each trading day, achieves an IRR comparable to the investment strategy \tb{Selected} under \tb{Top10}. This further justifies the competitivity of our proposed method.
\end{itemize}
}
\begin{table}[tb]
	\centering
	%\vspace{-0.2cm}
	\caption{{\color{blue}Performance of RSR\_I as compared to market indices and ideal portfolios.}}
	\vspace{-0.4cm}
	\label{tab:bt_all}
	\resizebox{0.55\textwidth}{!}{%
	\begin{tabular}{|c||c|c|c||c|c|c|}
		\hline
		\multirow{2}{*}{} & \multicolumn{3}{c||}{NASDAQ} & \multicolumn{3}{c|}{NYSE} \\ \cline{2-7} 
		& Top1 & Top5 & Top10 & Top1 & Top5 & Top10 \\ \hline \hline
		\textbf{Market} & \textbf{3.40} & \textbf{2.36} & \textbf{1.99} & \textbf{2.42} & \textbf{1.90} & \textbf{1.47} \\ \hline
		\textbf{Selected} & 1.63 & 0.81 & 1.10 & 2.24 & 1.78 & 1.39 \\ \hline
		\textbf{RSR\_I} & 1.19 & 0.40 & 0.27 & 1.06 & 0.18 & 0.26 \\ \hline \hline
		\textbf{S\&P 500} & \multicolumn{6}{c|}{0.17} \\ \hline
		\textbf{DJI} & \multicolumn{6}{c|}{0.22} \\ \hline
	\end{tabular}%
	}
	\vspace{-0.3cm}
\end{table}

%% file: 2_related.tex
\section{Related Work}
\label{sec:related}
Our work is directly related to the recent work on \textit{stock prediction}, \textit{graph-based learning}, and \textit{knowledge graph embedding}.

\subsection{Stock Prediction}
Recent work on stock prediction can be separated into two main categories: \textit{stock price regression} and \textit{stock trend classification}. On one hand, Bao \etal predicted the 1-day ahead closing price of stocks with historical prices as input. 
%They decomposed the historical prices with a Wavelet Transform, filtered noises with a Stacked Autoencoder, and applied a LSTM for prediction 
The authors viewed the historical price as a signal and decomposed the historical price into multiple frequencies with a Wavelet Transform. They then filtered out noises in the frequency domain with a Stacked Autoencoder (SAE) and fed the output of SAE to an LSTM to make prediction \cite{bao2017deep}. 
Zhang \etal devised an extension of LSTM, which decomposes the historical prices into frequency domain with a Discrete Fourier Transform and equips each frequency with a memory state to capture the patterns in different frequencies \cite{zhang2017stock}. 
Instead of directly modeling the stock prices, Alberg and Lipton used a combination of an LSTM and Multi-Layer Perception to predict the future trend of fundamental indicators of a company and trade the corresponding stock based on the predicted indicators \cite{alberg2017improving}.

On the other hand, Nguyen and Shirai proposed a stock trend classification solution, which learns a topic distribution representation of each stock from posts mentioning it on stock message boards, and fed the topic representation into a Support Vector Machine to make the trend classification \cite{nguyen2015topic}. 
{\color{blue}Under a similar classification setting, another line of research is classifying the trend of a stock from relevant financial news reports~\cite{schumaker2009textual,ding2015deep,ding2014using,zhao2017constructing,hu2018listening}. For instance, Zhao \etal achieve it via constructing a event causality network of news reports and learning news embeddings from the causality networks, which is fed into a classification layer \cite{zhao2017constructing}.} 
Taking financial news as input as well, Hu \etal devised a neural network-based solution, named \textit{Hybrid Attention Networks}, which leverages a hybrid attention mechanism to attentively fuse multiple news reports mentioning a stock into a joint representation \cite{hu2018listening}. {\color{blue}In addition, textual contents mentioning stocks in social medial are also used to forecast the movement of stocks~\cite{li2018web}.} 

{\color{blue}However, none of the existing work is able to incorporate the rank/relative order among stocks regarding the expected revenue, tending to lead to suboptimal stock selections. Moreover, the existing work either totally ignores stock relations or heuristically models such relations. For instance, an intuitive consideration of sector-industry relation is to separately train a predictor for stocks under each sector~\cite{schumaker2008evaluating}. To the best of our knowledge, our work is to first one to leverage techniques of learning-to-rank to solve the stock prediction task and inject the stock relations into the learning framework with a new neural network component.}

\subsection{Graph-based Learning}
In the literature of graph-based learning, it has been intensively studied that incorporating the relationship among entities into the learning procedure of the target task to achieve better performance. Work on graph-based learning are mainly in two fashions: \textit{graph regularization} and \textit{graph convolution}. On one hand, Zhu \etal proposed a regularization term based on directed and undirected simple graphs with pair-wise entity relations to smooth the predictions across the topology of the graph \cite{zhu2003semi}. 
Zhou \etal regularized the learning procedure of the target task with a hypergraph that captures the higher-order relations among entities \cite{zhou2007learning}. On the other hand, Bruna \etal proposed spectral graph convolutions to capture the local connection patterns in graphs and propagate information of locally connected vertices for better representations \cite{bruna2014spectral}. Upon the spectral graph convolution, several fast approximations have been proposed for accelerating \cite{kipf2017semi, defferrard2016convolutional}. However, most of the works of graph-based learning fail to handle the temporal evolution property of stock market. {\color{blue}Consequently, they would suffer from severe information loss and achieve limited improvement when directly applied to model stock relations.}

\subsection{Knowledge Graph Embedding}
In a similar line, modeling the relations of two entities (a.k.a. knowledge graph embedding) has been intensively studied in the literature of knowledge graphs. Bordes, \etal represented entities and relations with embedding vectors and transferred entity embeddings through adding relation embedding \cite{bordes2013translating}. Similarly, Socher \etal represented relations as matrices and transferred entity embeddings via matrix multiplication \cite{socher2013reasoning}. Such techniques mainly focus on solving knowledge graph-oriented problems such as knowledge graph completion, while we target on a different problem setting of stock ranking. The idea of embedding propagation that revises stock sequential embeddings through stock relations is partially inspired by TransE, our TGC is more generic in terms of jointly capturing temporal properties and topologies of relations.

%% file: 7_conclusion.tex
\section{Conclusions}
\label{sec:conclusion}
In this paper, we formulated stock prediction as a ranking task and demonstrated the potential of learning-to-rank methods for predicting stocks. To solve the problem, we proposed a \textit{Relational Stock Ranking} framework. The core of the framework is a neural network modeling component, named \textit{Temporal Graph Convolution}, which can handle the impact between different stocks by encoding stock relations in a time-sensitive way. Experimental results on NASDAQ and NYSE demonstrate the effectiveness of our solution --- with three different back-testing strategies, the RSR framework outperforms the S\&P 500 Index with significantly higher return ratio.

As mentioned in Section \ref{sec:experiment}, we will explore the potential of emphasizing top-ranked entities with more advanced learning-to-rank techniques. In addition, we will integrate risk management techniques in finance into the RSR framework to force the predictions to be risk sensitive. Furthermore, we will investigate the performance of RSR under multiple investment operations such as \textit{buy-hold-sell} (\aka long position) and \textit{borrow-sell-buy} (\aka short position). Moreover, we will integrate alternative data such as financial news and social media contents into the predictive model. {\color{blue}Lastly, considering that the proposed TGC is a general component to model relational data, especially structured domain knowledge, we would like to explore the potential of TGC in enhancing the neural network solutions for tasks with such relational data, such as recommender system and product search.}

%% file: 8_appendix.tex
\appendix
%\vspace{+10pt}
\section{Stock Relation}
In this appendix, we describe the details of stock relations (\ie \textit{sector-industry relations} and \textit{Wiki company-based relations}) in our collected data (Section \ref{sec:data}).

\subsection{Sector-Industry Relation}
\label{ss:sec_ind_rel}
We extract 112 and 130 types of industry relations from the {\color{blue} company classification} hierarchy structure of NASDAQ and NYSE stocks, respectively. Table \ref{tab:ind_rel_nasdaq} and \ref{tab:ind_rel_nyse} illustrates the specific industry relations in NYSDAQ and NYSE markets, respectively.

% \usepackage{longtable}
% Note: It may be necessary to compile the document several times to get a multi-page table to line up properly
\footnotesize
\begin{longtable}[c]{|c|l|c|}
	\caption{Industry relations among 1,026 selected stocks from the NASDAQ market.}
	\vspace{-0.3cm}
	\label{tab:ind_rel_nasdaq}\\
	\hline
	Sectors                                                         & \multicolumn{1}{c|}{Industries}                                                                                                                                                                                                                                                                                                                                                                                                                                                                                                                                                                                                       & \begin{tabular}[c]{@{}c@{}}Count of \\ Industries\end{tabular} \\ \hline
	\endfirsthead
	\endhead
	\begin{tabular}[c]{@{}c@{}}Consumer\\ Durables\end{tabular}     & \begin{tabular}[c]{@{}l@{}}Office Equipment/Supplies/Services, Consumer Specialties, \\ Specialty Chemicals, Metal Fabrications, Consumer \\ Electronics/Appliances, Building Products, Containers/\\ Packaging, Miscellaneous manufacturing industries, \\ Automotive Aftermarket\end{tabular}                                                                                                                                                                                                                                                                                                                                       & 9                                                              \\ \hline
	Transportation                                                  & \begin{tabular}[c]{@{}l@{}}Transportation Services, Air Freight/Delivery Services, \\ Trucking Freight/Courier Services, Oil Refining/Marketing\end{tabular}                                                                                                                                                                                                                                                                                                                                                                                                                                                                          & 4                                                              \\ \hline
	Finance                                                         & \begin{tabular}[c]{@{}l@{}}Specialty Insurers, Commercial Banks, Savings Institutions, \\ Real Estate, Major Banks, Investment Managers, Investment \\ Bankers/Brokers/Service, Life Insurance, Finance: Consumer \\ Services, Banks, Property-Casualty Insurers, Finance \\ Companies\end{tabular}                                                                                                                                                                                                                                                                                                                                   & 12                                                             \\ \hline
	\begin{tabular}[c]{@{}c@{}}Public\\ Utilities\end{tabular}      & \begin{tabular}[c]{@{}l@{}}Telecommunications Equipment, Environmental Services, \\ Natural Gas Distribution\end{tabular}                                                                                                                                                                                                                                                                                                                                                                                                                                                                                                             & 3                                                              \\ \hline
	Energy                                                          & Electric Utilities: Central, Coal Mining, Oil \& Gas Production                                                                                                                                                                                                                                                                                                                                                                                                                                                                                                                                                                       & 3                                                              \\ \hline
	Miscellaneous                                                   & Business Services, Publishing, Multi-Sector Companies                                                                                                                                                                                                                                                                                                                                                                                                                                                                                                                                                                                 & 3                                                              \\ \hline
	\begin{tabular}[c]{@{}c@{}}Consumer\\ Non-Durables\end{tabular} & \begin{tabular}[c]{@{}l@{}}Plastic Products, Meat/Poultry/Fish, Beverages (Production/\\ Distribution), Shoe Manufacturing, Packaged Foods, Package \\ Goods/Cosmetics, Apparel, Farming/Seeds/Milling, Food \\ Distributors, Specialty Foods, Recreational Products/Toys\end{tabular}                                                                                                                                                                                                                                                                                                                                                & 11                                                             \\ \hline
	Health Care                                                     & \begin{tabular}[c]{@{}l@{}}Ophthalmic Goods, Medical/Nursing Services, Hospital/\\ Nursing Management, Biotechnology: In Vitro \& In Vivo \\ Diagnostic Substances, Biotechnology: Commercial Physical\\ \& Biological Resarch, Other Pharmaceuticals, Major \\ Pharmaceuticals, Medical Specialities, Medical Electronics, \\ Biotechnology: Electromedical \& Electrotherapeutic \\ Apparatus, Biotechnology: Biological Products, Medical/\\ Dental Instruments, Industrial Specialties\end{tabular}                                                                                                                               & 13                                                             \\ \hline
	\begin{tabular}[c]{@{}c@{}}Consumer\\ Services\end{tabular}     & \begin{tabular}[c]{@{}l@{}}Other Consumer Services, Restaurants, Clothing/Shoe/\\ Accessory Stores, Marine Transportation, Television \\ Services, Consumer Electronics/Video Chains, Other \\ Specialty Stores, Home Furnishings, Diversified \\ Commercial Services, Paper, Professional Services, Hotels/\\ Resorts, Rental/Leasing Companies, Real Estate Investment \\ Trusts, Food Chains, Broadcasting, Books, Motor Vehicles, \\ Movies/Entertainment, RETAIL: Building Materials, \\ Advertising, Catalog/Specialty Distribution, Services-Misc. \\ Amusement \& Recreation, Department/Specialty Retail Stores\end{tabular} & 24                                                             \\ \hline
	\begin{tabular}[c]{@{}c@{}}Basic\\ Industries\end{tabular}      & \begin{tabular}[c]{@{}l@{}}Water Supply, Miscellaneous, Forest Products, Precious \\ Metals, Mining \& Quarrying of Nonmetallic Minerals, \\ Engineering \& Construction, Major Chemicals\end{tabular}                                                                                                                                                                                                                                                                                                                                                                                                                                & 7                                                              \\ \hline
	\begin{tabular}[c]{@{}c@{}}Capital\\ Goods\end{tabular}         & \begin{tabular}[c]{@{}l@{}}Military/Government/Technical, Biotechnology: Laboratory \\ Analytical Instruments, Electrical Products, Building Materials, \\ Railroads, Ordnance And Accessories, Homebuilding,\\ Electronic Components, Aerospace, Industrial Machinery/\\ Components, Construction/Ag Equipment/Trucks, Auto Parts:\\ O.E.M., Steel/Iron Ore, Auto Manufacturing\end{tabular}                                                                                                                                                                                                                                         & 14                                                             \\ \hline
	Technology                                                      & \begin{tabular}[c]{@{}l@{}}Computer Manufacturing, Radio And Television Broadcasting\\ And Communications Equipment, Computer Communications\\ Equipment, Computer peripheral equipment, EDP Services, \\ Computer Software: Programming, Data Processing, Computer\\ Software: Prepackaged Software, Semiconductors, Retail: \\ Computer Software \& Peripheral Equipment\end{tabular}                                                                                                                                                                                                                                               & 9                                                              \\ \hline
	N/A                                                             & N/A                                                                                                                                                                                                                                                                                                                                                                                                                                                                                                                                                                                                                                   & 1                                                              \\ \hline
\end{longtable}

% \usepackage{longtable}
% Note: It may be necessary to compile the document several times to get a multi-page table to line up properly
% \usepackage{longtable}
% Note: It may be necessary to compile the document several times to get a multi-page table to line up properly
\begin{longtable}[c]{|c|l|c|}
	\caption{Industry relations among 1,737 selected stocks from the NYSE market.}
	\vspace{-0.3cm}
	\label{tab:ind_rel_nyse}\\
	\hline
	Sectors                                                         & \multicolumn{1}{c|}{Industries}                                                                                                                                                                                                                                                                                                                                                                                                                                                                                                                                         & \begin{tabular}[c]{@{}c@{}}Count of \\ Industries\end{tabular} \\ \hline
	\endfirsthead
	\endhead
	\begin{tabular}[c]{@{}c@{}}Consumer\\ Durables\end{tabular}     & \begin{tabular}[c]{@{}l@{}}Electrical Products, Home Furnishings, Specialty Chemicals, Metal\\ Fabrications, Consumer Electronics/Appliances, Building Products,\\ Miscellaneous manufacturing industries, Containers/Packaging, \\ Publishing, Automotive Aftermarket, Industrial Specialties\end{tabular}                                                                                                                                                                                                                                                             & 11                                                             \\ \hline
	Transportation                                                  & \begin{tabular}[c]{@{}l@{}}Transportation Services, Air Freight/Delivery Services, Trucking \\ Freight/Courier Services, Railroads, Oil Refining/Marketing\end{tabular}                                                                                                                                                                                                                                                                                                                                                                                                 & 5                                                              \\ \hline
	Finance                                                         & \begin{tabular}[c]{@{}l@{}}Finance/Investors Services, Specialty Insurers, Commercial Banks,\\ Savings Institutions, Real Estate, Major Banks, Investment Managers,\\ Investment Bankers/Brokers/Service, Life Insurance, Diversified \\ Financial Services, Accident \&Health Insurance, Finance: Consumer\\ Services, Banks, Property-Casualty Insurers, Finance Companies\end{tabular}                                                                                                                                                                               & 15                                                             \\ \hline
	\begin{tabular}[c]{@{}c@{}}Public\\ Utilities\end{tabular}      & \begin{tabular}[c]{@{}l@{}}Electric Utilities: Central, Telecommunications Equipment, Oil/Gas\\ Transmission, Water Supply, Power Generation\end{tabular}                                                                                                                                                                                                                                                                                                                                                                                                               & 5                                                              \\ \hline
	Energy                                                          & \begin{tabular}[c]{@{}l@{}}Coal Mining, Oil \& Gas Production, Integrated oil Companies,\\ Oilfield Services/Equipment, Natural Gas Distribution\end{tabular}                                                                                                                                                                                                                                                                                                                                                                                                           & 5                                                              \\ \hline
	Miscellaneous                                                   & \begin{tabular}[c]{@{}l@{}}Business Services, Office Equipment/Supplies/Services, Multi-\\ Sector Companies\end{tabular}                                                                                                                                                                                                                                                                                                                                                                                                                                                & 3                                                              \\ \hline
	\begin{tabular}[c]{@{}c@{}}Consumer\\ Non-Durables\end{tabular} & \begin{tabular}[c]{@{}l@{}}Electronic Components, Plastic Products, Meat/Poultry/Fish, Shoe\\ Manufacturing, Beverages, Packaged Foods, Consumer Specialties,\\ Apparel, Farming/Seeds/Milling, Food Distributors, Specialty Foods,\\ Motor Vehicles, Recreational Products/Toys\end{tabular}                                                                                                                                                                                                                                                                           & 13                                                             \\ \hline
	Health Care                                                     & \begin{tabular}[c]{@{}l@{}}Ophthalmic Goods, Medical/Nursing Services, Hospital/Nursing\\ Management, Major Pharmaceuticals, Biotechnology: Commercial\\ Physical Resarch, Biotechnology: Electromedical Apparatus, Other\\ Pharmaceuticals, Medical/Dental Instruments, Medical Specialities\end{tabular}                                                                                                                                                                                                                                                              & 9                                                              \\ \hline
	\begin{tabular}[c]{@{}c@{}}Consumer\\ Services\end{tabular}     & \begin{tabular}[c]{@{}l@{}}Other Consumer Services, Restaurants, Clothing/Shoe/Accessory\\ Stores, Electronics/Video Chains, Other Specialty Stores, Home\\ Furnishings, Diversified Commercial Services, Paper, Professional\\ Services, Hotels/Resorts, Rental/Leasing Companies, Real Estate \\ Investment Trusts, Food Chains, Broadcasting, Books, Motor \\ Vehicles, Movies/Entertainment, RETAIL: Building Materials, \\ Advertising, Catalog/Specialty Distribution, Services-Misc. \\ Amusement \& Recreation, Department/Specialty Retail Stores\end{tabular} & 24                                                             \\ \hline
	\begin{tabular}[c]{@{}c@{}}Basic\\ Industries\end{tabular}      & \begin{tabular}[c]{@{}l@{}}Water Supply, Miscellaneous, Forest Products, Precious \\ Metals, Mining \& Quarrying of Nonmetallic Minerals, \\ Engineering \& Construction, Major Chemicals\end{tabular}                                                                                                                                                                                                                                                                                                                                                                  & 7                                                              \\ \hline
	\begin{tabular}[c]{@{}c@{}}Capital\\ Goods\end{tabular}         & \begin{tabular}[c]{@{}l@{}}Package Goods/Cosmetics, Forest Products, Precious Metals, \\ Environmental Services, Paper, Agricultural Chemicals, Mining \&\\ Quarrying of Nonmetallic Minerals, Engineering \& Construction, \\ General Bldg Contractors - Nonresidential Bldgs, Aluminum, \\ Major Chemicals, Paints/Coatings, Steel/Iron Ore, Textiles\end{tabular}                                                                                                                                                                                                    & 13                                                             \\ \hline
	Technology                                                      & \begin{tabular}[c]{@{}l@{}}Computer Manufacturing, Computer peripheral equipment, Computer\\ Software: Programming, Semiconductors, Data Processing, Computer\\ Software: Prepackaged Software, Diversified Commercial Services, \\ Professional Services, Computer Communications Equipment, EDP \\ Services, Retail: Computer Software \& Peripheral Equipment, Radio\\ And Television Broadcasting Equipment, Advertising\end{tabular}                                                                                                                               & 12                                                             \\ \hline
	N/A                                                             & N/A                                                                                                                                                                                                                                                                                                                                                                                                                                                                                                                                                                     & 1                                                              \\ \hline
\end{longtable}
\normalsize

\subsection{Wiki Company-based Relations}
\label{ss:wiki_com_rel}
From Wikidata, one of the biggest and most active open domain knowledge bases, we obtain 5 and 53 types of first-order (in the format of \textcircled{$A$} $\xrightarrow{R}$ \textcircled{$B$}) and second-order relations (in the format of \textcircled{$A$} $\xrightarrow{R_1}$ \textcircled{$C$} $\xleftarrow{R_2}$ \textcircled{$B$}) between companies corresponding to the selected stocks in NASDAQ and NYSE markets, respectively. Note that $A$ and $B$ denote entities in Wikidata corresponding to two companies; $C$ denotes another entity bridging two company-entities in a second-order relation; $R$, $R_1$, and $R_2$ denotes different types of entity relation defined in Wikidata\footnote{\href{https://www.wikidata.org/wiki/Wikidata:List_of_properties/all}{https://www.wikidata.org/wiki/Wikidata:List\_of\_properties/all}}. In Table \ref{tab:first_order_wiki} and \ref{tab:second_order_wiki}, we summarize the extracted first-order and second-order relations, respectively.

% \usepackage{longtable}
% Note: It may be necessary to compile the document several times to get a multi-page table to line up properly
\footnotesize
\begin{longtable}[c]{|c|c|l|}
	\caption{{\color{blue}First-order Wiki company-based relations in the format of \textcircled{$A$} $\xrightarrow{R}$ \textcircled{$B$}.}}
	\vspace{-0.3cm}
	\label{tab:first_order_wiki}\\
	\hline
	& \begin{tabular}[c]{@{}c@{}}Wikidata\\ Relation ($R$)\end{tabular}           & \multicolumn{1}{c|}{Relation Description}                                              \\ \hline
	\endfirsthead
	\endhead
	1 & \href{https://www.wikidata.org/wiki/Property:P127}{P127} & \textit{Owned by}: owner of the subject.                                                        \\ \hline
	2 & \href{https://www.wikidata.org/wiki/Property:P155}{P155} & \textit{Follows}: immediately prior item in a series of which the subject is a part.            \\ \hline
	3 & \href{https://www.wikidata.org/wiki/Property:P156}{P156} & \textit{Followed by}: immediately following item in a series of which the subject is a part.    \\ \hline
	4 & \href{https://www.wikidata.org/wiki/Property:P355}{P355} & \textit{Subsidiary}: subsidiary of a company or organization.                                   \\ \hline
	5 & \href{https://www.wikidata.org/wiki/Property:P749}{P749} & \textit{Parent organization}: parent organization of an organisation, opposite of subsidiaries. \\ \hline
\end{longtable}
% \usepackage{multirow}
% \usepackage{longtable}
% Note: It may be necessary to compile the document several times to get a multi-page table to line up properly
\begin{longtable}[c]{|c|c|l|}
	\caption{{\color{blue}Second-order Wiki company-based relations in the format of \textcircled{$A$} $\xrightarrow{R_1}$ \textcircled{$C$} $\xleftarrow{R_2}$ \textcircled{$B$}.}}
	\vspace{-0.3cm}
	\label{tab:second_order_wiki}\\
	\hline
	& \begin{tabular}[c]{@{}c@{}}Wikidata\\ Relations\end{tabular}                & \multicolumn{1}{c|}{Relation Descriptions}                                             \\ \hline
	\endfirsthead
	\endhead
	\multirow{2}{*}{1}	& $R_1$ = \href{https://www.wikidata.org/wiki/Property:P31}{P31}	& \textit{Instance of}: that class of which this subject is a particular example and member.\\ \cline{2-3}
	& $R_2$ = \href{https://www.wikidata.org/wiki/Property:P366}{P366}	& \textit{Use}: main use of the subject.\\ \hline
	
	\multirow{2}{*}{2}	& $R_1$ = \href{https://www.wikidata.org/wiki/Property:P31}{P31}	& \textit{Instance of}: that class of which this subject is a particular example and member.\\ \cline{2-3}
	& $R_2$ = \href{https://www.wikidata.org/wiki/Property:P452}{P452}	& \textit{Industry}: industry of company or organization.\\ \hline
	
	\multirow{2}{*}{3}	& $R_1$ = \href{https://www.wikidata.org/wiki/Property:P31}{P31}	& \textit{Instance of}: that class of which this subject is a particular example and member.\\ \cline{2-3}
	& $R_2$ = \href{https://www.wikidata.org/wiki/Property:P1056}{P1056}	& \textit{Product or material produced}: material or product produced by an agency.\\ \hline
	
	\multirow{2}{*}{4}	& $R_1$ = \href{https://www.wikidata.org/wiki/Property:P112}{P112}	& \textit{Founded by}: founder or co-founder of this organization.\\ \cline{2-3}
	& $R_2$ = \href{https://www.wikidata.org/wiki/Property:P112}{P112}	& \textit{Founded by}: founder or co-founder of this organization.\\ \hline
	
	\multirow{2}{*}{5}	& $R_1$ = \href{https://www.wikidata.org/wiki/Property:P112}{P112}	& \textit{Founded by}: founder or co-founder of this organization.\\ \cline{2-3}
	& $R_2$ = \href{https://www.wikidata.org/wiki/Property:P127}{P127}	& \textit{Owned by}: owner of the subject.\\ \hline
	
	\multirow{2}{*}{6}	& $R_1$ = \href{https://www.wikidata.org/wiki/Property:P112}{P112}	& \textit{Founded by}: founder or co-founder of this organization.\\ \cline{2-3}
	& $R_2$ = \href{https://www.wikidata.org/wiki/Property:P169}{P169}	& \textit{Chief executive officer}: the CEO within an organization.\\ \hline
	
	\multirow{2}{*}{7}	& $R_1$ = \href{https://www.wikidata.org/wiki/Property:P113}{P113}	& \textit{Airline hub}: airport that serves as a hub for an airline.\\ \cline{2-3}
	& $R_2$ = \href{https://www.wikidata.org/wiki/Property:P113}{P113}	& \textit{Airline hub}: airport that serves as a hub for an airline.\\ \hline
	
	\multirow{2}{*}{8}	& $R_1$ = \href{https://www.wikidata.org/wiki/Property:P114}{P114}	& \textit{Airline alliance}: alliance the airline belongs to.\\ \cline{2-3}
	& $R_2$ = \href{https://www.wikidata.org/wiki/Property:P114}{P114}	& \textit{Airline alliance}: alliance the airline belongs to.\\ \hline
	
	\multirow{2}{*}{9}	& $R_1$ = \href{https://www.wikidata.org/wiki/Property:P121}{P121}	& \textit{Item operated}: equipment, installation or service operated by the subject.\\ \cline{2-3}
	& $R_2$ = \href{https://www.wikidata.org/wiki/Property:P1056}{P1056}	& \textit{Product or material produced}: material or product produced by an agency.\\ \hline
	
	\multirow{2}{*}{10}	& $R_1$ = \href{https://www.wikidata.org/wiki/Property:P121}{P121}	& \textit{Item operated}: equipment, installation or service operated by the subject.\\ \cline{2-3}
	& $R_2$ = \href{https://www.wikidata.org/wiki/Property:P121}{P121}	& \textit{Item operated}: equipment, installation or service operated by the subject.\\ \hline
	
	\multirow{2}{*}{11}	& $R_1$ = \href{https://www.wikidata.org/wiki/Property:P127}{P127}	& \textit{Owned by}: owner of the subject.\\ \cline{2-3}
	& $R_2$ = \href{https://www.wikidata.org/wiki/Property:P112}{P112}	& \textit{Founded by}: founder or co-founder of this organization.\\ \hline
	
	\multirow{2}{*}{12}	& $R_1$ = \href{https://www.wikidata.org/wiki/Property:P127}{P127}	& \textit{Owned by}: owner of the subject.\\ \cline{2-3}
	& $R_2$ = \href{https://www.wikidata.org/wiki/Property:P127}{P127}	& \textit{Owned by}: owner of the subject.\\ \hline
	
	\multirow{2}{*}{13}	& $R_1$ = \href{https://www.wikidata.org/wiki/Property:P127}{P127}	& \textit{Owned by}: owner of the subject.\\ \cline{2-3}
	& $R_2$ = \href{https://www.wikidata.org/wiki/Property:P169}{P169}	& \textit{Chief executive officer}: the CEO within an organization.\\ \hline
	
	\multirow{2}{*}{14}	& $R_1$ = \href{https://www.wikidata.org/wiki/Property:P127}{P127}	& \textit{Owned by}: owner of the subject.\\ \cline{2-3}
	& $R_2$ = \href{https://www.wikidata.org/wiki/Property:P355}{P355}	& \textit{Subsidiary}: subsidiary of a company or organization.\\ \hline
	
	\multirow{2}{*}{15}	& $R_1$ = \href{https://www.wikidata.org/wiki/Property:P127}{P127}	& \textit{Owned by}: owner of the subject.\\ \cline{2-3}
	& $R_2$ = \href{https://www.wikidata.org/wiki/Property:P749}{P749}	& \textit{Parent organization}: parent organization of an organisation.\\ \hline
	
	\multirow{2}{*}{16}	& $R_1$ = \href{https://www.wikidata.org/wiki/Property:P127}{P127}	& \textit{Owned by}: owner of the subject.\\ \cline{2-3}
	& $R_2$ = \href{https://www.wikidata.org/wiki/Property:P1830}{P1830}	& \textit{Owner of}: entities owned by the subject.\\ \hline
	
	\multirow{2}{*}{17}	& $R_1$ = \href{https://www.wikidata.org/wiki/Property:P127}{P127}	& \textit{Owned by}: owner of the subject.\\ \cline{2-3}
	& $R_2$ = \href{https://www.wikidata.org/wiki/Property:P3320}{P3320}	& \textit{Board member}: member(s) of the board for the organization.\\ \hline
	
	\multirow{2}{*}{18}	& $R_1$ = \href{https://www.wikidata.org/wiki/Property:P155}{P155}	& \textit{Follows}: immediately prior item in a series of which the subject is a part.\\ \cline{2-3}
	& $R_2$ = \href{https://www.wikidata.org/wiki/Property:P155}{P155}	& \textit{Follows}: immediately prior item in a series of which the subject is a part.\\ \hline
	
	\multirow{2}{*}{19}	& $R_1$ = \href{https://www.wikidata.org/wiki/Property:P155}{P155}	& \textit{Follows}: immediately prior item in a series of which the subject is a part.\\ \cline{2-3}
	& $R_2$ = \href{https://www.wikidata.org/wiki/Property:P355}{P355}	& \textit{Subsidiary}: subsidiary of a company or organization.\\ \hline
	
	\multirow{2}{*}{20}	& $R_1$ = \href{https://www.wikidata.org/wiki/Property:P166}{P166}	& \textit{Award received}: award or recognition received by a person, organisation.\\ \cline{2-3}
	& $R_2$ = \href{https://www.wikidata.org/wiki/Property:P166}{P166}	& \textit{Award received}: award or recognition received by a person, organisation.\\ \hline

	\multirow{2}{*}{21}	& $R_1$ = \href{https://www.wikidata.org/wiki/Property:P169}{P169}	& \textit{Chief executive officer}: the CEO within an organization.\\ \cline{2-3}
	& $R_2$ = \href{https://www.wikidata.org/wiki/Property:P112}{P112}	& \textit{Founded by}: founder or co-founder of this organization.\\ \hline
	\multirow{2}{*}{22}	& $R_1$ = \href{https://www.wikidata.org/wiki/Property:P169}{P169}	& \textit{Chief executive officer}: the CEO within an organization.\\ \cline{2-3}
	& $R_2$ = \href{https://www.wikidata.org/wiki/Property:P127}{P127}	& \textit{Owned by}: owner of the subject.\\ \hline
	\multirow{2}{*}{23}	& $R_1$ = \href{https://www.wikidata.org/wiki/Property:P169}{P169}	& \textit{Chief executive officer}: the CEO within an organization.\\ \cline{2-3}
	& $R_2$ = \href{https://www.wikidata.org/wiki/Property:P169}{P169}	& \textit{Chief executive officer}: the CEO within an organization.\\ \hline
	
	\multirow{2}{*}{24}	& $R_1$ = \href{https://www.wikidata.org/wiki/Property:P169}{P169}	& \textit{Chief executive officer}: the CEO within an organization.\\ \cline{2-3}
	& $R_2$ = \href{https://www.wikidata.org/wiki/Property:P3320}{P3320}	& \textit{Board member}: member(s) of the board for the organization.\\ \hline
	
	\multirow{2}{*}{25}	& $R_1$ = \href{https://www.wikidata.org/wiki/Property:P199}{P199}	& \textit{Business division}: divisions of this organization.\\ \cline{2-3}
	& $R_2$ = \href{https://www.wikidata.org/wiki/Property:P355}{P355}	& \textit{Subsidiary}: subsidiary of a company or organization.\\ \hline
	
	\multirow{2}{*}{26}	& $R_1$ = \href{https://www.wikidata.org/wiki/Property:P306}{P306}	& \textit{Operating system}: operating system (OS) on which a software works.\\ \cline{2-3}
	& $R_2$ = \href{https://www.wikidata.org/wiki/Property:P1056}{P1056}	& \textit{Product or material produced}: material or product produced by an agency.\\ \hline
	
	\multirow{2}{*}{27}	& $R_1$ = \href{https://www.wikidata.org/wiki/Property:P355}{P355}	& \textit{Subsidiary}: subsidiary of a company or organization.\\ \cline{2-3}
	& $R_2$ = \href{https://www.wikidata.org/wiki/Property:P127}{P127}	& \textit{Owned by}: owner of the subject.\\ \hline
	
	\multirow{2}{*}{28}	& $R_1$ = \href{https://www.wikidata.org/wiki/Property:P355}{P355}	& \textit{Subsidiary}: subsidiary of a company or organization.\\ \cline{2-3}
	& $R_2$ = \href{https://www.wikidata.org/wiki/Property:P155}{P155}	& \textit{Follows}: immediately prior item in a series of which the subject is a part.\\ \hline

	\multirow{2}{*}{29}	& $R_1$ = \href{https://www.wikidata.org/wiki/Property:P355}{P355}	& \textit{Subsidiary}: subsidiary of a company or organization.\\ \cline{2-3}
	& $R_2$ = \href{https://www.wikidata.org/wiki/Property:P199}{P199}	& \textit{Business division}: divisions of this organization.\\ \hline
	\multirow{2}{*}{30}	& $R_1$ = \href{https://www.wikidata.org/wiki/Property:P355}{P355}	& \textit{Subsidiary}: subsidiary of a company or organization.\\ \cline{2-3}
	& $R_2$ = \href{https://www.wikidata.org/wiki/Property:P355}{P355}	& \textit{Subsidiary}: subsidiary of a company or organization.\\ \hline
	\multirow{2}{*}{31}	& $R_1$ = \href{https://www.wikidata.org/wiki/Property:P361}{P361}	& \textit{Part of}: object of which the subject is a part.\\ \cline{2-3}
	& $R_2$ = \href{https://www.wikidata.org/wiki/Property:P361}{P361}	& \textit{Part of}: object of which the subject is a part.\\ \hline
	
	\multirow{2}{*}{32}	& $R_1$ = \href{https://www.wikidata.org/wiki/Property:P366}{P366}	& \textit{Use}: main use of the subject.\\ \cline{2-3}
	& $R_2$ = \href{https://www.wikidata.org/wiki/Property:P31}{P31}	& \textit{Instance of}: that class of which this subject is a particular example and member.\\ \hline
	
	\multirow{2}{*}{33}	& $R_1$ = \href{https://www.wikidata.org/wiki/Property:P400}{P400}	& \textit{Platform}: platform for which a work was developed or released.\\ \cline{2-3}
	& $R_2$ = \href{https://www.wikidata.org/wiki/Property:P1056}{P1056}	& \textit{Product or material produced}: material or product produced by an agency.\\ \hline
	
	\multirow{2}{*}{34}	& $R_1$ = \href{https://www.wikidata.org/wiki/Property:P452}{P452}	& \textit{Industry}: industry of company or organization.\\ \cline{2-3}
	& $R_2$ = \href{https://www.wikidata.org/wiki/Property:P31}{P31}	& \textit{Instance of}: that class of which this subject is a particular example and member.\\ \hline
	
	\multirow{2}{*}{35}	& $R_1$ = \href{https://www.wikidata.org/wiki/Property:P452}{P452}	& \textit{Industry}: industry of company or organization.\\ \cline{2-3}
	& $R_2$ = \href{https://www.wikidata.org/wiki/Property:P452}{P452}	& \textit{Industry}: industry of company or organization.\\ \hline
	
	\multirow{2}{*}{36}	& $R_1$ = \href{https://www.wikidata.org/wiki/Property:P452}{P452}	& \textit{Industry}: industry of company or organization.\\ \cline{2-3}
	& $R_2$ = \href{https://www.wikidata.org/wiki/Property:P1056}{P1056}	& \textit{Product or material produced}: material or product produced by an agency.\\ \hline
	
	\multirow{2}{*}{37}	& $R_1$ = \href{https://www.wikidata.org/wiki/Property:P452}{P452}	& \textit{Industry}: industry of company or organization.\\ \cline{2-3}
	& $R_2$ = \href{https://www.wikidata.org/wiki/Property:P2770}{P2770}	& \textit{Source of income}: source of income of an organization or person.\\ \hline
	
	\multirow{2}{*}{38}	& $R_1$ = \href{https://www.wikidata.org/wiki/Property:P463}{P463}	& \textit{Member of}: organization or club to which the subject belongs.\\ \cline{2-3}
	& $R_2$ = \href{https://www.wikidata.org/wiki/Property:P463}{P463}	& \textit{Member of}: organization or club to which the subject belongs.\\ \hline
	
	\multirow{2}{*}{39}	& $R_1$ = \href{https://www.wikidata.org/wiki/Property:P749}{P749}	& \textit{Parent organization}: parent organization of an organisation.\\ \cline{2-3}
	& $R_2$ = \href{https://www.wikidata.org/wiki/Property:P127}{P127}	& \textit{Owned by}: owner of the subject.\\ \hline
	\multirow{2}{*}{40}	& $R_1$ = \href{https://www.wikidata.org/wiki/Property:P749}{P749}	& \textit{Parent organization}: parent organization of an organisation.\\ \cline{2-3}
	& $R_2$ = \href{https://www.wikidata.org/wiki/Property:P1830}{P1830}	& \textit{Owner of}: entities owned by the subject.\\ \hline
	
	\multirow{2}{*}{41}	& $R_1$ = \href{https://www.wikidata.org/wiki/Property:P1056}{P1056}	& \textit{Product or material produced}: material or product produced by an agency.\\ \cline{2-3}
	& $R_2$ = \href{https://www.wikidata.org/wiki/Property:P31}{P31}	& \textit{Instance of}: that class of which this subject is a particular example and member.\\ \hline
	
	\multirow{2}{*}{42}	& $R_1$ = \href{https://www.wikidata.org/wiki/Property:P1056}{P1056}	& \textit{Product or material produced}: material or product produced by an agency.\\ \cline{2-3}
	& $R_2$ = \href{https://www.wikidata.org/wiki/Property:P121}{P121}	& \textit{Item operated}: equipment, installation or service operated by the subject.\\ \hline
	
	\multirow{2}{*}{43}	& $R_1$ = \href{https://www.wikidata.org/wiki/Property:P1056}{P1056}	& \textit{Product or material produced}: material or product produced by an agency.\\ \cline{2-3}
	& $R_2$ = \href{https://www.wikidata.org/wiki/Property:P306}{P306}	& \textit{Operating system}: operating system (OS) on which a software works.\\ \hline
	
	\multirow{2}{*}{44}	& $R_1$ = \href{https://www.wikidata.org/wiki/Property:P1056}{P1056}	& \textit{Product or material produced}: material or product produced by an agency.\\ \cline{2-3}
	& $R_2$ = \href{https://www.wikidata.org/wiki/Property:P400}{P400}	& \textit{Platform}: platform for which a work was developed or released.\\ \hline
	
	\multirow{2}{*}{45}	& $R_1$ = \href{https://www.wikidata.org/wiki/Property:P1056}{P1056}	& \textit{Product or material produced}: material or product produced by an agency.\\ \cline{2-3}
	& $R_2$ = \href{https://www.wikidata.org/wiki/Property:P452}{P452}	& \textit{Industry}: industry of company or organization.\\ \hline
	
	\multirow{2}{*}{46}	& $R_1$ = \href{https://www.wikidata.org/wiki/Property:P1056}{P1056}	& \textit{Product or material produced}: material or product produced by an agency.\\ \cline{2-3}
	& $R_2$ = \href{https://www.wikidata.org/wiki/Property:P1056}{P1056}	& \textit{Product or material produced}: material or product produced by an agency.\\ \hline
	
	\multirow{2}{*}{47}	& $R_1$ = \href{https://www.wikidata.org/wiki/Property:P1344}{P1344}	& \textit{Participant of}: event a person or an organization was a participant in.\\ \cline{2-3}
	& $R_2$ = \href{https://www.wikidata.org/wiki/Property:P1344}{P1344}	& \textit{Participant of}: event a person or an organization was a participant in.\\ \hline
	
	\multirow{2}{*}{48}	& $R_1$ = \href{https://www.wikidata.org/wiki/Property:P1830}{P1830}	& \textit{Owner of}: entities owned by the subject.\\ \cline{2-3}
	& $R_2$ = \href{https://www.wikidata.org/wiki/Property:P127}{P127}	& \textit{Owned by}: owner of the subject.\\ \hline
	\multirow{2}{*}{49}	& $R_1$ = \href{https://www.wikidata.org/wiki/Property:P1830}{P1830}	& \textit{Owner of}: entities owned by the subject.\\ \cline{2-3}
	& $R_2$ = \href{https://www.wikidata.org/wiki/Property:P749}{P749}	& \textit{Parent organization}: parent organization of an organisation.\\ \hline
	
	\multirow{2}{*}{50}	& $R_1$ = \href{https://www.wikidata.org/wiki/Property:P2770}{P2770}	& \textit{Source of income}: source of income of an organization or person.\\ \cline{2-3}
	& $R_2$ = \href{https://www.wikidata.org/wiki/Property:P452}{P452}	& \textit{Industry}: industry of company or organization.\\ \hline

	\multirow{2}{*}{51}	& $R_1$ = \href{https://www.wikidata.org/wiki/Property:P3320}{P3320}	& \textit{Board member}: member(s) of the board for the organization.\\ \cline{2-3}
	& $R_2$ = \href{https://www.wikidata.org/wiki/Property:P127}{P127}	& \textit{Owned by}: owner of the subject.\\ \hline
	\multirow{2}{*}{52}	& $R_1$ = \href{https://www.wikidata.org/wiki/Property:P3320}{P3320}	& \textit{Board member}: member(s) of the board for the organization.\\ \cline{2-3}
	& $R_2$ = \href{https://www.wikidata.org/wiki/Property:P169}{P169}	& \textit{Chief executive officer}: the CEO within an organization.\\ \hline
	
\end{longtable}
\normalsize

\begin{acks}
This research is part of NExT++ project, supported by the National Research Foundation, Prime Minister\'s Office, Singapore under its IRC@Singapore Funding Initiative.
\end{acks}

%% file: 0_main.bbl

\begin{thebibliography}{47}

%%% the Web address.

\ifx \showCODEN    \undefined \def \showCODEN     #1{\unskip}     \fi
\ifx \showDOI      \undefined \def \showDOI       #1{#1}\fi
\ifx \showISBNx    \undefined \def \showISBNx     #1{\unskip}     \fi
\ifx \showISBNxiii \undefined \def \showISBNxiii  #1{\unskip}     \fi
\ifx \showISSN     \undefined \def \showISSN      #1{\unskip}     \fi
\ifx \showLCCN     \undefined \def \showLCCN      #1{\unskip}     \fi
\ifx \shownote     \undefined \def \shownote      #1{#1}          \fi
\ifx \showarticletitle \undefined \def \showarticletitle #1{#1}   \fi
\ifx \showURL      \undefined \def \showURL       {\relax}        \fi
% invisible to TeX
\providecommand\bibfield[2]{#2}
\providecommand\bibinfo[2]{#2}
\providecommand\natexlab[1]{#1}
\providecommand\showeprint[2][]{arXiv:#2}

\bibitem[\protect\citeauthoryear{Adebiyi}{Adebiyi}{2014}]%
        {adebiyi2014comparison}
\bibfield{author}{\bibinfo{person}{Ayodele Ariyo et~al. Adebiyi}.}
  \bibinfo{year}{2014}\natexlab{}.
\newblock \showarticletitle{Comparison of ARIMA and artificial neural networks
  models for stock price prediction}.
\newblock \bibinfo{journal}{{\em Journal of Applied Mathematics\/}}
  \bibinfo{volume}{2014} (\bibinfo{year}{2014}).
\newblock


\bibitem[\protect\citeauthoryear{Aggarwal and Reddy}{Aggarwal and
  Reddy}{2013}]%
        {aggarwal2013data}
\bibfield{author}{\bibinfo{person}{Charu~C Aggarwal} {and}
  \bibinfo{person}{Chandan~K Reddy}.} \bibinfo{year}{2013}\natexlab{}.
\newblock \bibinfo{booktitle}{{\em Data clustering: algorithms and
  applications}}.
\newblock \bibinfo{publisher}{CRC press}.
\newblock


\bibitem[\protect\citeauthoryear{Alberg and Lipton}{Alberg and Lipton}{2017}]%
        {alberg2017improving}
\bibfield{author}{\bibinfo{person}{John Alberg} {and}
  \bibinfo{person}{Zachary~C Lipton}.} \bibinfo{year}{2017}\natexlab{}.
\newblock \showarticletitle{Improving Factor-Based Quantitative Investing by
  Forecasting Company Fundamentals}.
\newblock \bibinfo{journal}{{\em arXiv preprint arXiv:1711.04837\/}}
  (\bibinfo{year}{2017}).
\newblock


\bibitem[\protect\citeauthoryear{Bao}{Bao}{2017}]%
        {bao2017deep}
\bibfield{author}{\bibinfo{person}{Wei et~al. Bao}.}
  \bibinfo{year}{2017}\natexlab{}.
\newblock \showarticletitle{A deep learning framework for financial time series
  using stacked autoencoders and long-short term memory}.
\newblock \bibinfo{journal}{{\em PloS one\/}} \bibinfo{volume}{12},
  \bibinfo{number}{7} (\bibinfo{year}{2017}), \bibinfo{pages}{e0180944}.
\newblock


\bibitem[\protect\citeauthoryear{Bordes, Usunier, Garcia-Duran, Weston, and
  Yakhnenko}{Bordes et~al\mbox{.}}{2013}]%
        {bordes2013translating}
\bibfield{author}{\bibinfo{person}{Antoine Bordes}, \bibinfo{person}{Nicolas
  Usunier}, \bibinfo{person}{Alberto Garcia-Duran}, \bibinfo{person}{Jason
  Weston}, {and} \bibinfo{person}{Oksana Yakhnenko}.}
  \bibinfo{year}{2013}\natexlab{}.
\newblock \showarticletitle{Translating embeddings for modeling
  multi-relational data}. In \bibinfo{booktitle}{{\em NIPS}}.
  \bibinfo{pages}{2787--2795}.
\newblock


\bibitem[\protect\citeauthoryear{Bruna}{Bruna}{2014}]%
        {bruna2014spectral}
\bibfield{author}{\bibinfo{person}{Joan et~al. Bruna}.}
  \bibinfo{year}{2014}\natexlab{}.
\newblock \showarticletitle{Spectral networks and locally connected networks on
  graphs}. In \bibinfo{booktitle}{{\em ICLR}}.
\newblock


\bibitem[\protect\citeauthoryear{Cho}{Cho}{2014}]%
        {cho2014learning}
\bibfield{author}{\bibinfo{person}{Kyunghyun et~al. Cho}.}
  \bibinfo{year}{2014}\natexlab{}.
\newblock \showarticletitle{Learning phrase representations using RNN
  encoder-decoder for statistical machine translation}.
\newblock \bibinfo{journal}{{\em arXiv preprint arXiv:1406.1078\/}}
  (\bibinfo{year}{2014}).
\newblock


\bibitem[\protect\citeauthoryear{Defferrard, Bresson, and
  Vandergheynst}{Defferrard et~al\mbox{.}}{2016}]%
        {defferrard2016convolutional}
\bibfield{author}{\bibinfo{person}{Micha{\"e}l Defferrard},
  \bibinfo{person}{Xavier Bresson}, {and} \bibinfo{person}{Pierre
  Vandergheynst}.} \bibinfo{year}{2016}\natexlab{}.
\newblock \showarticletitle{Convolutional neural networks on graphs with fast
  localized spectral filtering}. In \bibinfo{booktitle}{{\em NIPS}}.
  \bibinfo{pages}{3844--3852}.
\newblock


\bibitem[\protect\citeauthoryear{Dixon}{Dixon}{2016}]%
        {dixon2016classification}
\bibfield{author}{\bibinfo{person}{Matthew et~al. Dixon}.}
  \bibinfo{year}{2016}\natexlab{}.
\newblock \showarticletitle{Classification-based financial markets prediction
  using deep neural networks}.
\newblock \bibinfo{journal}{{\em Algorithmic Finance\/}}
  \bibinfo{number}{Preprint} (\bibinfo{year}{2016}), \bibinfo{pages}{1--11}.
\newblock


\bibitem[\protect\citeauthoryear{Donnat, Zitnik, Hallac, and Leskovec}{Donnat
  et~al\mbox{.}}{2017}]%
        {donnat2017spectral}
\bibfield{author}{\bibinfo{person}{Claire Donnat}, \bibinfo{person}{Marinka
  Zitnik}, \bibinfo{person}{David Hallac}, {and} \bibinfo{person}{Jure
  Leskovec}.} \bibinfo{year}{2017}\natexlab{}.
\newblock \showarticletitle{Spectral Graph Wavelets for Structural Role
  Similarity in Networks}.
\newblock \bibinfo{journal}{{\em arXiv preprint arXiv:1710.10321\/}}
  (\bibinfo{year}{2017}).
\newblock


\bibitem[\protect\citeauthoryear{Feng}{Feng}{2017}]%
        {feng2017computational}
\bibfield{author}{\bibinfo{person}{Fuli et~al. Feng}.}
  \bibinfo{year}{2017}\natexlab{}.
\newblock \showarticletitle{Computational social indicators: a case study of
  chinese university ranking}. In \bibinfo{booktitle}{{\em SIGIR}}. ACM,
  \bibinfo{pages}{455--464}.
\newblock


\bibitem[\protect\citeauthoryear{Goller and Kuchler}{Goller and
  Kuchler}{1996}]%
        {goller1996learning}
\bibfield{author}{\bibinfo{person}{Christoph Goller} {and}
  \bibinfo{person}{Andreas Kuchler}.} \bibinfo{year}{1996}\natexlab{}.
\newblock \showarticletitle{Learning task-dependent distributed representations
  by backpropagation through structure}. In \bibinfo{booktitle}{{\em
  International Conference on Neural Networks}}, Vol.~\bibinfo{volume}{1}.
  IEEE, \bibinfo{pages}{347--352}.
\newblock


\bibitem[\protect\citeauthoryear{Graves, Mohamed, and Hinton}{Graves
  et~al\mbox{.}}{2013}]%
        {graves2013speech}
\bibfield{author}{\bibinfo{person}{Alex Graves}, \bibinfo{person}{Abdel-rahman
  Mohamed}, {and} \bibinfo{person}{Geoffrey Hinton}.}
  \bibinfo{year}{2013}\natexlab{}.
\newblock \showarticletitle{Speech recognition with deep recurrent neural
  networks}. In \bibinfo{booktitle}{{\em ICASSP}}. IEEE,
  \bibinfo{pages}{6645--6649}.
\newblock


\bibitem[\protect\citeauthoryear{Hammond, Vandergheynst, and Gribonval}{Hammond
  et~al\mbox{.}}{2011}]%
        {hammond2011wavelets}
\bibfield{author}{\bibinfo{person}{David~K Hammond}, \bibinfo{person}{Pierre
  Vandergheynst}, {and} \bibinfo{person}{R{\'e}mi Gribonval}.}
  \bibinfo{year}{2011}\natexlab{}.
\newblock \showarticletitle{Wavelets on graphs via spectral graph theory}.
\newblock \bibinfo{journal}{{\em Applied and Computational Harmonic
  Analysis\/}} \bibinfo{volume}{30}, \bibinfo{number}{2}
  (\bibinfo{year}{2011}), \bibinfo{pages}{129--150}.
\newblock


\bibitem[\protect\citeauthoryear{He}{He}{2017}]%
        {he2017neural}
\bibfield{author}{\bibinfo{person}{Xiangnan et~al. He}.}
  \bibinfo{year}{2017}\natexlab{}.
\newblock \showarticletitle{Neural collaborative filtering}. In
  \bibinfo{booktitle}{{\em WWW}}. International World Wide Web Conferences
  Steering Committee, \bibinfo{pages}{173--182}.
\newblock


\bibitem[\protect\citeauthoryear{Hochreiter and Schmidhuber}{Hochreiter and
  Schmidhuber}{1997}]%
        {hochreiter1997long}
\bibfield{author}{\bibinfo{person}{Sepp Hochreiter} {and}
  \bibinfo{person}{J{\"u}rgen Schmidhuber}.} \bibinfo{year}{1997}\natexlab{}.
\newblock \showarticletitle{Long short-term memory}.
\newblock \bibinfo{journal}{{\em Neural computation\/}} \bibinfo{volume}{9},
  \bibinfo{number}{8} (\bibinfo{year}{1997}), \bibinfo{pages}{1735--1780}.
\newblock


\bibitem[\protect\citeauthoryear{Hu}{Hu}{2018}]%
        {hu2018listening}
\bibfield{author}{\bibinfo{person}{Ziniu et~al. Hu}.}
  \bibinfo{year}{2018}\natexlab{}.
\newblock \showarticletitle{Listening to Chaotic Whispers: A Deep Learning
  Framework for News-oriented Stock Trend Prediction}. In
  \bibinfo{booktitle}{{\em WSDM}}. ACM, \bibinfo{pages}{403--412}.
\newblock


\bibitem[\protect\citeauthoryear{Jiang}{Jiang}{2016}]%
        {jiang2016learning}
\bibfield{author}{\bibinfo{person}{Shan et~al. Jiang}.}
  \bibinfo{year}{2016}\natexlab{}.
\newblock \showarticletitle{Learning query and document relevance from a
  web-scale click graph}. In \bibinfo{booktitle}{{\em SIGIR}}. ACM,
  \bibinfo{pages}{185--194}.
\newblock


\bibitem[\protect\citeauthoryear{Kingma and Ba}{Kingma and Ba}{2014}]%
        {kingma2014adam}
\bibfield{author}{\bibinfo{person}{Diederik~P Kingma} {and}
  \bibinfo{person}{Jimmy Ba}.} \bibinfo{year}{2014}\natexlab{}.
\newblock \showarticletitle{Adam: A method for stochastic optimization}.
\newblock \bibinfo{journal}{{\em arXiv preprint arXiv:1412.6980\/}}
  (\bibinfo{year}{2014}).
\newblock


\bibitem[\protect\citeauthoryear{Kipf and Welling}{Kipf and Welling}{2017}]%
        {kipf2017semi}
\bibfield{author}{\bibinfo{person}{Thomas~N Kipf} {and} \bibinfo{person}{Max
  Welling}.} \bibinfo{year}{2017}\natexlab{}.
\newblock \showarticletitle{Semi-supervised classification with graph
  convolutional networks}.
\newblock \bibinfo{journal}{{\em ICLR\/}} (\bibinfo{year}{2017}).
\newblock


\bibitem[\protect\citeauthoryear{Krizhevsky}{Krizhevsky}{2012}]%
        {krizhevsky2012imagenet}
\bibfield{author}{\bibinfo{person}{Alex et~al. Krizhevsky}.}
  \bibinfo{year}{2012}\natexlab{}.
\newblock \showarticletitle{Imagenet classification with deep convolutional
  neural networks}. In \bibinfo{booktitle}{{\em NIPS}}.
  \bibinfo{pages}{1097--1105}.
\newblock


\bibitem[\protect\citeauthoryear{Kumar and Ravi}{Kumar and Ravi}{2016}]%
        {kumar2016survey}
\bibfield{author}{\bibinfo{person}{B~Shravan Kumar} {and}
  \bibinfo{person}{Vadlamani Ravi}.} \bibinfo{year}{2016}\natexlab{}.
\newblock \showarticletitle{A survey of the applications of text mining in
  financial domain}.
\newblock \bibinfo{journal}{{\em Knowledge-Based Systems\/}}
  \bibinfo{volume}{114} (\bibinfo{year}{2016}), \bibinfo{pages}{128--147}.
\newblock


\bibitem[\protect\citeauthoryear{Li}{Li}{2016}]%
        {li2016tensor}
\bibfield{author}{\bibinfo{person}{Qing et~al. Li}.}
  \bibinfo{year}{2016}\natexlab{}.
\newblock \showarticletitle{A tensor-based information framework for predicting
  the stock market}.
\newblock \bibinfo{journal}{{\em TOIS\/}} \bibinfo{volume}{34},
  \bibinfo{number}{2} (\bibinfo{year}{2016}), \bibinfo{pages}{11}.
\newblock


\bibitem[\protect\citeauthoryear{Lo and MacKinlay}{Lo and MacKinlay}{2002}]%
        {lo2002non}
\bibfield{author}{\bibinfo{person}{Andrew~W Lo} {and} \bibinfo{person}{A~Craig
  MacKinlay}.} \bibinfo{year}{2002}\natexlab{}.
\newblock \bibinfo{booktitle}{{\em A non-random walk down Wall Street}}.
\newblock \bibinfo{publisher}{Princeton University Press}.
\newblock


\bibitem[\protect\citeauthoryear{Maas, Hannun, and Ng}{Maas
  et~al\mbox{.}}{2013}]%
        {maas2013rectifier}
\bibfield{author}{\bibinfo{person}{Andrew~L Maas}, \bibinfo{person}{Awni~Y
  Hannun}, {and} \bibinfo{person}{Andrew~Y Ng}.}
  \bibinfo{year}{2013}\natexlab{}.
\newblock \showarticletitle{Rectifier nonlinearities improve neural network
  acoustic models}. In \bibinfo{booktitle}{{\em Proc. icml}},
  Vol.~\bibinfo{volume}{30}. \bibinfo{pages}{3}.
\newblock


\bibitem[\protect\citeauthoryear{Mei, Rui, Li, and Tian}{Mei
  et~al\mbox{.}}{2014}]%
        {mei2014multimedia}
\bibfield{author}{\bibinfo{person}{Tao Mei}, \bibinfo{person}{Yong Rui},
  \bibinfo{person}{Shipeng Li}, {and} \bibinfo{person}{Qi Tian}.}
  \bibinfo{year}{2014}\natexlab{}.
\newblock \showarticletitle{Multimedia search reranking: A literature survey}.
\newblock \bibinfo{journal}{{\em ACM Computing Surveys (CSUR)\/}}
  \bibinfo{volume}{46}, \bibinfo{number}{3} (\bibinfo{year}{2014}),
  \bibinfo{pages}{38}.
\newblock


\bibitem[\protect\citeauthoryear{Musgrave}{Musgrave}{1997}]%
        {musgrave1997random}
\bibfield{author}{\bibinfo{person}{Gerald~L Musgrave}.}
  \bibinfo{year}{1997}\natexlab{}.
\newblock \showarticletitle{A Random Walk Down Wall Street}.
\newblock \bibinfo{journal}{{\em Business Economics\/}} \bibinfo{volume}{32},
  \bibinfo{number}{2} (\bibinfo{year}{1997}), \bibinfo{pages}{74--76}.
\newblock


\bibitem[\protect\citeauthoryear{Nassirtoussi}{Nassirtoussi}{2014}]%
        {nassirtoussi2014text}
\bibfield{author}{\bibinfo{person}{Arman Khadjeh et~al. Nassirtoussi}.}
  \bibinfo{year}{2014}\natexlab{}.
\newblock \showarticletitle{Text mining for market prediction: A systematic
  review}.
\newblock \bibinfo{journal}{{\em Expert Systems with Applications\/}}
  \bibinfo{volume}{41}, \bibinfo{number}{16} (\bibinfo{year}{2014}),
  \bibinfo{pages}{7653--7670}.
\newblock


\bibitem[\protect\citeauthoryear{Nguyen and Shirai}{Nguyen and Shirai}{2015}]%
        {nguyen2015topic}
\bibfield{author}{\bibinfo{person}{Thien~Hai Nguyen} {and}
  \bibinfo{person}{Kiyoaki Shirai}.} \bibinfo{year}{2015}\natexlab{}.
\newblock \showarticletitle{Topic modeling based sentiment analysis on social
  media for stock market prediction}. In \bibinfo{booktitle}{{\em ACL}},
  Vol.~\bibinfo{volume}{1}. \bibinfo{pages}{1354--1364}.
\newblock


\bibitem[\protect\citeauthoryear{Omari}{Omari}{2016}]%
        {omari2016novelty}
\bibfield{author}{\bibinfo{person}{Adi et~al. Omari}.}
  \bibinfo{year}{2016}\natexlab{}.
\newblock \showarticletitle{Novelty Based Ranking of Human Answers for
  Community Questions}. In \bibinfo{booktitle}{{\em SIGIR}}.
  \bibinfo{publisher}{ACM}, \bibinfo{pages}{215--224}.
\newblock


\bibitem[\protect\citeauthoryear{Page}{Page}{1999}]%
        {page1999pagerank}
\bibfield{author}{\bibinfo{person}{Lawrence et~al. Page}.}
  \bibinfo{year}{1999}\natexlab{}.
\newblock \bibinfo{booktitle}{{\em The PageRank citation ranking: Bringing
  order to the web.}}
\newblock \bibinfo{type}{{T}echnical {R}eport}. \bibinfo{institution}{Stanford
  InfoLab}.
\newblock


\bibitem[\protect\citeauthoryear{Preethi and Santhi}{Preethi and
  Santhi}{2012}]%
        {preethi2012stock}
\bibfield{author}{\bibinfo{person}{G Preethi} {and} \bibinfo{person}{B
  Santhi}.} \bibinfo{year}{2012}\natexlab{}.
\newblock \showarticletitle{STOCK MARKET FORECASTING TECHNIQUES: A SURVEY.}
\newblock \bibinfo{journal}{{\em Journal of Theoretical \& Applied Information
  Technology\/}} \bibinfo{volume}{46}, \bibinfo{number}{1}
  (\bibinfo{year}{2012}).
\newblock


\bibitem[\protect\citeauthoryear{Schwert}{Schwert}{2002}]%
        {schwert2002stock}
\bibfield{author}{\bibinfo{person}{G~William Schwert}.}
  \bibinfo{year}{2002}\natexlab{}.
\newblock \showarticletitle{Stock volatility in the new millennium: how wacky
  is Nasdaq?}
\newblock \bibinfo{journal}{{\em Journal of Monetary Economics\/}}
  \bibinfo{volume}{49}, \bibinfo{number}{1} (\bibinfo{year}{2002}),
  \bibinfo{pages}{3--26}.
\newblock


\bibitem[\protect\citeauthoryear{Socher}{Socher}{2013}]%
        {socher2013reasoning}
\bibfield{author}{\bibinfo{person}{Richard et~al. Socher}.}
  \bibinfo{year}{2013}\natexlab{}.
\newblock \showarticletitle{Reasoning with neural tensor networks for knowledge
  base completion}. In \bibinfo{booktitle}{{\em NIPS}}.
  \bibinfo{pages}{926--934}.
\newblock


\bibitem[\protect\citeauthoryear{Srivastava, Mansimov, and
  Salakhudinov}{Srivastava et~al\mbox{.}}{2015}]%
        {srivastava2015unsupervised}
\bibfield{author}{\bibinfo{person}{Nitish Srivastava}, \bibinfo{person}{Elman
  Mansimov}, {and} \bibinfo{person}{Ruslan Salakhudinov}.}
  \bibinfo{year}{2015}\natexlab{}.
\newblock \showarticletitle{Unsupervised learning of video representations
  using lstms}. In \bibinfo{booktitle}{{\em ICML}}. \bibinfo{pages}{843--852}.
\newblock


\bibitem[\protect\citeauthoryear{Tu}{Tu}{2016}]%
        {tu2016investment}
\bibfield{author}{\bibinfo{person}{Wenting et~al. Tu}.}
  \bibinfo{year}{2016}\natexlab{}.
\newblock \showarticletitle{Investment recommendation using investor opinions
  in social media}. In \bibinfo{booktitle}{{\em SIGIR}}. ACM,
  \bibinfo{pages}{881--884}.
\newblock


\bibitem[\protect\citeauthoryear{Vrande{\v{c}}i{\'c} and
  Kr{\"o}tzsch}{Vrande{\v{c}}i{\'c} and Kr{\"o}tzsch}{2014}]%
        {vrandevcic2014wikidata}
\bibfield{author}{\bibinfo{person}{Denny Vrande{\v{c}}i{\'c}} {and}
  \bibinfo{person}{Markus Kr{\"o}tzsch}.} \bibinfo{year}{2014}\natexlab{}.
\newblock \showarticletitle{Wikidata: a free collaborative knowledgebase}.
\newblock \bibinfo{journal}{{\it Commun. ACM}} \bibinfo{volume}{57},
  \bibinfo{number}{10} (\bibinfo{year}{2014}), \bibinfo{pages}{78--85}.
\newblock


\bibitem[\protect\citeauthoryear{Weimer}{Weimer}{2007}]%
        {Weimer:2007}
\bibfield{author}{\bibinfo{person}{Markus et~al. Weimer}.}
  \bibinfo{year}{2007}\natexlab{}.
\newblock \showarticletitle{COFIRANK Maximum Margin Matrix Factorization for
  Collaborative Ranking}. In \bibinfo{booktitle}{{\em NIPS}}.
  \bibinfo{pages}{1593--1600}.
\newblock


\bibitem[\protect\citeauthoryear{Xu and Zhang}{Xu and Zhang}{2015}]%
        {xu2015application}
\bibfield{author}{\bibinfo{person}{Yan Xu} {and} \bibinfo{person}{Guosheng
  Zhang}.} \bibinfo{year}{2015}\natexlab{}.
\newblock \showarticletitle{Application of Kalman Filter in the Prediction of
  Stock Price}. In \bibinfo{booktitle}{{\em KAM}}.
\newblock


\bibitem[\protect\citeauthoryear{Yan, Song, and Wu}{Yan et~al\mbox{.}}{2016}]%
        {yan2016learning}
\bibfield{author}{\bibinfo{person}{Rui Yan}, \bibinfo{person}{Yiping Song},
  {and} \bibinfo{person}{Hua Wu}.} \bibinfo{year}{2016}\natexlab{}.
\newblock \showarticletitle{Learning to respond with deep neural networks for
  retrieval-based human-computer conversation system}. In
  \bibinfo{booktitle}{{\em Proceedings of the 39th ACM SIGIR}}. ACM,
  \bibinfo{pages}{55--64}.
\newblock


\bibitem[\protect\citeauthoryear{Yu, Qiu, Wen, Lin, and Liu}{Yu
  et~al\mbox{.}}{2016}]%
        {yu2016survey}
\bibfield{author}{\bibinfo{person}{Rose Yu}, \bibinfo{person}{Huida Qiu},
  \bibinfo{person}{Zhen Wen}, \bibinfo{person}{ChingYung Lin}, {and}
  \bibinfo{person}{Yan Liu}.} \bibinfo{year}{2016}\natexlab{}.
\newblock \showarticletitle{A survey on social media anomaly detection}.
\newblock \bibinfo{journal}{{\em SIGKDD\/}} \bibinfo{volume}{18},
  \bibinfo{number}{1} (\bibinfo{year}{2016}), \bibinfo{pages}{1--14}.
\newblock


\bibitem[\protect\citeauthoryear{Zhang}{Zhang}{2017a}]%
        {zhang2017stock}
\bibfield{author}{\bibinfo{person}{Liheng et~al. Zhang}.}
  \bibinfo{year}{2017}\natexlab{a}.
\newblock \showarticletitle{Stock Price Prediction via Discovering
  Multi-Frequency Trading Patterns}. In \bibinfo{booktitle}{{\em SIGKDD}}. ACM,
  \bibinfo{pages}{2141--2149}.
\newblock


\bibitem[\protect\citeauthoryear{Zhang}{Zhang}{2017b}]%
        {zhang2017learning}
\bibfield{author}{\bibinfo{person}{Yao et~al. Zhang}.}
  \bibinfo{year}{2017}\natexlab{b}.
\newblock \showarticletitle{Learning Node Embeddings in Interaction Graphs}. In
  \bibinfo{booktitle}{{\em CIKM}}. \bibinfo{publisher}{ACM},
  \bibinfo{pages}{397--406}.
\newblock


\bibitem[\protect\citeauthoryear{Zhao}{Zhao}{2017}]%
        {zhao2017constructing}
\bibfield{author}{\bibinfo{person}{Sendong et~al. Zhao}.}
  \bibinfo{year}{2017}\natexlab{}.
\newblock \showarticletitle{Constructing and embedding abstract event causality
  networks from text snippets}. In \bibinfo{booktitle}{{\em WSDM}}. ACM,
  \bibinfo{pages}{335--344}.
\newblock


\bibitem[\protect\citeauthoryear{Zheng}{Zheng}{2007}]%
        {zheng2007regression}
\bibfield{author}{\bibinfo{person}{Zhaohui et~al. Zheng}.}
  \bibinfo{year}{2007}\natexlab{}.
\newblock \showarticletitle{A regression framework for learning ranking
  functions using relative relevance judgments}. In \bibinfo{booktitle}{{\em
  SIGIR}}. ACM, \bibinfo{pages}{287--294}.
\newblock


\bibitem[\protect\citeauthoryear{Zhou}{Zhou}{2007}]%
        {zhou2007learning}
\bibfield{author}{\bibinfo{person}{Denny et~al. Zhou}.}
  \bibinfo{year}{2007}\natexlab{}.
\newblock \showarticletitle{Learning with hypergraphs: Clustering,
  classification, and embedding}. In \bibinfo{booktitle}{{\em NIPS}}.
  \bibinfo{pages}{1601--1608}.
\newblock


\bibitem[\protect\citeauthoryear{Zhu}{Zhu}{2003}]%
        {zhu2003semi}
\bibfield{author}{\bibinfo{person}{Xiaojin et~al. Zhu}.}
  \bibinfo{year}{2003}\natexlab{}.
\newblock \showarticletitle{Semi-supervised learning using gaussian fields and
  harmonic functions}. In \bibinfo{booktitle}{{\em ICML}}.
  \bibinfo{pages}{912--919}.
\newblock


\end{thebibliography}
